\pdfoutput=1
\documentclass{article}

\usepackage{ifthen}
\usepackage{xcolor}

\newboolean{arxiv}
\newboolean{sisc}
\newboolean{toms}
\newboolean{appendSupplementMaterialAsAppendix}
\newboolean{markNewContentA}
\newboolean{markNewContentB}
\newboolean{markNewContentC}

\setboolean{arxiv}{true}
\setboolean{sisc}{false}
\setboolean{toms}{false}
\setboolean{appendSupplementMaterialAsAppendix}{true}
\setboolean{markNewContentA}{false}
\setboolean{markNewContentB}{false}
\setboolean{markNewContentC}{false}

\ifthenelse{\boolean{markNewContentA}}{
  \newcommand{\newA}[1]{\textcolor{blue}{#1}}
}{
  \newcommand{\newA}[1]{#1}
}

\ifthenelse{\boolean{markNewContentB}}{
  \newcommand{\newB}[1]{\textcolor{blue}{#1}}
}{
  \newcommand{\newB}[1]{#1}
}

\ifthenelse{\boolean{markNewContentC}}{
  \newcommand{\newC}[1]{\textcolor{cyan}{#1}}
}{
  \newcommand{\newC}[1]{#1}
}

\usepackage{graphicx}
\usepackage[numbers,sort&compress]{natbib}

\newcommand{\TheTitle}{The Peano software---parallel, automaton-based, dynamically adaptive grid traversals}

\newcommand{\TheAuthors}{T.~Weinzierl}

\usepackage{algorithm}
\usepackage{algpseudocode}

\usepackage{enumitem}

\newcounter{designdecisioncounter} 
\newenvironment{designdecision}[1][Design decision]
{\vspace{0.2cm}\noindent\textbf{#1 \arabic{designdecisioncounter}. }\
\stepcounter{designdecisioncounter} }{\vspace{0.2cm}}

\ifthenelse{\boolean{toms}}{
}{
 \newcounter{examplecounter} 
 \setcounter{examplecounter}{1}
 \newenvironment{example}[1][Example]
 {\vspace{0.2cm}\noindent\textbf{#1 \arabic{examplecounter}. }\
 \stepcounter{examplecounter} }{\vspace{0.2cm}}
}

\newcounter{observationcounter} 
\newenvironment{observation}[1][Observation]
{\vspace{0.2cm}\noindent\textbf{#1 \arabic{observationcounter}. }\
\stepcounter{observationcounter} }{\vspace{0.2cm}}

\ifthenelse{\boolean{arxiv}}{
  \usepackage{geometry}
\geometry{left=4.1cm,textwidth=14.2cm,top=2cm,textheight=24cm}
\title{\TheTitle}

\author{
  Tobias Weinzierl
  \thanks{
    Department of Computer Science, Durham University,
    UK
    (tobias.weinzierl@durham.ac.uk)
  }
}

}{} 
\ifthenelse{\boolean{sisc}}{
\headers{Parallel, automaton-based AMR}{\TheAuthors}

\title{\TheTitle}

\Crefname{ALC@unique}{Line}{Lines}

\author{
  Tobias Weinzierl
  \thanks{
    School of Engineering and Computing Sciences, Durham University
    (\email{tobias.weinzierl@dur.ac.uk})
  }
}

\slugger{sisc}{xxxx}{xx}{x}{x--x}%slugger should be set to mms, siap, sicomp,

\setlength\floatsep{0.4\baselineskip}
\setlength\textfloatsep{0.4\baselineskip}

}{} 
\ifthenelse{\boolean{toms}}{
\doi{0000001.0000001}

\issn{1234-56789}

\title{\TheTitle}

\markboth{\TheAuthors}{\TheTitle}

\author{
  TOBIAS WEINZIERL
  \affil{
    Department of Computer Science, 
    Durham University
  }
}

\acmformat{
  Tobias Weinzierl, 2017. \TheTitle
 }

\category{}{Mathematics of computing}{Mathematical software}
\category{}{Computing methodologies}{Modeling and simulation}[Multiscale systems]

}{}

\begin{document}

%
% ACM needs it the other way round than arXiv
%
\ifthenelse{\boolean{toms}}{
  \begin{abstract}
We discuss the design decisions, design alternatives and rationale behind
the third generation of Peano, a framework for dynamically
adaptive Cartesian meshes derived from spacetrees.
Peano ties the mesh traversal to the mesh storage and supports only one
element-wise traversal order resulting from space-filling curves.
The user is not free to choose a traversal order herself. 
The traversal can exploit regular
grid subregions and shared memory as well as distributed memory systems with
almost no modifications to a serial application code.
\newB{We formalize}
the software design \newB{by means} of two
interacting automata---one automaton for the multiscale grid traversal and one
for the application-specific algorithmic steps.
\newB{This yields a} callback-based programming paradigm.
\newB{We 
 \newC{further}
sketch the} supported application
types and the two data storage schemes realized, before we detail high-performance computing
aspects and lessons learned.
Special emphasis is put on observations regarding the
used programming \newB{idioms} and algorithmic concepts.
This transforms our report from a ``one way to implement things''
code description into a generic 
\newC{discussion and summary of some}
 alternatives, rationale \newC{and}
design decisions to be made for any tree-based adaptive
mesh refinement software.
\end{abstract}

  \maketitle
}{
  \maketitle
  \begin{abstract}
We discuss the design decisions, design alternatives and rationale behind
the third generation of Peano, a framework for dynamically
adaptive Cartesian meshes derived from spacetrees.
Peano ties the mesh traversal to the mesh storage and supports only one
element-wise traversal order resulting from space-filling curves.
The user is not free to choose a traversal order herself. 
The traversal can exploit regular
grid subregions and shared memory as well as distributed memory systems with
almost no modifications to a serial application code.
\newB{We formalize}
the software design \newB{by means} of two
interacting automata---one automaton for the multiscale grid traversal and one
for the application-specific algorithmic steps.
\newB{This yields a} callback-based programming paradigm.
\newB{We 
 \newC{further}
sketch the} supported application
types and the two data storage schemes realized, before we detail high-performance computing
aspects and lessons learned.
Special emphasis is put on observations regarding the
used programming \newB{idioms} and algorithmic concepts.
This transforms our report from a ``one way to implement things''
code description into a generic 
\newC{discussion and summary of some}
 alternatives, rationale \newC{and}
design decisions to be made for any tree-based adaptive
mesh refinement software.
\end{abstract}

} 

\ifthenelse{\boolean{sisc}}{
 
}
{}

\ifthenelse{\boolean{toms}}{
 \begin{bottomstuff}
%  All underlying software is open source and available at
%  \cite{Software:Peano}.
%
   Author's addresses: 
   T. Weinzierl, \newA{Department of Computer Science}, Durham University,
   DH1 3LE Durham, United Kingdom
 \end{bottomstuff}
}{}

\section{Introduction}

%
% Problem field; also the keywords from the call
%
Dynamically adaptive grids are mortar and catalyst of mesh-based scientific
computing and thus important to a large range of scientific and engineering
applications.
They enable scientists and engineers to solve problems with high
accuracy as they invest grid entities and computational effort
where they pay off most.
Naturally, these regions may change \newA{over} time for time-dependent
problems and may change throughout a solve process due to error estimators.
\newB{The}
design of meshing software for dynamically adaptive grids \newB{is non-trivial
since it}
has
to support a magnitude of advanced algorithmic building blocks such as
discretization, refinement criteria, solver steps or visualization.
It has to facilitate functional diversity and be accessible.
At the same time, mesh storage, administration and processing have to meet
efficiency and concurrency requirements of high-performance computing (HPC).
One specific yet popular meshing paradigm that tries to meet these criteria
and is subject of study here are spacetrees.
\newB{We use the term as generalization for quadtrees, octrees} \newC{and
related data structures.} 
They yield adaptive Cartesian
grids
\cite{Burstedde:11:p4est,Gadeschi:15:HierachicalCartesianGrids,Griebel:99:SFCAndMultigrid,Khokhlov:98:FullyThreadedTree,Lashuk:12:MultipoleOnHeterogeneous,Robey:13:Hashing,Sampath:08:Dendro,Sundar:08:BalancedOctrees,Tumblin:15:CompactHashing,Weinzierl:11:Peano}.

%
% Ziel
%
This paper \newA{orbits around} the spacetree software Peano
\cite{Software:Peano} that is available in its third generation. 
The first generation \newB{was} a set of proof-of-concept codes tackling various
challenges with different core code routines
(\cite{Bungartz:06:Parallel,Mehl:06:MG}, e.g.).
\newB{A} second generation \newB{fused} these concepts into one code base made
primarily for CFD-type applications though some generic software rationale
\newB{have been} reported
(\cite{Bungartz:10:PDEFramework,Bungartz:11:Multiphase,Weinzierl:09:Diss,Weinzierl:11:Peano} and others).
The third generation is a complete rewrite that focuses on an abstraction from
any application, offers different memory organization schemes and emphasizes
scalability.
Our objective is two-fold:
On the one hand, we want to present this third generation of the software as
well as the underlying rationale and, thus, sketch its potential.
While Peano is used as \newA{workhorse} for our own projects
(\cite{Eckhardt:10:Blocking,Reps:16:Helmholtz,Schreiber:13:sfc-based,Software:ExaHyPE,Weinzierl:14:BlockFusion,Weinzierl:16:PIC,Weinzierl:17:BoxMG},
e.g.), making it freely available allows other groups to benefit from the
development effort, too.
For this, however, its design philosophy has to be described explicitly.
\newB{Academics} 
have to be enabled to assess whether it suits particular needs
and \newB{whether} certain design decisions make it inferior or superior to
other codes.
On the other hand, we want to identify and outline design and
realization patterns \cite{Gamma:94:DesignPatterns} for tree-based adaptive mesh
refinement (AMR).
Given the popularity of this meshing paradigm, our manuscript brings together
and compares fundamental decisions to be made by any developer.
Many of these comparisons and classifications have, to the best of our
knowledge, not been done before in a concentrated effort.

%
% Was machen wir, was machen andere
%
While we work application-generic, \newB{Peano does} not strive for 
the flexibility and diversity along the lines of 
other solutions such as 
\cite{Software:Chombo,Software:dealii,Bastian:08:Dune1,Burstedde:11:p4est,Davison:00:Uintah,Feichtinger:11:Walberla,Sampath:08:Dendro,Teunissen:17:Afivo}
or \newB{other projects such as described}
\newC{and cited in \cite{Deiterding:05:AMR,Dreher:05:Racoon,Dubey:16:SAMR}.
This list is not comprehensive.}
%\cite{Software:Chombo,Software:dealii,Software:Uintah,Bastian:08:Dune1,Burstedde:11:p4est,Feichtinger:11:Walberla,Sampath:08:Dendro,Teunissen:17:Afivo}
% and others}.
Our approach ties the mesh traversal and programming model to the mesh
structure:
the user is not allowed to navigate through the grid freely.
Instead, the grid traversal, i.e.~the sequence in which
grid entities are processes, is prescribed.
This poses a restriction
that may force algorithms to be redesigned, but it allows \newB{users} to focus 
\newC{on} which algorithmic steps are mapped onto the processing of which grid
entity (vertex or cell, e.g.) and which temporal and data dependencies between these steps
exist.
How the processing is realized is hidden.
This picks up recent trends in task-based computing and
has successfully been used in various spacetree
and non-spacetree codes
such as \cite{Meister:12:Software,Weinbub:14:ViennaX}.
A minimalist set of constraints 
\newB{prescribing a}
partial order allows us to tune the grid traversal.
\newA{It further enables us} to hide concurrent execution from the
application codes.
It is one goal of the manuscript to highlight how such a
restrictive programming model interacts with other well-known concepts such as 
spacetree linearization based upon space-filling curves
\cite{Sundar:08:BalancedOctrees},
various discretization and data modeling choices,
persistent data storage,
domain decomposition and task processing.
Since the traversal invokes user-defined operations and disallows \newA{users}
to control the traversal, our programming paradigm can be summarizes by the Hollywood principle: 
Don't call us, we call you \cite{Sweet:85:HollywoodPrinciple}.

%
% What we do should not be in the cover letter but in the introduction
%
% A discussion of the programming model requires us to uncover and categorise the
% underlying design choices made \newB{with respect to}~data structures, ordering and access.
\newB{Many other implementation}
choices would have been possible, so a
description of the chosen approaches and the motivation is of use to
other spacetree code developers facing similar challenges.
\newB{
Our paper contains a brief introduction of the grid data structure and the
two grid enumeration schemes we rely on.
The main part of the manuscript starts from a description of our
callback-based programming model (Section \ref{section:api}).
This allows us to clarify which classes of applications are supported by
the code as well as application limitations.
In Section \ref{section:data-management}, we review some data storage
paradigms for spacetrees and classify the two storage variants offered by Peano:
stream- and heap-based persistency.
}
Total linearization of the tree through space-filling curves (SFCs)
enables us to encode the grid data and adjacency with minimal memory and to stream it 
%continuously 
through the processor.
Holding a tree as stream rather than a pointer data structure is a popular
technique and, thus, can be called a design pattern \cite{Gamma:94:DesignPatterns}.
Our manuscript applies the well-known linearization \newB{pattern} to multilevel
grids and details the relation to depth-first (DFS) and breadth-first (BFS) searches.
\newB{While depth-first} linearization yields vertically integrated
\cite{Adams:15:Segmental} grid traversals---multiple levels are processed
\newB{by one mesh sweep}---it
% ---which yield excellent memory access characteristics through data
% accesses with oblivious spatial and temporal access locality \cite{Kowarschik:03:CacheTechniquesOverview}, while
% most tree codes focus on the handling of the finest tessellation resolutions only.
lacks concurrency for traditional stencil codes and we thus derive a
hybrid of DFS and BFS.
Three sections are dedicated to a discussion of memory movement and
administration minimization (Section \ref{section:overhead-reduction})
as well as the supported \newB{shared memory (Section
\ref{section:shared-memory}) and
distributed (Section \ref{section:mpi}) parallelization}.
Special emphasis here is put on the interplay of dependencies with
spacetree traversal strategies as well as the programming interface.
Notably, we discuss two competing multiscale data splitting strategies,
synchronization-avoiding application interfaces and a constraint technique that
can guide a task-based \newA{parallelization}.
% \newB{Yet,} we never set up any task
% graph explicitly.
Some experiments highlight properties of the proposed realization ideas
before an outlook closes the discussion.

% We detail from hereon further design decisions, notably the
% embedding or linking of application data with the tree structure (our results
% clarify that a decision which variant is superior depends on the
% data cardinality per grid entity), variants of tree traversal orders and their
% concurrency (while DFS yields vertically integrated traversals,
% it lacks obvious concurrency for traditional stencil codes and we thus derive a
% hybrid of DFS and BFS) and the decomposition of multilevel tree data over
% multiple compute ranks.
% Our \newB{decomposition} avoids, different to other published approaches, the
% replication of coarse resolution data.

\section{Spacetree definition and enumeration paradigms}
\label{section:spacetree}

This paper stands on the shoulders of spacetrees \newB{as generalization of the
quadtree/octree idea
\cite{Bader:13:SFCs,Weinzierl:09:Diss,Weinzierl:11:Peano}}.
The computational domain is suitably scaled and dilated to fit into a unit hypercube of dimension $d$.
All ingredients introduced in this paper as well as the underlying software
work for any dimension $d \geq 2$.
We cut the hypercube into $k$ equidistant slices along each coordinate axis and
end up with $k^d$ small cubes.
They form a Cartesian grid while the cut process describes a relation
$\sqsubseteq_{child\ of}$ between the newly generated children and their parent
cube.
It yields an embedding of the finer into the coarser grid.
We continue recursively yet independently for each new cell and end up with a 
\newA{set of cubes $\mathcal{T}$}, where the tree leaves, i.e.~the cells not
refined further, span an adaptive Cartesian grid.
\newA{The parent-child relation $\sqsubseteq_{child\ of}$ or its inverse,
respectively, induce a hierarchy on $\mathcal{T}$.
It renders the data structure into a tree.}
%\newA{This set is rendered a} tree of cubes with cells $c \in \mathcal{T}$
The overall construction process yields a cascade of \newB{disconnected}
Cartesian grids.
\newB{They are embedded into each other but each individual resolution's
Cartesian grid might be disconnected and not covering the whole domain.
} 
\newC{
 An ``embed disconnected Cartesian grids'' formalism is widely used among
 tree codes in fluid dynamics
 \cite{Bungartz:10:PDEFramework,Deiterding:05:AMR,Dreher:05:Racoon,Dubey:16:SAMR}---notably
 if they rely on the classic Berger-Colella patch language.
%   (see historic
%  remarks in \cite{Deiterding:05:AMR,Dubey:16:SAMR}, e.g.).
 % 89
 Prior to this, early work on adaptive multigrid methods already relies on such
 data structures
 \cite{Brandt:73:MLAT,Brandt:77:MLAT,McMormick:89:FAC}.
}
While geometric multigrid \newC{and -level} algorithms benefit from
the nested grids,
applications requiring only the finest tessellation ignore the
coarser levels.

 \begin{figure}[htb]
   \begin{center}
   \includegraphics[width=.75\textwidth]{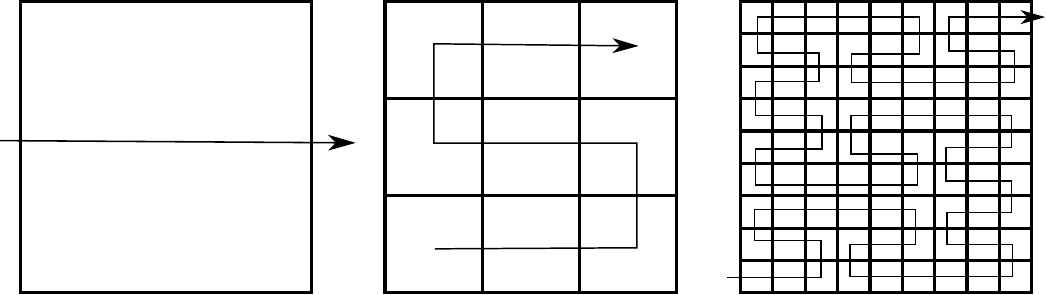}
   \hspace{0.2cm}
   \includegraphics[width=.2\textwidth]{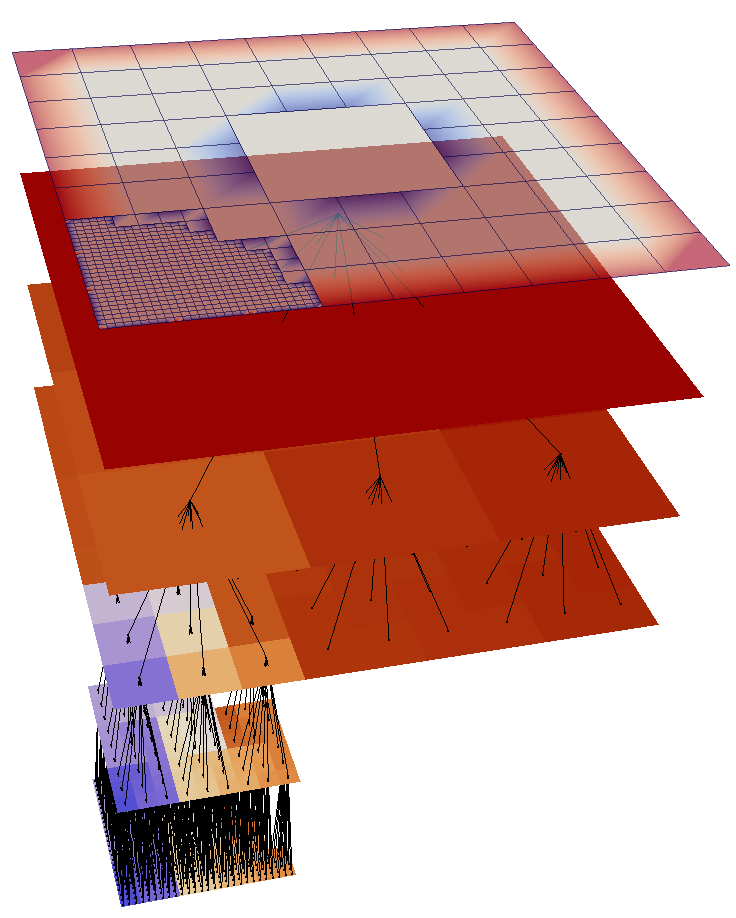}
   \caption{
     We embed the computational domain into a square (left).
     The Peano SFC motif (2nd from left) induces an ordering on a Cartesian grid
     (next) that results from successive refinement starting from the 
     square (left).
     Illustration \newB{following} \cite{Weinzierl:09:Diss}.
     \newA{The sequence depicts three regular grids $\Omega _{h,0}$, $\Omega
     _{h,1}$ and $\Omega _{h,2}$ which can be embedded into each other}.
     The construction of an adaptive Cartesian grid 
     (\newB{right, top layer}) yields a spacetree 
     \newB{as visualized in the layer below}
     (from \cite{Weinzierl:14:BlockFusion}).
    }
   \end{center}
   \label{figure:spacetrees:Peano}
 \end{figure}

Two classic enumerations of $\mathcal{T}$ follow depth-first search
(DFS) and breadth-first search (BFS).
\newA{
 Yet, $\sqsubseteq_{child\ of}$ plus DFS \newC{or}
 BFS remain partial orders as long as no enumeration is defined on the children of the
refined cells.}
If we use one leitmotif for all parent-children relations, we
end up with a space-filling curve (SFC) \cite{Bader:13:SFCs}.
If properly equipped with rotations and mirroring, the leitmotif yields 
the Hilbert or Lebesgue curve for $k=2$ or the Peano curve for $k=3$.
In our code base, we rely on \newB{a} Peano curve
(Figure~\ref{figure:spacetrees:Peano}) which motivates our
choice of three-partitioning $k=3$.

% , while
% Figure~\ref{figure:spacetrees:tree-vs-grid} and the following example use
% bipartitioning.
% SFCs induce an ordering on
% the leaves of the spacetree, i.e.~of the adaptive fine grid,
% but an SFC motif also induces a total
% order on $\mathcal{T}$ once we choose BFS or DFS.

\begin{observation}
Tree and grid language for spacetrees are equivalent once we read an
adaptive Cartesian grid as composite of \newB{disconnected} regular Cartesian
grids.
\end{observation}

\noindent
Let the level $\ell $ of a tree's node equal the \newB{length} of the path to
the root of the \newB{tree}, i.e.~the original bounding box.
The nodes of one level span a grid
\newB{
\[
  \Omega _{h,\ell} = \{ c \in \mathcal{T}: level(c)=\ell \} 
  \quad  \mbox{with}  \quad
  \Omega _{h} = \bigcup _\ell  \Omega _{h,\ell}.
\]
}

% \begin{eqnarray*}
%   \Omega _{h,\ell} & = & \{ c \in \mathcal{T}: level(c)=\ell \} \qquad
%   \mbox{with the tessellation} \\
%   \Omega _{h} & = & \bigcup _\ell  \Omega _{h,\ell}.
% \end{eqnarray*}

\begin{designdecision}
 \newA{
  Cells and vertices are made unique through their position and
  space plus the level.
  Thus, multiple vertices may coincide spatially but `live'
  on different levels.
 }
\end{designdecision}

\noindent
Our approach offers complete level grids $\Omega _{h,\ell}$ and thus 
permits applications to hold data on each level separately.
\newB{Multigrid and multiscale algorithms for example can exploit this fact 
\cite{Weinzierl:16:PIC,Reps:16:Helmholtz,Weinzierl:17:BoxMG}.} 
If an application exploits solely the finest
grid $\Omega _{h}$, vertices and cells are held redundantly and the design decision imposes an overhead.
\newB{Coarse, ``spatially redundant'' data then are simply ignored.}
The overhead is bounded by a factor of 
\newB{
$\sum _{\ell =0}^{\infty}\left(
\frac{1}{k^d}\right)^\ell = \frac{1}{1-k^{-d}}$.
}
\newC{
 All aforementioned approaches working with embedded Cartesian grids---the
 present software historically is inspired by the cited multigrid
 work---conceptionally hold partially spatial redundant data.
}

Given our multilevel grid representation, we define the term {\em hanging
vertex} in a multilevel sense, too:
Most vertices of \newC{a} level have $2^d$ adjacent cells on the same
level.
Some vertices have fewer adjacent cells. 
They are hanging.
Multiple hanging vertices \newB{from different resolution levels} may coincide.

Any SFC-based DFS or
BFS ordering linearizes the spacetree \cite{Sundar:08:BalancedOctrees}.
It enumerates all cells of the tree.
Peano's core routines never use this enumeration explicitly.
Instead, the enumeration orders all data structures used and thus is
exploited \newB{implicitly}.
Yet, one data administration approach built on top of the core routines uses
explicit numbering.

Within the total DFS order, cell numbers (identifiers) 
can be derived from the whole path from the root to any cell.
The root has entry 0. 
Children of a refined node inherit the node's identifier and append \newB{a}
number along the SFC---similar to adding additional digits after the 
\newB{decimal point}
if numbers were taken from $[0,1[$.
This way, it is possible to derive a cell's position and size from its
identifier, and to search the linearization of neighboring cells.
% For BFS, no such statement is possible for arbitrary adaptive grids as the
% Cartesian grid of one level might be ragged, i.e.~disconnected.
The code formalism is equivalent to a formalization of a recursive function
running DFS through the tree:
Each call stack entry then implicitly encodes one edge within the tree graph
from the spacetree's root to the current spacetree node.
For a plain, continuous numbering of cells along the natural numbers this is not
possible.
Efficient numbering \newB{constrains the tree depth
\newC{\cite{Grandin:15:HighDimAMR,Sampath:08:Dendro}} and} stores each
cell identifier with a fixed number of bits.
\newA{
DFS and BFS yield the same cell ordering if an algorithm solely works with
$\Omega _h$ and neglects/omits coarse levels. 
}
\newC{
 SFC codes implicitly encode through their code length that multiple grid
 entities may coexist on different resolution levels, even though a user might
 decide not to store data on coarse levels overlapping finer resolutions
 (cf.~application data storage in Section~\ref{section:data-management}). 
}
%
% Bla bla bla
%

\begin{designdecision}
We use tripartitioning instead of the predominant bipartitioning, and we use
the Peano space-filling curve \newB{($k=3$)}.
\end{designdecision}

\noindent
While Hilbert and Peano yield face-connected DFS enumerations, i.e.~any two subsequent
cells along the SFC enumeration share one face, only Peano and Lebesgue are
straightforward to extend from $d=2$ to $d\geq 3$.
\newB{
See for example \cite{Bader:13:SFCs} for curve illustrations and discussions of
their properties, 
\cite{Haverkort:16:SFCs} and citations therein for further formalization,
classification and variants, as well as
\cite{Skilling:04:HighDimensionalHilbert} for a high-dimensional generalization of Hilbert with links to Gray codes.
}
Lebesgue uses a tensor-product approach, Peano extends its motif from $d$ into
dimension $d+1$ by mirroring the $d$ motif once, appending it along the
dimension $d+1$, and then appending on top the original motif again.

As we stick to Peano, 
\newB{
our software uses $k=3$.
The concepts of this manuscript however work for any $k\geq 2$.
}
Our code base supports any \newB{$2 \leq d \leq 7$}.
\newB{$d \geq 8$ requires manual extension of a few macros,}
though it runs into the curse of dimensionality
\cite{Bellman:61:CurseOfDimensionality}.

\section{Spacetree application programming interface}
\label{section:api}

\begin{table}[htb]
 \ifthenelse{\boolean{toms}}{
  \tbl{
   A classification of spacetree user interfaces at hands of two orthogonal
   metrics. 
   \label{table:classification}
  }
 }{
  \caption{
   A classification of spacetree user interfaces at hands of two orthogonal
   metrics. 
  }
  \label{table:classification}
  \vspace{-0.2cm}
  \begin{center}
 }
  { \footnotesize
  \begin{tabular}{p{0.2\textwidth}||p{0.3\textwidth}|p{0.3\textwidth}}
    & restrictive, constrained (cell-wise) data access
    & read and write of (non-local) grid data (RAM)
    \\
    \hline
    \hline    
    prescribed grid traversal order & 
    \center{Peano} & 
    \newA{
     Examples are deal.II or Dune (with user-defined large grid overlaps) if
     solely iterators are used.
     Yet, they allow users to navigate through the grid starting from a iterator position \ldots} 
    \\
    \hline
    user controls grid run-through order &
    \newA{
     p4est, e.g., supports arbitrary run-through orders while its ghost layer
     of width one constrains/localizes data access. If users use solely p4est's
     iterators, the code fits into the rubric above.
    } 
    &
    \ldots or to fuse/nest multiple iterators into each other. The exact
    classification of both examples depends on how the \newB{
    iterators are used.
    }
%     y are used
%     configured (which underlying Dune grid is used, e.g.).
  \end{tabular}
  }
 \ifthenelse{\boolean{toms}}{
 }{
 \end{center}
 }
\end{table}

AMR code interfaces can be classified along various metrics. 
We use two
\newB{(Table \ref{table:classification})}.
Both assume that the most important operation on an AMR grid is to run over all
entities. 
An interface either can permit the user code to arbitrarily navigate through the
spacetree, or a code can prescribe the traversal order of the grid entities.
Orthogonal to this decision, a user interface has to define which data the user
is allowed to read and write.
Strict element-wise traversals allow a code to access
solely the cell data itself plus data of its adjacent vertices
while they march through the grid.
The other extreme of a cell-based API allows an application to read and write
any data transitively associated to a cell: 
A code can process adjacent vertices of a cell, or any neighboring cell
associated through a face, or adjacent vertices of the neighbors, and so forth.
The latter scheme supports, from the traversal's point of view, random access to
the memory (RAM).

\begin{designdecision}
Peano sticks to a strict element-wise multiscale tree traversal.
All
cells of each tree level are processed per grid sweep.
The process order is determined by the tree traversal code.
The user has no influence on this order and has to program agnostic of it.
However, many temporal constraints are guaranteed, i.e.~there is a 
\newB{guaranteed}
partial order on the traversal's transitions.
At any time solely cell data, the vertices
adjacent to a cell, the cell's parent plus the adjacent vertices of the parent are exposed to the application
code.
\end{designdecision}

\noindent
A restrictive programming model where the user is not in control of
the access order allows us to hide how the data is held and maintained.
It is thus our method of choice for a separation-of-concerns software
architecture.

% \begin{figure}
%  \begin{center}
%   \includegraphics[width=0.28\textwidth]{applications/boxmg.png}
%   \hspace{1.6cm}
%   \includegraphics[width=0.3\textwidth]{applications/helmholtz.png}
%  \end{center}
%  \caption{
%   \newA{
%   Left: Studies on the stability of Deuterium described by cascades of Helmholtz
%   equations (illustration from \cite{Reps:16:Helmholtz}) which are solved by a
%   low-order, complex-valued additive multigrid solver.
%   Right: Studies on the robustness and cost of multiplicative
%   geometric-algebraic multigrid for convection-diffusion equations
%   (illustration from \cite{Weinzierl:17:BoxMG}).
%   \label{figure:use-cases:multigrid}
%   }
%  }
% \end{figure}

\subsection{Supported application types}
\label{section:api:supported-application-types}

Different applications fit to 
our strict element-wise traversal.
\newA{We detail some applications in the appendix and summarize the key
characteristics here.}
Naturally, any stencil that decomposes
additively over $2^d$ cells arranged in a cube can be realized.
$d$-linear finite element codes  
% \newA{(Figure \ref{figure:use-cases:multigrid})} 
fall into this class as well as
low order finite volume and finite difference schemes.
We may assemble system matrices explicitly through PETSc \cite{Software:PETSc},
e.g., or 
\newB{
 make the grid traversal realize matrix-free matrix-vector products
 \cite{Reps:16:Helmholtz,Weinzierl:17:BoxMG}. 
}
Arbitrary adaptivity \newC{is} supported.
\newC{We can code it} recursively
by writing two-level interaction operations.
As Peano holds all grid levels, also different stencil types \newC{or} even PDEs
can be hosted by different resolutions.

% \begin{figure}
%  \begin{center}
%   \includegraphics[width=0.3\textwidth]{applications/fv.jpg}
%   \includegraphics[width=0.45\textwidth]{applications/dem.jpg}
%  \end{center}
%  \caption{
%   \newA{
%   Left: Snapshot of a shallow water Finite Volumes solver which is applied to an
%   initial water height profile initialized through the Durham code of arms.
%   Each spacetree cell here hosts a patch of Finite Volumes
%   (illustration from \cite{Weinzierl:14:BlockFusion}).
%   Right: Snapshot of a Discrete Element code where rigid particles are embedded
%   into the Peano grid.
%   \label{figure:use-cases:fv-dem}
%   }
%  }
% \end{figure}

A spacetree cell can host a whole patch of unknowns rather than only
single unknowns.
% (Figure \ref{figure:use-cases:fv-dem}).
\newA{The same concept allows us to host higher order shape functions.} 
In this context, it is reasonable to weaken the
notion of element\newB{-wise}:
If we augment, similar to
\cite{Khokhlov:98:FullyThreadedTree}, a vertex by pointers to its $2^d$ adjacent
cells, we can construct the inverse of the directed connectivity graph constructed by the spacetree.
\newB{
 Through the vertices,
 a cell then \newC{has} access \newC{to} its neighbor cells or an adjacent
 coarser cell if there is no neighbor on the same level.
 Inter-grid transfer operators still can be realized through the traversal
 events.
}

\begin{designdecision}
  By default, Peano offers strict element-wise multilevel data access.
  Yet, we allow to weaken these data access permissions: 
  At the price of $2^d$ pointers per vertex, we make each vertex point
  to its adjacent cell data.
\end{designdecision}

\noindent
This additional adjacency information is maintained, also for dynamically
adaptive grids, by the \newA{traversal code}:
There is memory overhead to hold adjacency properties that allow us to weaken
the strictness of \newB{the term cell-wise}.
There is no additional algorithmic cost as all links, also for hanging
vertices, are kept consistent on-the-fly
\footnote{
 \newB{
 The augmentation of the grid with pointers to neighbor cells
 \newC{as well as}
 the particle (-in-cell) management are available as optional Peano extensions.
 }
}.

% Notably, adjacency information for hanging nodes can be propagated top-down
% throughout the traversal.
% As the automaton has access to the $2^d$ vertices of any processed cell, the
% links allow us to reconstruct the $3^d-1$ neighbour cells of any cell.
\newA{Patch-based} codes \newA{(block-structured AMR)} can \newA{either} fill
ghost layers 
\newB{with copied data from neighbors}
or make a patch interact directly with neighbor patches \cite{Weinzierl:14:BlockFusion}.
If we embed \newA{$n \times n$ or $n \times n \times n$ or, in general, $n^d$}
patches into each cell, we can through the inverse adjacency information equip each patch with a ghost
layer of a width of up to $n$.
\newC{
 We come back to the impact of patches on runtimes in
 \ref{subsection:persistency-model-and-data-management}.
}
Besides low order ansatz spaces and patches, 
higher order discretizations fit to the traversal as long as their support is
localized.
Classic Discontinuous Galerkin
where the shape function support covers solely one cell \newB{yields}
admissible stencils.
Here, multiple degrees of freedom have to be assigned
to vertices or edges---edge unknowns always can be mapped onto vertex locations
which implies that solely vertices have to be kept persistent from a data
structure point of view---or degree of freedoms have to be stored within the
cell.
Higher order approaches with increased global smoothness are more difficult to
realize:
B-spline shapes spanning multiple cells for example seem \newB{not to fit} the
element-wise concept.
Yet, \newB{we may use} $n^d$ patch data structures with a
ghost layer of width $n$ carrying higher order shape functions such as B-splines of order
$2n-1$.
 \newA{Support overlaps from one spline with others then can be evaluated.}
% On the other hand, we can store helper vectors within the vertices.
% Once we add $d$ values per vertex that hold the average values of the solution
% along the $d$ coordinate axes, each cell may reconstruct the value of
% $2d\cdot 2^{d-1}$ vertices that are adjacent to cells connected to the current
% cell via a $d-1$-dimensional face.
% Such a helper-reconstruction technique can be extended to larger distances.

Particle-grid formalisms such as Particle-In-Cell (PIC) fit to our concept
as long as the particle-grid interactions do not span more than one cell:
If particles fall into a cell, they may interact solely with the $2^d$ adjacent
vertices of this cell.
Alternatively, we can assign each particle to its closest vertex and thus store
them in a dual tree grid \cite{Weinzierl:16:PIC}.
This facilitates wider interaction areas.
%This widens the interaction radius by half a mesh width.

Particle-particle interactions are straightforward to realize if we rely on the
aforementioned neighbor cell links \cite{Eckhardt:15:SPHCompression} and if the
interaction radius fits the mesh width.
Again, particles can live on different spacetree levels
(Figure \ref{figure:use-cases:automata}).
Moving them through various levels facilitates tunneling where particles move
more than one cell per time step \cite{Weinzierl:16:PIC}.

\begin{algorithm}[htb]
 \caption{
  \newA{Depth-first traversal of spacetree. It is invoked on the trees root and
   is passed the linearised tree as $S_{in}$. It yields the output stream
   $S_{out}$ subject to dynamic grid refinement and coarsening which is used for
   the next traversal. $S$ is the traversal automaton state encoding the
   space-filling curve (level, orientation, \ldots) and the cell's position and
   size $h$.
  } 
 }
 \label{algorithm:dfs}
 {\footnotesize
  \begin{algorithmic}[1]
   \Function{traverse}{$S_{in},S_{out}$,$S$} 
    \State $currentCell \gets pop(S_{in})$
    \For{adjacent vertices $v$ of $currentCell$}
     \If{$v$ is hanging vertex on level $\ell$}
      \State \texttt{createHangingVertex}($v$,\ldots)
     \ElsIf{$v$ used for the very first time}
      \State $v \gets $ load from input stream
      \State \texttt{touchVertexFirstTime}($v$,\ldots)
     \Else
      \State $v \gets $ load from temporary data container
     \EndIf
    \EndFor
    \State \texttt{enterCell}
    \If{$currentCell$ is refined or to be refined}
      \For{$i \in \{0,\ldots,3^d-1\}$ along SFC($S$)}
        \Comment Meander along SFC motif
        \State \Call{traverse}{$S_{in},S_{out}$,$S$}
        \Comment Pass down mirrored, scaled and translated $S$ 
      \EndFor
    \EndIf
    \State \texttt{leaveCell}
    \For{adjacent vertices $v$ of $currentCell$}
     \If{$v$ is hanging vertex on level $\ell$}
      \State \texttt{destroyHangingVertex}
     \ElsIf{$v$ used for the very last time}
      \State \texttt{touchVertexLastTime}
      \State store $v$ on output stream for next traversal
     \Else
      \State store $v$ in temporary data container 
     \EndIf
    \EndFor
    \If{part of grid not to be erased}
      \State $push(currentCell,S_{out})$
    \EndIf    
   \EndFunction
  \end{algorithmic}
 }
\end{algorithm}

\begin{table}[htb]
 \ifthenelse{\boolean{toms}}{
  \tbl{
    Table of events defined by Peano that act as
    plug-in points for the application.
    \label{table:events}
  }
 }
 {
  \caption{
    Table of events defined by Peano that act as
    plug-in points for the application.
  }
  \label{table:events}
  \vspace{-0.2cm}
  \begin{center}
 }
  {\footnotesize
  \begin{tabular}{l|p{0.6\textwidth}}
    {\bf Event} & {\bf Semantics} \\
    \hline 
    \texttt{beginIteration} & Is called once per tree traversal prior to any
    other event.
    \\
    \texttt{endIteration} & Is called once per tree traversal in the very end. 
    \\
    \hline 
    \texttt{createVertex} & Creational event \cite{Gamma:94:DesignPatterns} that allows proper
    intialisation of vertices. 
    \newA{Is invoked only once per vertex if the grid is refined 
    by the traversal automaton.}
    \\
    \texttt{destroyVertex} & Counterpart of \texttt{createVertex} invoked just
    before the memory of a vertex is released. 
    \newA{Is invoked by
    the traversal automaton once before it erases parts of the spacetree.}
    \\
    \hline
    \texttt{createHangingVertex} & Hanging vertices are never held persistently
    but (re-)created on-the-fly whenever they are required, i.e.~whenever an
    adjacent cell on the respective level is traversed. This implies that a
    hanging vertex might be created up to $2^d-1$ times. 
    \\
    \texttt{destroyHangingVertex} & Counterpart of \texttt{createHangingVertex}.
    \\
    \texttt{createCell} & Creational event for cells. \\
    \texttt{destroyCell} & Counterpart of \texttt{createCell}.
    \\
    \hline
    \texttt{touchVertexFirstTime} & Event invoked on a vertex once per
    traversal just before it is used for the very first time. 
    \\
    \texttt{touchVertexLastTime} & Event invoked for a vertex after all adjacent
    cells have been traversed. 
    \\
    \texttt{enterCell} & Whenever the traversal automaton enters a spacetree
    cell, it invokes an \texttt{enterCell} event.
    \\
    \texttt{leaveCell} & Counterpart of \texttt{enterCell} that is invoked
    throughout the automaton's backtracking. 
    \\
    \texttt{descend} & Variant of \texttt{enterCell} that is offered to
    simplify multigrid algorithms. Passes a refined cell plus its adjacent
    vertices to the application-specific code as well as all $3^d$ child cells
    and their $4^d$ vertices. 
    \\
    \texttt{ascend} & Counterpart of \texttt{descend}. \\
  \end{tabular}
  }
 \ifthenelse{\boolean{toms}}{
 }{
  \end{center}
 }
\end{table}

% Besides the
%     application-specific cell data and spatial information, the event also
%     hands over the data of all adjacent vertices as well as the parent cell
%     and its vertices. 

\subsection{An automaton-based traversal}

\begin{figure}[ht]
  \begin{center}
  \includegraphics[width=0.42\textwidth]{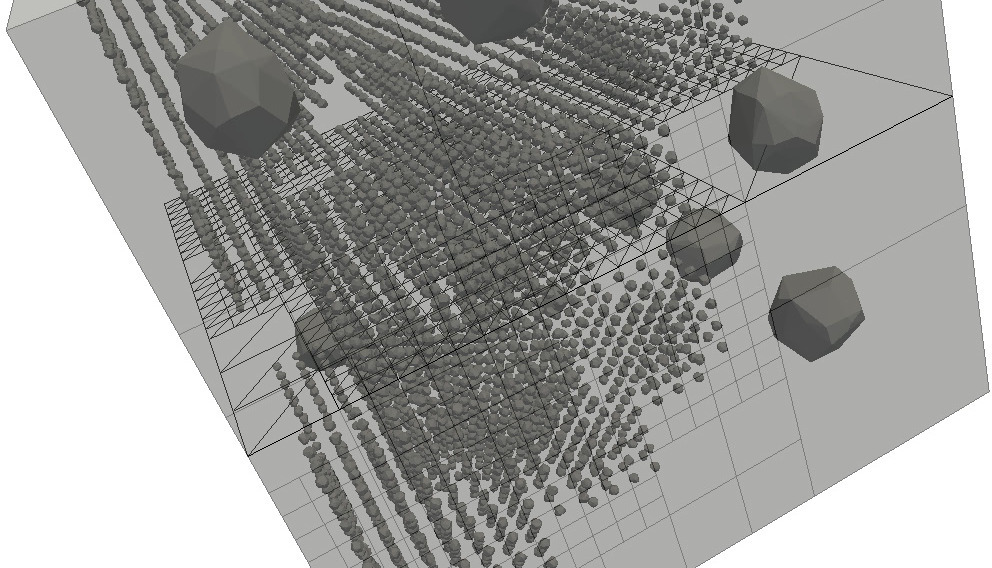}
  \hspace{0.35cm}
  \includegraphics[width=0.4\textwidth]{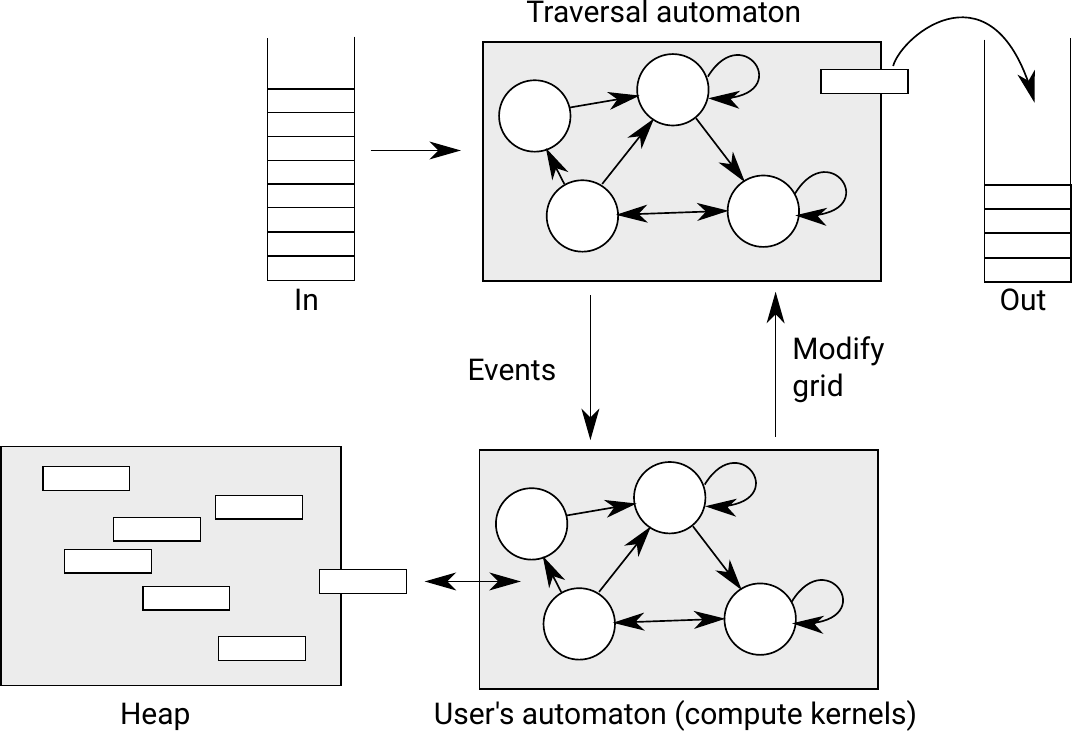}
  \caption{
    Left: Snapshot of a Discrete Element code where rigid particles are embedded
    into the Peano grid.
    Right: The traversal automaton runs through the grid (top) reading in
    streams and piping out streams. 
    Each transition triggers kernels in the application-specific automaton
    (bottom) that may use or may not use heap data. 
    \label{figure:use-cases:automata}
  }
 \end{center}
\end{figure}

As we disallow the user to navigate through the spacetree herself (Table
\ref{table:classification}), we read the multiscale Cartesian grid
traversal as a deterministic push-back automaton\newA{---the formal 
equivalent to a recursive function}---that
reads the tree structure from an input stream while 
adaptivity criteria insert or remove elements.
It runs from cell to cell \newA{(Algorithm \ref{algorithm:dfs})}.

Without the \newA{optimizations} from Section
\ref{section:shared-memory}, we make the automaton run through the tree along a
\newA{modified} DFS.
Such a convention facilitates a memory-efficient realization of the automaton
(\newB{compare} Section \ref{section:data-management}) and makes the automaton 
run through each spacetree cell twice:
once throughout 
\newB{the steps down,}
%steps the down, 
once when it backtracks bottom-up.
This equals an element-wise
multiscale adaptive Cartesian grid traversal which is formalized by 
a sequence of transitions such as `move from one cell into another cell'.
In our code, these transitions act as plug-in points for the application-specific functions
(compute kernels) \cite{Meister:12:Software}.
We refer to them as {\em events} (Table \ref{table:events}).
\newA{Events are defined on cells, vertices or traversal start and end.}

\begin{designdecision}
Peano does not impose any balancing condition. \newC{Each} hyperface in the 
grid may host an arbitrary number of hanging vertices
\cite{Isaac:12:Balancing,Sampath:08:Dendro,Sundar:08:BalancedOctrees}.
% \newB{Optionally}, balancing can be \newB{added.} 
\newC{Balancing can be added optionally as an extension, i.e.~without the need
to code it manually (Appendix \ref{section:appendix:balancing}).}
% enforced by the user if favoured by the application
% domain.
\end{designdecision}

\noindent
A user writes routines that are invoked on grid entities
through the transitions.
\newA{The routines' semantics may depend on a state.} 
Such a system can be read as a combination of two automata (Figure
\ref{figure:use-cases:automata}):
One runs through the spacetree. 
Its transitions \newA{trigger stimuli that} feed the other automaton
implementing the solver's behavior as reaction.
Stimuli comprise both automaton properties such as position in space,
level, grid statistics (such as the total number of vertices) and the data
associated to the event.
Data circumscribes vertex attributes for vertex-based events, cell attributes 
plus all attributes of adjacent vertices for cell-based events, and always the
same type of data associated to the next coarser level, i.e.~to the parent grid
entity.
The latter facilitates the realization of multiscale applications. 
Level and spatial properties are held within the automaton state and thus are
not stored within the vertices and cells.
This is memory efficient.
The event-based programming model forces the user to express all
algorithms in local operations, 
\newB{i.e.~element\newB{-wise} or per vertex with both having access to
respective local coarse-grid counterparts}.
The application code may express the wish to refine or coarsen the grid
to the traversal through the return value of the events.

Let $a,b \in \mathcal{T}$ with $a \sqsubseteq _{child\ of} b$.
Further, $v_a, v_b$ are vertices adjacent to $a$ or $b$, respectively. 
\newB{Finally, $\sqsubseteq _{pre}$ is a temporal relation, i.e.~clarifies
which steps run in which partial order.} 
While the processing order of cells and
vertices is hidden, all applications can rely on the invariants 
\begin{eqnarray}
  \mbox{\texttt{touchVertexFirstTime}}(v_b) & \sqsubseteq _{pre} &
    \mbox{\texttt{touchVertexFirstTime}}(v_a), 
    \nonumber \\
  \mbox{\texttt{touchVertexFirstTime}}(v_b) & \sqsubseteq _{pre} &
    \mbox{\texttt{enterCell}}(b), 
    \nonumber \\
  \mbox{\texttt{enterCell}}(b) & \sqsubseteq _{pre} &
    \mbox{\texttt{enterCell}}(a), 
    \nonumber \\
  \mbox{\texttt{enterCell}}(a) & \sqsubseteq _{pre} &
    \mbox{\texttt{leaveCell}}(a), 
    \nonumber \\
  \mbox{\texttt{leaveCell}}(a) & \sqsubseteq _{pre} &
    \mbox{\texttt{leaveCell}}(b), 
    \nonumber \\
  \mbox{\texttt{leaveCell}}(a) & \sqsubseteq _{pre} &
    \mbox{\texttt{touchVertexLastTime}}(v_a),
    \nonumber \\
  \mbox{\texttt{touchVertexLastTime}}(v_a) & \sqsubseteq _{pre} &
    \mbox{\texttt{touchVertexLastTime}}(v_b).
  \label{equation:order-on-events}
\end{eqnarray}
It identifies which event is
invoked prior to another event.
It is easy to verify that DFS and
BFS both suit (\ref{equation:order-on-events}).

\begin{designdecision}
We allow the application automaton to flag operations
from Table \ref{table:events} as empty. They then are  automatically
skipped and thus removed from (\ref{equation:order-on-events}).
\end{designdecision}

\begin{observation}
In the context of trees in object-oriented languages, our inversion of
control---the application kernels do not determine how the data structures are
processed---can be read as a composite pattern in combination with a visitor
pattern \cite{Gamma:94:DesignPatterns}.
In the context of \newA{finite element (FEM)} solvers, this paradigm is used by
various codes \cite{Burstedde:11:p4est,Meister:12:Software,Weinbub:14:ViennaX}.
In the context of general programming, it mirrors higher-order functional
programming where the application's function set is passed to the traversal
function.
\end{observation}

\noindent
We favor to call the programming pattern Hollywood principle
\cite{Sweet:85:HollywoodPrinciple}:
Don't call us, we call you.
The user's implementation is unaware of when events are invoked---though the constraints
(\ref{equation:order-on-events}) hold---it is unaware where events are invoked
in a distributed environment, and it is unaware of any other
events invoked concurrently.

% However, a software infrastructure as introduced here also would work with a
% graph-based spacetree storage where pointers represent parent-child relations.

\subsection{Limitations of the approach}
\label{section:api:limitations}

Our API poses limitations. 
First, the grid's structuredness does not offer the flexibility of 
unstructured meshes  though users can weaken the structuredness by embedding
unstructured/point data into the spacetree cells.
Second, the hiding of traversal details makes the code not a black-box library that can be used within any  
code base through few library invocations.
It requires users to break down their algorithm's workflow into events and
dependencies first.
The ``loss of control'' might require algorithm designers to rethink
code snippets and algorithmic realizations.
Finally, a very strict element-wise mindset is not the standard mindset of many
application developers.
This steepens the learning curve.

\section{Persistency model and data management}
\label{section:data-management}

How to store the spacetree is an important design decision for a software.
Also, we have to clarify how to hold data associated
to the grid.

\subsection{Spacetree storage schemes}
To hold the spacetree, we distinguish two paradigms: 
The tree either is
mapped onto a graph---typically structures/classes with pointers---or it is held 
linearized.
Hybrids exist.
Graph-based approaches are flexible but 
suffer from pointer overhead. 
% If a tree requires bidirectional links, we have to store $2\ |\mathcal{T}|-2$
% pointers where $|\mathcal{T}|$ denotes the number of cells. 
Furthermore, the tree records scatter in memory and induce non-local memory
accesses for a traversal code once the grid is subject to strong dynamic
adaptivity.
This may cause poor memory usage profiles.
Yet, a spacetree mapped onto a graph data structure still yields a
lower memory footprint than arbitrary unstructured grids, as associativity
between cells is encoded in the tree relations: a cell neighbor either
is a sibling, or a particular child of the sibling of the parent node, and so
forth.
We note that some graph approaches exploit structuredness,
i.e.~combine the graph with linearization: 
siblings can be stored as one block
\cite{Gadeschi:15:HierachicalCartesianGrids,Khokhlov:98:FullyThreadedTree} or
regular grids are used on certain grid levels
\cite{Feichtinger:11:Walberla} and thus
increase data access locality while they reduce the memory footprint.

As alternative to the graph, we may hold the spacetree in a stream.
This stream holds all cells of the spacetree according to a total order on
$\mathcal{T}$.
A natural approach stores the \newB{DFS-plus-SFC} code within the stream
\newC{\cite{Sampath:08:Dendro}.}
As the code comprises the cell's level plus position, lookups for neighbors
within the stream are efficient---a neighbor's code can directly be computed
by bit-wise index manipulations.
% In practice, it is convenient to constain the tree depth and work with codes of
% fixed length, i.e.~fixed byte count.
\newB{
 If BFS determines the total storage order, the stream consists of
 chunks.
 Each holds data from one tree levels
 \cite{BangerthEtAl:07:dealII,Bangerth:11:dealiiwithp4est}.
 If DFS determines the total storage order, the individual resolution levels are
 interwoven, i.e.~vertically integrated \cite{Adams:15:Segmental}.
}

\begin{observation}
  Linearized spacetrees are superior to graph-based data structures,
  i.e.~structs connected via pointers, in terms of memory as they only require
  one code \newA{(identifier)} to be stored per cell.
\end{observation}

\noindent
As we forbid control over the traversal order in our code and
stick to a DFS/SFC combination, we can \newA{implement} the tree traversal as
recursive function \newA{(Algorithm \ref{algorithm:dfs})}.
It is a push-back automaton relying on the system's call stack.
The automaton knows at any time its level and position and thus solely has to
know from an input linearization whether to recurse further or not.
No maximum tree depth constraints are imposed while 
two bits per cell 
(unrefined, refined, to be refined, to be erased) are sufficient to store all
the dynamic adaptivity information \cite{Weinzierl:11:Peano}.
For static adaptive grids, one bit is sufficient.
\newA{In Algorithm \ref{algorithm:dfs}, these bits are encoded within $S_{in}$.}

\begin{designdecision}
  Peano linearizes the tree along the Peano SFC.
  It sticks to a DFS and the traversal thus reads in the spacetree as bit stream. 
\end{designdecision}

\noindent
To support arbitrary adaptivity, it is advantageous to make the
automaton read one bit stream and output another stream that acts as input to
the subsequent traversal.
We hence avoid data movements due to insertion and deletion while
the stream read/writes are advantageous in terms of
memory access characteristics.
They yield high temporal and spatial data
access locality \cite{Kowarschik:03:CacheTechniquesOverview}.
Yet, for reasonably structured grids we abandon the strict linearization
(Section \ref{section:overhead-reduction}).

\subsection{Application data management}

Automaton-based spacetree traversals do
not prescribe how the actual application records are held.
Besides graph-based descriptions---spacetree entities hold pointers
to cell or vertex properties---two storage strategies \newB{are convenient}.
On the one hand, spacetrees can be both a technique to
encode the grid structure and a container for the data itself.
On each and every level, the user assigns data to vertices and cells, 
and the stream of the spacetree is enriched with this application-specific
data.
The spacetree then acts as both organizational and compute data
structure.
On the other hand, we can rely on a heap.
All data are stored within a (hash) map.
The cell's DFS/SFC identifier contained in the automaton state is a natural
candidate to provide a key to this hash map:
\newB{
 If we compute the hash code from
 both the spatial position along the SFC plus its level,
 dynamic mesh refinement does not require frequent hash map
 reordering or induces many hashing conflicts.
}

% Vertex keys can be derived from the cell codes.
\newA{Alternatively, we may use pointers from grid entities to map
entries}.
With a semantic separation of the tree/grid data container from an application 
container, the realization resembles \citet{Bangerth:11:dealiiwithp4est}.
However, our implementations are originally inspired by
\citet{Griebel:99:SFCAndMultigrid,Griebel:98:HashStorage}.
\newA{
  If the grid does not change too frequently, it is, as a variation of the hash
  storage, convenient to flatten also the application data along the spacetree
  linearization and hold it as one big chunk of data accessed along the SFC.
  This is particularly interesting for hardware suffering from indirect
  and scattered memory access.
  We omit this variant \newC{for the heap} as we assume that our grid changes in
  each and every grid sweep.
  \newC{Yet, we also point out that embedding user data into the linearized tree
  stream, i.e.~enriching the tree stream, delivers} such data
  flattening/reordering \newC{implicitly}.
}

\begin{designdecision}
  Both data storage strategies---embedding data into the spacetree and holding
  it in a separate hash map---are available in Peano:
  We refer to them as stack-based or heap-based data storage.
\end{designdecision}

\noindent
Both variants come along with pros and cons. 
If we embed PDE-specific data into the read/write streams, data associated
to the respective cell as well as vertex data are immediately available to the
automaton.
Records associated to the spacetree cells are directly interwoven with the 
spacetree's bitstream.
Records associated to the spacetree vertices 
\newB{
are ordered along \texttt{touchVertexFirstTime} and are
conveniently held in a separate stream.
Yet, they could be merged into the cell
data structure.}
We show in \citet{Weinzierl:09:Diss,Weinzierl:11:Peano} that we can write out
the vertex data as stream following \texttt{touchVertexLastTime} and use this stream
as input the subsequent iteration even if the grid is dynamically adaptive.
Throughout the grid traversal, vertices temporarily have to be stored on stacks.
The stacks serve \newB{as} temporary data containers in Algorithm
\ref{algorithm:dfs}. 
$2 \cdot d$ such stacks are required and their maximum size is bounded by the
spacetree depth.
The number of stacks is fixed, small, and their size is small, too.
Cell-vertex associativity is encoded in the grid traversal automaton,
i.e.~the automaton knows at any time from which data container (stream or stack) to take all vertex
data from, \newB{and it knows} where to write vertices to before the next
cell is entered.
The whole scheme comes for free in terms of user source code---the automaton
simply hands over references to the stacks to the user automaton---the total
memory footprint is minimal and all data remains read and written in a stream/stack fashion.
This yields excellent memory access characteristics
\cite{Bungartz:10:PDEFramework,Mehl:06:MG,Reps:16:Helmholtz,Weinzierl:09:Diss,Weinzierl:11:Peano,Weinzierl:16:PIC}.

The method falls short if the data cardinality per vertex and cell
varies---our cell or vertex stack entries all have to have the same byte
count---or if the data per cell/vertex is massive and thus
moving it from one stack to another is expensive.
\newB{
 The former case covers the frequent situation that some data such as matrix
 entries are not required persistently over the whole simulation workflow.
}
%  or if the algorithm does not require to
% read all data on every level all the time in each traversal.
In this case, heap-based storage is advantageous though
it requires additional coding, induces hash bookkeeping overhead and may
introduce scattered data with non-uniform data
access cost.
Yet, efficient hash codes exist
\cite{Robey:13:Hashing,Tumblin:15:CompactHashing}.
Notably DFS/SFC codes yield high quality hash codes or preimages to a hash
function due to their H\"older continuity
\cite{Bader:13:SFCs,Bungartz:06:Parallel,Gotsman:96:MetricSFCs,Griebel:99:SFCAndMultigrid,Griebel:98:HashStorage,Hungershoefer:02:SFCQuality}.
\newB{
If a mesh remains stationary or remeshing cost are negligible compared to the
compute load, it finally is convenient to make the ``hash code'' store all user
data continuously in memory \cite{Tu:2005:Octor}.
The user data is linearised.
We do not follow up on this option as we focus on changing meshes.
}

\section{Reduction of tree traversal overhead}
\label{section:overhead-reduction}

DFS tree traversals are a powerful tool to facilitate arbitrary dynamic
adaptivity for meshes that change frequently.
As we realize the tree's linearization through a recursive function, i.e.~a
pushback automaton, we however end up with a code that requires a certain amount
of integer arithmetics and callstack administration (recursion overhead).
\newA{
This overhead is not negligible if (i) we study codes with low arithmetic
intensity per grid entity and (ii) the user cannot mask out operations a
priori.
The latter feature is enabled by additional callbacks that can be used to
clarify that \texttt{enterCell} for example in some algorithmic phases never is
invoked on refined cells.
In this case, Peano omits them automatically, and it obtains performance closer
to codes that work on the fine mesh $\Omega _h$ only---though the performance gap
never will be closed as our code is inherently written for multiscale.
}
%This is problematic for applications with low arithmetic intensity.

%We follow-up two superior approaches.
This cost can be reduced if we embed regular subgrids (patches) into
the cells \cite{Weinzierl:14:BlockFusion}.
Traversing Cartesian arrays is among the best-understood and cheapest traversal
variants. A code with a spacetree hosting regular patches thus
yields an administrative overhead that is in-between the minimalist cost of Cartesian
grids and the cost of unconstrained dynamically adaptive spacetree meshes.
% The resulting grid layout then resembles the classic \newB{``hyperbolic''} AMR
% grids \cite{Software:Chombo}.
%However, 
A refinement criterion then is not free to adopt arbitrarily accurate
to a feature anymore.
It has to overlap features of interest with patches.
As alternative to patches, one can temporarily or locally disallow the
tree to coarsen or refine and then skip many logical checks.

\newB{Uniform meshes are a special type of spacetree meshes:}
Let a balanced spacetree be a tree where any path from the root to a leaf
has the same length.
A balanced spacetree yields a cascade of regular Cartesian grids.
We therefore refer to such trees as regular spacetrees.
Their BFS traversal first running from coarse to fine and then backtracking
from fine to coarse combines the efficiency of regular Cartesian access with 
the constraints from (\ref{equation:order-on-events}).
We thus identify regular subtrees within the spacetree that do no
change and do not accommodate any hanging vertices---they
trigger additional events---on-the-fly and to switch from the DFS event invocation on this spacetree to BFS \cite{Eckhardt:10:Blocking}.
% If a refinement criterion changes the regular subtree, it is broken up and
% processed DFS in the next grid sweep where the actual refinement or erasing is
% implemented.

\subsection{Transformation of DFS into BFS}
To find the regular subtrees \newB{Peano relies} on an analyzed tree grammar
\cite{Knuth:90:AttributeGrammar}.
Let each spacetree cell hold a marker $f$ with 
\begin{equation}
  \forall c \in \mathcal{T}: \ f(c) = \left\{
    \begin{array}{rcl}
      0 & & \mbox{$c$ is a leaf with no hanging adjacent vertices,} \\
      \hat f & \mbox{if} & \mbox{$c$ is refined and}\ \forall a \sqsubseteq
      _{child\ of }c:\\
      && f(a) = \hat f-1, \mbox{or} \\
      \bot && \mbox{otherwise.} 
    \end{array}
  \right.
  \label{equation:recursion-unrolling-grammar}
\end{equation}

\noindent
\newB{
 $\bot $ marks spacetree nodes which root adaptive, non-regular subtrees
 or regular subtrees framed by hanging vertices.  
}
Equation (\ref{equation:recursion-unrolling-grammar}) is accompanied by some veto mechanisms
\newA{overruling the outcome of (\ref{equation:recursion-unrolling-grammar})}:
If refinement
or grid coarsening is triggered, all markers of surrounding cells are  
cleared to $\bot$.
$f$ is an augmentation of the spacetree bit stream and can be
maintained on-the-fly:
we take $f$ from the input stream for our optimization while we
concurrently redetermine its new value for the subsequent traversal.
In any iteration following the identification of an $f>1$, the traversal
automaton can modify its event invocation and data processing
\cite{Eckhardt:10:Blocking,Schreiber:13:sfc-based}:
If it encounters a cell with label $f>1$, it knows that from hereon a regular
subtree of depth $f$ is to be traversed.
We know how much data for this subtree is to be read from all input streams, and
we create a temporary buffer in the memory than can accommodate the whole
regular subtree as a cascade of regular Cartesian grids.
This buffer is \newB{filled}.
Following the load, the automaton invokes all the \texttt{touchVertexFirstTime},
\texttt{enterCell}, and so forth events in a BFS order, before the tree is
streamed to the stacks again.
Formally, the reordering is local recursion unrolling.

If the BFS is fed a refinement or coarsening request by the application
automaton, the markers in the regular subtree are set to $\bot$ and we bookkeep
the erase or refine request.
It is realized in the subsequent iteration when the subtree is not treated as a
regular one anymore due to the invalidated $f$.

\begin{table}[htb]
 \ifthenelse{\boolean{toms}}{
  \tbl{
   \newA{
   One characterization of various levels of structuredness and
   flexibility for locally regular adaptive grids.
   While the simplicity reduces form left to right, the flexibility (where are
   we allowed to invest grid entities) increases.
  }
   \label{table:blockstructured}
  }
 }{
  \caption{
   One characterization of various levels of structuredness and
   flexibility for locally regular adaptive grids.
   While the simplicity reduces form left to right, the flexibility (where are
   we allowed to invest grid entities) increases.
  }
  \label{table:blockstructured}
  \vspace{-0.2cm}
  \begin{center}
 }
  { \footnotesize
  \newA{
  \begin{tabular}{p{0.2\textwidth}|p{0.2\textwidth}|p{0.2\textwidth}|p{0.2\textwidth}}
    Regular Cartesian grid & 
    Block-structured AMR (patch-based) with logical tree topology between
    patches& On-the-fly block identification as realized in Peano &
    Pure tree codes 
    \\
    \hline
    \hline
    Globally simple, i.e.~simple administration and data access & 
    Simple per entity (block/ptach) & Simple per identified
    block; remaining (skeleton) grid complex &
    No regular data accesses and simple programming (without indirect
    addressing, e.g.)
    \\
    \hline
    No adaptivity & Adaptivity constrained & 
    No constraints on adaptive patterns &
    No constraints on adaptive patterns
 \end{tabular}
  }
  }
 \ifthenelse{\boolean{toms}}{
 }{
 \end{center}
 }
\end{table}

\subsection{Persistent regular subtrees}
\label{section:traversal-optimization:regular-subtrees}

By default, the DFS-BFS transformation is applied locally,
on-the-fly and temporarily, while we preserve the DFS/SFC input and output order
on all streams.
%The transformation has no impact on the order of any persistent data structure.
% This is important if the user refines or coarsens.
% In this case, we always immediately fall back to DFS,
% process all streams sequentially as new data might have to be inserted or chunks
% removed, and finally reevaluate (\ref{equation:recursion-unrolling-grammar}).
Though our approach eliminates many case checks and allows for an efficient
triggering of events, it retains recursive code parts for the loads/stores from the streams to the Cartesian buffers and back.

\begin{figure}[ht]
  \begin{center}
    \includegraphics[width=0.36\textwidth]{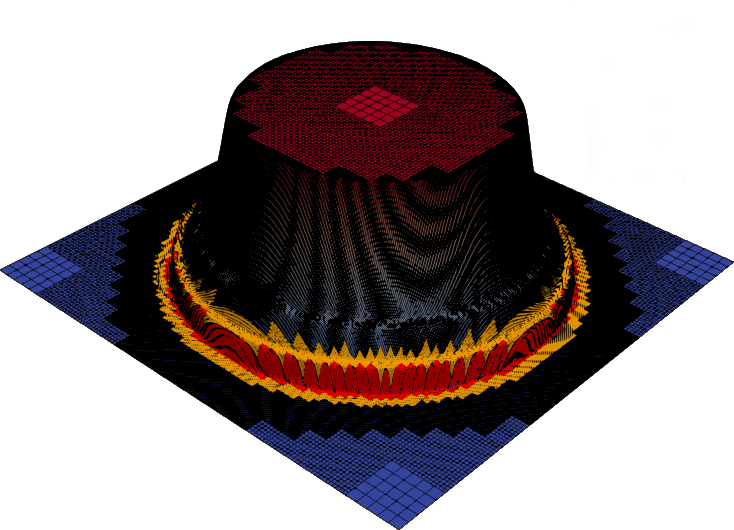}
    \hspace{0.8cm}
    \includegraphics[width=0.25\textwidth]{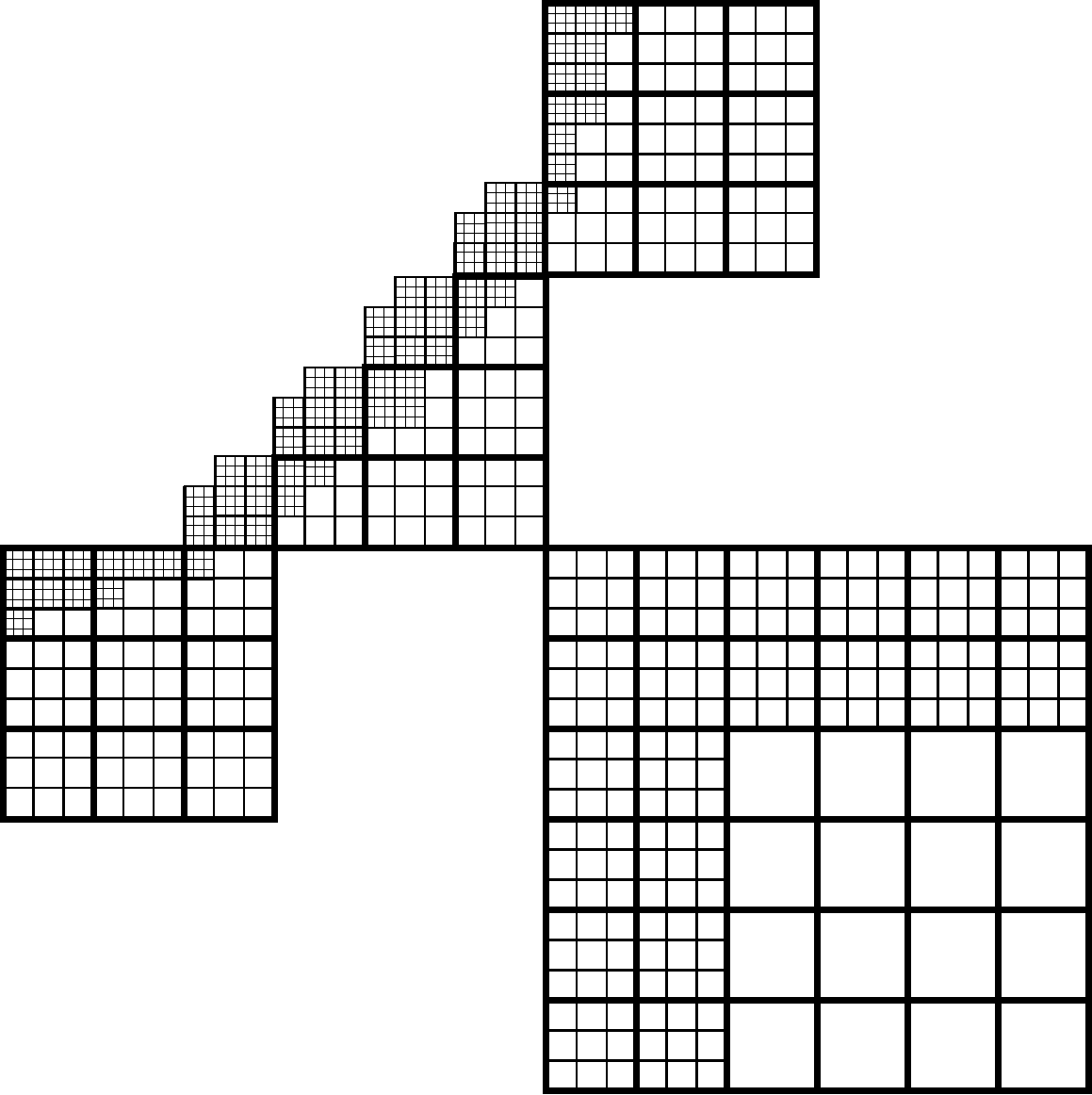}
  \end{center}
  \caption{
   \newA{
    Left: Euler equation simulation of a point shock from the ExaHyPE project
    \cite{Software:ExaHyPE}.
    Right: Illustration of the skeleton grid of the bottom corner from the left
    simulation, i.e.~the grid that is held as a tree if all regular subtrees of
    dimension $9 \times 9 $ or bigger are held as separate regular grids.
   }
  \label{figure:overhead-reduction:blocks}
  }
\end{figure}

Provided that regular subtrees remain regular, we can remove them
from the tree stream and hold them as cascade of Cartesian mesh separately
(Figure \ref{figure:overhead-reduction:blocks}).
Vertices from the regular subtree that are adjacent to the remainder of the
(adaptive) grid are replicated.
Prior to entering a regular subtree, these vertices are updated with the
most recent vertex version from the spacetree stream.
Once a subtree is processed, vertices are mirrored back.
These consistency updates can be done efficiently as they affect a 
lower-dimensional submanifold only and as 
a projection of the Peano SFC onto the face of the cube spanned by the regular
subtree yields a dimension-reduced Peano SFC on the face again
\cite{Weinzierl:09:Diss,Weinzierl:11:Peano}.
The projection yields a total order that matches the inverse of the SFC's
total order on the remaining grid's entities \cite{Schreiber:13:sfc-based}.
Our linearized spacetree spans a holed grid\newA{, a skeleton grid,} with links
to regular subtrees.
Though this is a pointer-linearization hybrid, consistency data exchange all
follows the SFCs through stacks.

If an application code wants to modify the grid within a regular subtree, we
postpone the grid modification and first reintegrate the linearised subtree
into the spacetree stream.
The follow-up grid sweep then refines or coarsens.
The technique follows the cluster-based AMR of
\citet{Schreiber:13:Cluster} but combines it with
(\ref{equation:recursion-unrolling-grammar}) to identify stationary clusters
on-the-fly, applies it solely to regular subtrees, and augments it by
the multiresolution grid notion.

It is similar to approaches composing the AMR grid as assembly of regular
patches.
However, we do not start from regular patches as building blocks.
Instead, we identify regular subregions on-the-fly and treat them then more
efficiently than the remainder of the grid.
\newA{
\newB{Therefore, the}
adaptivity of the spacetree mesh that can be constructed is not
restricted at all (Table \ref{table:blockstructured}).
Regular subgrids are held separately and the linearised tree encodes only the
skeleton in-between or those regions that change frequently (Figure
\ref{figure:overhead-reduction:blocks}).
The grid is decomposed.
We notice the similarity of this decomposition to a generalized variant of
enclave partitioning \cite{Sundar:15:Enclave} though we use our technique
solely to reduce administrative overhead rather than to parallelize.
}

\section{Shared memory concurrent traversals}
\label{section:shared-memory}

% While data decomposition with separated data spaces and data replication are
% reasonable strategies for distributed memory parallelization, native shared
% memory or MPI+X codes often 
% 
\newB{With dynamically adaptive grids,}
shared memory parallelization shall be lightweight \newB{(i.e.~avoid
expensive setup phases)}, shall not synchronise much data and enable work
stealing to \newB{adapt} seamlessly to workload \newB{changes}.
Task-based systems promise this.

\begin{observation}
Our event-based programming
model can be casted into a task language: 
the user implements a fixed set of task types (event implementations), the
constraint set (\ref{equation:order-on-events}) describes task dependencies, and the tree instantiates the tasks.
\end{observation}

\subsection{Dependency-based programming interface}
The observation implies that we may throw any pair of traversal and user 
automata directly into a task management system.
However, not all codes allow for a concurrent invocation of all
events from (\ref{equation:order-on-events}).
They impose additional constraints on the data accesses.

A matrix-free matrix-vector product for example may not allow two
adjacent cells on the same level to add their residual contributions to the
vertex concurrently.
Red-black Gau\ss-Seidel-type colouring of cells with $2^d$ colours
ensures that no vertex is accessed simultaneously in this case.
Peano allows user codes to specify per
event per algorithmic step which colouring would ensure on a regular
Cartesian two-resolution grid that no data races occur.
We support a range of colouring choices:

\begin{itemize}[leftmargin=0.5cm]
  \item A complete serialization which is for example important to
  events that run IO and thus have to veto any concurrency.
  \item A degenerated colouring (one colour) implies that all cells or vertices,
    respectively, can be processed concurrently.
  \item $2^d$ colouring of cells ensures that no two
  cells access a shared vertex simultaneously.
  \newB{This implies that no face between two cells is accessed simultaneously
  by its two adjacent cells.}
  \item \newA{$3^d$} colouring on cells ensures that no two cells with the same
  parent cell are handled concurrently. This is useful for multiscale algorithms.
  \item $4^d$ colouring \newA{generalizes} this idea \newA{to} vertices.
  \item $6^d$ colouring ensures that whenever two cells are handled in parallel,
    their parent cells do not share any vertex.
  \item $7^d$ colouring \newA{generalizes} this idea \newA{to} vertices.
\end{itemize}

\begin{designdecision}
We allow the user code to specify per event which concurrent data writes have to
be avoided on a regular grid.
It is Peano's responsibility to schedule a well-suited parallel execution
of all tasks.
\end{designdecision}

\noindent
For a given spacetree $\mathcal{T}$, the constraints
(\ref{equation:order-on-events}) plus the colouring per event define a
race-free task graph.
While we ask the user to model colouring constraints in terms of a cascade of
regular grids and hide the complexity of multicore processing of a tree,
all constraints translate to the dynamically adaptive grid.

\subsection{Task parallelization without task graph assembly}
Once a task graph is determined, there are two main possibilities to
issue the tasks:
The task graph can be assembled and handed over to a scheduler. 
For static grids, this can be done in a preprocessing step.
For dynamically adaptive grids, it has to be done once per grid sweep, and
the obtained concurrency has to make up for assembly cost.
Alternatively, a grid traversal order which 
accommodates the task graph can be chosen: 
continuous chunks of 
grid entities within the traversal order describe independent tasks and can be
in parallel, before the next chunk of
grid constituents is traversed.

\begin{designdecision}
In Peano, the traversal order follows the task dependencies
and the task
dependency graph is never set up explicitly.
\end{designdecision}

\noindent
We notice that the DFS chosen for the spacetree \newA{linearization (Algorithm
\ref{algorithm:dfs})} exhibits a poor concurrency.
All cell accesses are \newA{serialized}.
Solely a few vertex-based events such as \texttt{touchVertexFirstTime} can be
evaluated in parallel.
If the user code adds a dependency between these vertices, even this
negligible concurrency is eliminated.
While we realize concurrent event invocations in the DFS traversal and also
parallelize the automaton code itself,  
BFS is a better traversal for \newA{parallel fors/stencil codes}.

\begin{table}[htb]
 \ifthenelse{\boolean{toms}}{
  \tbl{
    If ran on shared memory, Peano introduces two additional events.
    \label{table:shared-memory-events}
  }
 }
 {
  \caption{
    If ran on shared memory, Peano introduces two additional events.
  }
  \label{table:shared-memory-events}
  \vspace{-0.2cm}
  \begin{center}
 }
  {\footnotesize
  \begin{tabular}{l|p{0.6\textwidth}}
    {\bf Event} & {\bf Semantics} \\
    \hline
    Copy constructor & 
    The class holding all events, i.e.~the automaton, is replicated in parallel
    sections per thread. The copy constructor allows the user to plug into the
    replication.
    \\
    \texttt{mergeWithWorkerThread} & If the traversal automaton leaves a grid
    region handled concurrently, all adapter replicates are merged into a master
    copy and destroyed afterwards. 
    \newB{This merger is realised as tree reduction as found in Intel's
    Threading Building Blocks (TBB).}
  \end{tabular}
  }
 \ifthenelse{\boolean{toms}}{
 }{
  \end{center}
  }
\end{table}

We accept the DFS' limited concurrency in
general, and use the recursion unrolling's BFS to issue tasks in parallel if
possible.
\newB{
Within regular, unrolled subtrees, we iterate through the levels and rely on 
parallel fors with colouring to issue user tasks in parallel. 
}
\newC{
 The colouring is guided by the user through her colouring choices describing
 data dependencies.
}
Reordering of spacetree cells and the concurrent triggering
of events impose bulk-synchronous programming (BSP) on the
application automaton:
The application normally is passed events sequentially. 
When the traversal automaton runs into a regular subtree, it forks the
application automaton.
The BFS event invocations subsequently are reordered to suit the dependencies.
When the automaton leaves a regular subtree, all application automata are
merged again.
We reveal this by two additional events (Table
\ref{table:shared-memory-events}) establishing an API that hides the realization
with OpenMP or Intel's Threading Building Blocks.
They allow the application developer to focus on data
dependencies and data reduction.
% With its colouring and on-the-fly reordering, our concept resembles
% task-based schemes though Peano never sets up any task graph explicitly.
% Neither do we restrict the dynamic mesh refinement.

\subsection{Limitations of the approach}
BSP as currently realized in Peano does not support data access patterns where some cells depend
on their neighbours or parents while other cell pairs induce no constraint. 
We assume homogeneous dependencies.
The parallelization kicks in for regular subgrids, while
unstructured grid regions, \newA{i.e.~the skeleton}, the MPI domain boundaries
or start and wrap-up phases of single grid traversals do not really benefit from multiple cores.
If the grid changes globally per grid traversal or does not exhibit
regular subregions, our approach does not exhibit lots of concurrency.
\newB{Our results uncover some of these effects.}
Finally, we note that sophisticated multithreading exploits concurrency spanning
multiple iterative sweeps. 
Such a \newB{temporal} blocking is not built in.

\section{Tree decomposition}
\label{section:mpi}
 
We discuss our non-overlapping spacetree decompositions in this
section.
Non-overlapping refers to the fine grid here, i.e.~we assume that each fine grid cell/spacetree leaf is assigned
to exactly one rank.
% The term rank follows the MPI jargon and is an abstraction from computing
% process. 
Extensions of our techniques to overlapping decompositions are beyond scope.

We emphasize that our goal is to retain the spacetree concept in a distributed
environment.
Despite the \newA{parallelization}, we make each rank still hold a spacetree
and, thus, we allow each rank to continue to traverse its spacetree as discussed so
far.
The traversal logic \newA{from a serial code (Table \ref{table:events})} is
agnostic regarding the \newA{parallelization}.
However, there are additional MPI-specific events.
\newB{They ensure that each cell is traversed only by one rank and that the
constraints (\ref{equation:order-on-events}) are globally preserved;
unless this is explicitly not required by the user code.}
% No local data flow or workflow changes arise.

%
% Fire stuff out of the window
%

\begin{figure}
  \begin{center}
    \includegraphics[width=0.28\textwidth]{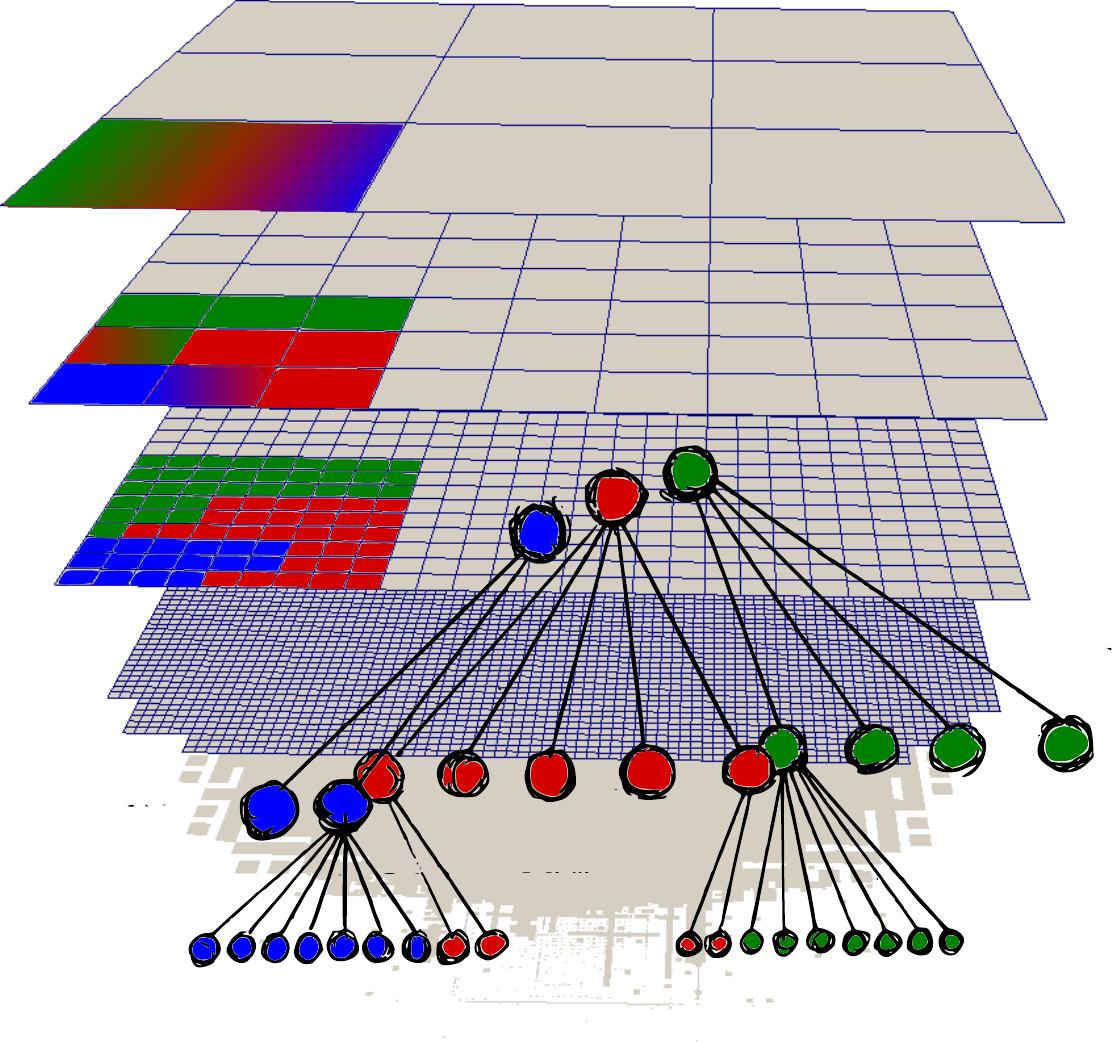}
    \hspace{1cm}
    \includegraphics[width=0.28\textwidth]{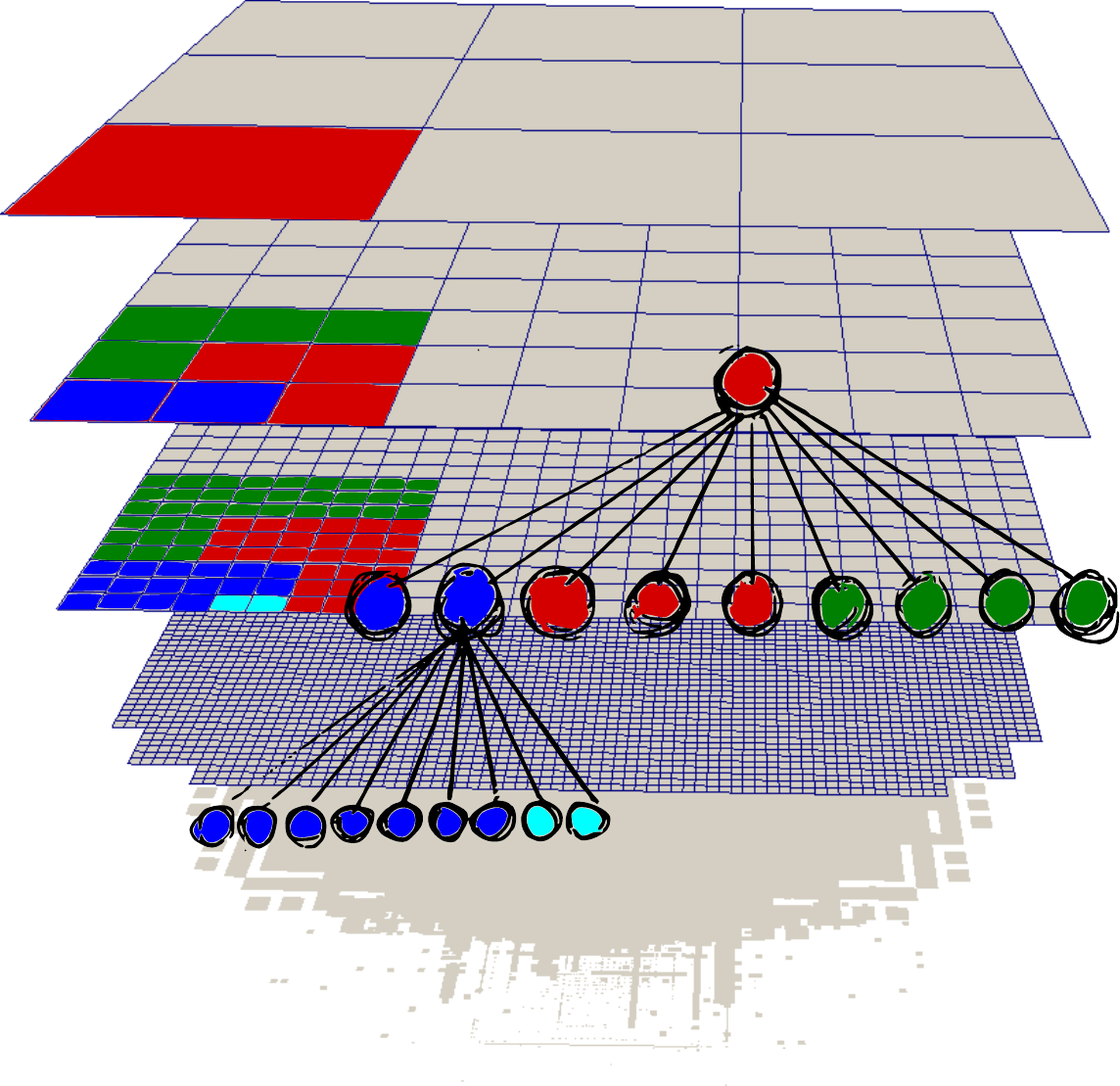}
  \end{center}
  \caption{
    Bottom-up spacetree decomposition (left) vs.~top-down approach (right).
  }
  \label{figure:mpi:spacetree-decomposition-types}
\end{figure}

\subsection{Bottom-up and top-down tree partitioning}
If the finest grid level is non-overlappingly distributed among ranks, we have
to clarify which rank holds which coarser grid entities.
Two options exist (Figure \ref{figure:mpi:spacetree-decomposition-types}):  
We can either assign the responsibility for every cell on every level to a
unique rank or we can replicate coarser grid levels on multiple ranks.
Such a distinction is important even for codes working solely with the finest
grid, as it clarifies whether individual ranks are aware of
the overall spacetree decomposition or just manage local decomposition
knowledge.
\newA{
 For the actual data flow however, both tree partitioning approaches
 become the same if solely the fine grid $\Omega _h$ is of interest.
}

Spacetree replication results from a classic bottom-up scheme
\newB{(compare \cite{Bangerth:11:dealiiwithp4est} for example)}: 
We start from the adaptive fine grid and decompose its
cells into chunks.
From hereon, we construct each rank's spacetree recursively: 
let a cell $a \in \mathcal{T}$ be held by a particular rank with $a \sqsubseteq
_{child\ of} c$.
Then, $c$ is held by this rank, too.
Obviously, $c$ is replicated on several ranks if its children are
replicated or distributed among several ranks.
The coarsest spacetree cell is available on all ranks.
Bottom-up schemes often use the term local tree for the tree held on
a particular rank.
The coarser the level the more redundant the data.
This implies
that (fragments of) global knowledge about the chosen domain splitting is
available per subtree.

The opposite approach without data replication is a top-down
splitting \cite{Weinzierl:09:Diss}: 
We assign the global spacetree root to one rank.
For each $a,c \in \mathcal{T}, a \sqsubseteq _{child\ of} c$ with $c$ associated
to a rank $r_1$, $a$ is either held by this rank $r_1$ as well, or it is
deployed to another rank $r_2$ that has not been employed on a coarser or the
same level yet.
We exempt siblings along the SFC from the latter constraint.
The rule applies recursively.
It introduces a logical tree topology among the compute ranks:
$r_2$ serves as worker to a master $r_1$.
Whenever a child of a refined cell is assigned to another rank, this child 
acts as root of a remote tree. We cut out subtrees from the 
global spacetree. 
% If sibling subtree roots can be assigned to one new rank to
% facilitate better load balancing.
% Individual ranks then work on a forest of octrees though all trees of this
% forest should be neighbours along the space-filling curves to preserve the
% connectivity and advantageous surface-to-volume ratios. 
% The logical master-worker topology on the ranks is preserved by such a
% modification. 

\begin{observation}
The majority of spacetree codes favor, to the best of our knowledge, the
partial replication with SFC cuts.
\end{observation}

\noindent
Using SFCs to obtain an appropriate initial splitting of the finest
grid level is popular.
SFC partitions exhibit an advantageous
\newB{surface-to-volume} 
ratio\footnote{To the best of our knowledge, the good ratio is
an empirical observation that can be backed-up by proofs only for regular grids where it results directly from the H\"older
continuity (cf.~\cite{Bader:13:SFCs,Bungartz:06:Parallel} and references
therein).
\newB{
Further advantageous connectivity properties are discussed in
\cite{Burstedde:17:Morton} and notably \cite{Haverkort:16:SFCs}.
}
} 
and result from a straightforward splitting of the SFC's
one-dimensional preimage, i.e.~the enumeration of cells.
We note that SFCs also can be used for the top-down splitting as the SFC motif
orders all levels: As long as any rank forks subtrees only along the SFC
to other ranks, the resulting splitting is an SFC-splitting, too.

\begin{designdecision}
Peano works with a non-replicating data layout, \newB{i.e.~with the top-down
tree partitioning.}
\newC{
 This tree distribution induces a logical master-worker topology on the ranks.
}
\end{designdecision}

\noindent
% While the availability of global partitioning knowledge in the replicated case
% is appealing, 
Ranks in Peano are aware only of their multiscale neighbors as
well as their master and worker ranks.
No global information is held per rank.

\subsection{Parallel tree traversal}

A parallel tree traversal on replicated trees yields one automaton per rank,
each traversing its local linearised tree.
All start their tree traversal at the same time.
\newB{In this approach,}
tree nodes are labeled as replicated, local or empty.
Empty spacetree nodes are required if only some children of a
refined node are processed locally, i.e., they are nodes purely
required to complete the spacetree.
As multiscale data is held redundantly, all information flow from coarser to
\newA{finer} grids can be \newA{realized} without communication.
After or throughout the backtracking, all replicated data has to be
\newA{synchronized}.
\newB{Peano however works without replication.}

A parallel tree traversal for the non-replicated tree yields one automaton per
rank, too.
However, the automata may not run in parallel right from the start since they
are \newA{synchronized} with each other through (\ref{equation:order-on-events}).
If a child of a refined cell is assigned to a remote rank, the remote rank's
traversal automaton is triggered to start \newA{traversing} `its' tree by the
automaton traversing the refined cell.
This motivates the term worker.
%Only the local spacetree is linearised.
Throughout the bottom-up steps, an automaton in return has to wait for workers
to finish prior to further backtracking.
Both master-worker and \newB{worker-master} communication are point-to-point
communication keeping the two-grid interface between masters and
workers consistent.
%Typically, these are multiscale data.

\begin{algorithm}[htb]
 \caption{
  \newA{Blueprint of our level-wise depth-first traversal.
  } 
 }
 \label{algorithm:lw-dfs}
 {\footnotesize
  \begin{algorithmic}[1]
   \Function{traverse}{$S_{in},S_{out}$,$S$} 
    \For{$i \in \{0,\ldots,3^d-1\}$ along SFC($S$)}
     \State $currentCell_i \gets pop(S_{in})$
     \Comment Load $3^d$ children in one rush
    \EndFor
    \For{$i \in \{0,\ldots,3^d-1\}$ along SFC($S$)}
     \For{adjacent vertices $v$ of $currentCell_i$}
      \Comment Load their vertices
      \If{$v$ is hanging vertex on level $\ell$}
       \Comment \newB{Traversal automaton identifies (and marks) } 
       \State \texttt{createHangingVertex}($v$,\ldots)
       \Comment \newB{hanging vertices autonomously}
      \ElsIf{$v$ used for the very first time}
       \State $v \gets $ load from input stream
       \State \texttt{touchVertexFirstTime}($v$,\ldots)
      \Else
       \State $v \gets $ load from temporary data container
      \EndIf
     \EndFor
    \EndFor
    \For{$i \in \{0,\ldots,3^d-1\}$ along SFC($S$)}
     \If{$currentCell_i$ is root of deployed subtree}
      \State trigger remote traversal of $currentCell_i$ and continue
     \Else
      \State \texttt{enterCell}( $currentCell_i$  )
     \EndIf
    \EndFor
    \For{$i \in \{0,\ldots,3^d-1\}$ along SFC($S$)}
     \If{$currentCell_i$ is refined or to be refined and not remote}
       \State \Call{traverse}{$S_{in},S_{out}$,$S$}
       \Comment Pass down mirrored, scaled and translated $S$ 
     \EndIf
    \EndFor
    \For{$i \in \{0,\ldots,3^d-1\}$ along SFC($S$)}
     \If{$currentCell_i$ is root of deployed subtree}
      \State wait for remote traversal on $currentCell_i$ to terminate and
      reduce data (if reduction enabled)
     \Else
      \State \texttt{leaveCell}( $currentCell_i$  )
     \EndIf
    \EndFor
    \State \ldots
   \EndFunction
  \end{algorithmic}
 }
\end{algorithm}

DFS is problematic for parallel codes without
replication \newA{unless communication is explicitly eliminated as discussed
below}:
It is strictly sequential.
Peano therefore applies 
one-step recursion unrolling \cite{Eckhardt:10:Blocking,Weinzierl:09:Diss} on the DFS:
In each refined node, the automaton reads in all children.
After the $k^d$ children are processed, they are put on the call stack and the
automaton descends along the children's order.
Once all $k^d$ recursive descends have terminated, $k^d$ cells are taken from
the call stack and the code backtracks.
This is a one-step breadth-first traversal within the depth-first framework.
We call it {\em level-wise depth-first} \newA{(Algorithm
\ref{algorithm:lw-dfs})}.
It allows us to trigger remote subtree traversals before we descend locally.
Though \newB{it} resembles
(\ref{equation:recursion-unrolling-grammar}), it does not rely on $f$
and is applied always; even if the tree is subject to change.
Level-wise DFS is not literally a third traversal paradigm realized: 
It evolves from DFS through a one-step recursion unrolling, while BFS can be
read
as transitive hull of one-step recursion unrolling steps over DFS.
Peano's mesh traversal thus can be formalized through DFS plus one-step
recursion unrolling.

The replicating
scheme comes along with a higher memory overhead than the top-down splitting.
There is consequently more data to keep consistent, and 
data exchange involves typically more than two ranks---notably on
the global root of which all ranks hold a replica.
The top-down approach requires solely point-to-point data exchange and minimizes
data redundancy. 
However, top-down induces a tighter, latency-sensitive coupling
\cite{Weinzierl:16:PIC}:
data of coarse cells is propagated into finer grid resolutions which might act
as coarsest resolutions to remote trees throughout the \newB{descent} of the
automaton. 
This on-the-fly information propagation has to integrate into the wake-up calls
of traversal of worker ranks.
Similar observations \newA{hold} for the bottom-up information flow.
% 
% While a scheme without data replication saves memory and 
% reduces the amount of data that has to be synchronised,
% this advantage is bought by a tighter coupling of
% the data traversals if (\ref{equation:order-on-events}) is not stripped of
% coarse grid data.
% It is thus, to the best of our knowledge, unclear which variant is better in
% which context.s

\subsection{Data exchange}

Both decomposition schemes distinguish two types of data exchange: 
Vertices that are adjacent to cells handled by different ranks are replicated
among all ranks and are subject \newB{to} horizontal data exchange
\cite{Reiter:13:GeometricMultigrid} to keep them consistent.
Vertices and cells that are held on two ranks due to a master-worker
decomposition 
are subject \newB{to} vertical data exchange.

Vertical data exchange is synchronously \newA{realized}.
Data is sent from the master to the worker upon the wake-up call and
coarse information from a traversal thus prolongs immediately, i.e.~in the
same traversal, to the worker.
Data is sent from the worker back to the master when the worker terminates.
Fine-to-coarse data propagates immediately in the same traversal.

Horizontal data exchange is asynchronously  \newA{realized} by non-blocking MPI.
Vertex copies and their data are sent out from one rank to all other ranks
holding a replica once they have been processed by the traversal.
It is received prior to the first re-read of a vertex in the
subsequent traversal.
We therefore throttle the refinement with a marker-refine scheme: 
Ranks may trigger a refinement or coarsening, respectively, in one iteration.
It however is not realized along a parallel boundary before the subsequent
traversal where all ranks holding a replica of a vertex have received the
grid modification request.
% Data exchanged vertically typically is small compared to data exchanged
% horizontally.
% However, horizontal data exchange can be hidden behind computations: vertex data
% can be sent out when a vertex has been written for the last time.
% It has to be received before it is read for the first time in the subsequent
% iteration.
Both data exchange patterns apply to stacks and heaps.

The usage of the Peano SFC simplifies and speeds up the  \newA{realization} of
the horizontal data exchange.
Let $v_a$ and $v_b$ be two vertices held both on rank \newB{$r_1$} and
\newB{$r_2$}.
We align the traversal orders on both \newB{$r_1$} and \newB{$r_2$} such that
$v_a$ is used for the last time before $v_b$ on both ranks.
Each rank thus can send out $v_a$ immediately to the other rank once $v_a$ has
been used for the last time.
The exchange of $v_a$ is automatically hidden behind the remainder of the
traversal (finishing work on $v_b$, e.g.).
In practice, multiple vertex sends are grouped into one chunk of data to reduce
MPI overhead.
% We nevertheless retain the automated data exchange hiding as long as partitions
% do not degenerate to single cells.

We reiterate that Peano's projection onto the surface of a partition yields a
Peano curve of a reduced dimensionality \cite{Weinzierl:11:Peano} and thus
totally orders all vertices/faces on any subpartition.
As a result, the data exchange between any two ranks can be modeled by one
channel/stream and no reordering of any incoming data is required as long as we
invert the traversal on all ranks after each grid sweep
\cite{Bungartz:06:Parallel}.
$v_a$ is sent out before $v_b$. 
The send of $v_a$ is hidden behind the treatment of $v_b$.
$v_b$ is read in the subsequent iteration from the remote rank before $v_a$.

\begin{designdecision}
\newA{
Peano exploits the fact that the Peano SFC projected onto a partition boundary
yields a $d-1$-dimensional Peano SFC \cite{Weinzierl:11:Peano}.
This makes all data exchange streams without any reordering.
To stick to this paradigm, we disable all DFS-BFS transformations along MPI
boundaries.
}
\end{designdecision}

\noindent
\newA{
 To realize this \newB{disabling}, we set the markers in 
 (\ref{equation:recursion-unrolling-grammar}) for all cells along an MPI 
 boundary to $\bot $.
 A hybrid MPI+X code runs MPI globally, the grid skeleton per rank remains
 single core, and it triggers shared memory
 parallelization only on interior enclaves \cite{Sundar:15:Enclave}.
}

\begin{table}[htb]
 \ifthenelse{\boolean{toms}}{
  \tbl{
    Additional events that are available in Peano if code is compiled with MPI.
    \label{table:mpi-events}
  }
 }{
  \caption{
    Additional events that are available in Peano if code is compiled with MPI.
  }
  \label{table:mpi-events}
  \vspace{-0.2cm}
  \begin{center}
 }
  {\footnotesize
  \begin{tabular}{l|p{0.66\textwidth}}
    {\bf Event} & {\bf Semantics} \\
    \hline
    \texttt{mergeWithNeighbour} & Called for a vertex per neighbouring rank
    before \texttt{touchVertexFirstTime} is invoked. Passes a copy, i.e.~the
    received replicate, from the neighbour alongside with the vertex data. \\
    \texttt{prepareSendToNeighbour} & Counterpart of
    \texttt{mergeWithNeighbour} that is called per neighbouring rank after
    \texttt{touchVertexLastTime} to produce the copy of the vertex that is
    then sent away. \\
    
    \hline
    
    \texttt{prepareSendToWorker} & Plug-in point to transfer data from master
    to worker just before the worker's traversal is invoked. \\
    \texttt{prepareSendToMaster} & Plug-in point to transfer data from worker
    to master just before the worker quits a traversal. \\
    \texttt{mergeWithWorker} & Counterpart of \texttt{prepareSendToWorker}
    invoked on the worker. \\
    \texttt{mergeWithMaster} & Counterpart of \texttt{prepareSendToMaster}
    invoked on the master. \\
    
    \hline
        
    \texttt{prepareCopyToRemoteNode} & Event invoked just before data is
    migrated to another rank due to dynamic load balancing. 
    \\
    \texttt{mergeWithRemoteDataDue-} \\
    \phantom{xxx} \texttt{ToForkOrJoin} & Counterpart of
    \texttt{prepareCopyToRemoteNode}. Rebalancing comprises two steps: a
    replicate of the tree parts is created on the new remote worker and then
    the data is transferred through these two events to the new worker. This
    way, data that can be regenerated on-the-fly does not have to pass through
    the network. \\
  \end{tabular}
  }
 \ifthenelse{\boolean{toms}}{
 }{
  \end{center}
 }
\end{table}

\subsection{Parallel programming interface}
Peano's non-overlapping strategy with its logic tree topology is mirrored by 
additional events (Table \ref{table:mpi-events}).
MPI Peano applications are strict extensions of a serial code base.
No
behavior of serial events is altered.
The main control loop is 
\newB{run} 
only on rank 0 (global master), and
any choice of a particular 
\newB{event set}
to be ran is automatically \newB{broadcast} to all
working ranks.
Upon an \texttt{iterate}, the global master's traversal automaton starts to
run through its tree and recursively triggers traversals on all
other ranks' spacetrees.
Per rank the same event set is invoked.

Dynamic load balancing is hidden from the events.
A distinct set of event-like plugin points does exist.
\newB{
 They allow the user to decide where to cut the tree, i.e.~which subtrees to 
 outsource to other ranks.
 It also allows to join two trees.
}
Due to \newB{these routines}, we may realize various load decomposition schemes. 
% controlling which subtrees are deployed to (which) new ranks or which
% master-worker decompositions shall be removed due to a tree merger.
As out-of-the-box, proof-of-concept solution, Peano comes along with a
greedy spacetree decomposition.
\newB{
 It assumes that each tree node has the same computational cost.
 The critical 
 \newC{
  path(s) 
 }
 within the global spacetree are identified, and the code
 then searches for the biggest subtree along this path that can be deployed to
 another rank. 
 The algorithm continues until no further ranks are left or the tree is 
 distributed completely.
}

\subsection{Communication reduction and elimination}

One showstopper in obtaining parallel scalability is any
algorithmic \newA{synchronization} (lock stepping).
\newA{Synchronization} in Peano materializes notably as vertical data exchange.
To streamline the master-worker communication, we allow the global master to
run a fixed number of grid traversals.
All MPI ranks are then informed beforehand about this fixed number of
traversals.
At the same time, users may specify that local automata do not require
information from coarser levels in these sweeps.
Each rank thus runs a fixed number of iterations and couples to its multiscale
neighbor, but the ranks are not globally \newA{synchronized} with each other.
\newA{We call one set of grid sweeps a batch.}

The other way round, users can specify whether and which data is to be sent from
workers to masters upon completion.
This mechanism unfolds its full potential once we clarify that the
wake-up call from a master decides whether vertical data transfer is required on a per rank per grid sweep base.
Every time a worker traversal is started, the aligned event
\texttt{prepareSendToWorker} on the master returns a flag, Peano memorizes this
flag and, depending on it, skips the reduction from this worker.
\newB{
 Users can explicitly disable inter-level (vertical) constraints from
 (\ref{equation:order-on-events}) between MPI ranks. 
}

Besides vertical information, the code also offers routines to switch off
horizontal data exchange via the stacks, 
users can \newA{minimize} the horizontal data transfer through the heaps by
sending out only those vertices/attributes that have changed, and we support algorithms that send out
heap data in one iteration but receive it $n$ iterations later.
This allows us to interweave grid traversals into a communication-demanding
scheme: data sent out in one iteration is allowed to run through the network
while other grid traversals are executed.

\begin{designdecision}
 \newA{Peano's default is a}
  non-replicating scheme where all traversal automata are
  \newA{synchronized} vertically both top-down and bottom-up.
  Yet, we allow the user codes to skip either \newA{synchronization}.
  This decision can be made autonomously by each master per worker per grid
  traversal.
\end{designdecision}

\noindent
If all master-worker data flow is masked out, Peano's communication patterns
resemble replicating schemes. 
All tree automata start to traverse, though with level shifts, at the same time.
If all worker-master data flow is eliminated, too, and global synchronization is
realized by the user within the events manually through MPI calls, the overall
communication scheme is exactly the same as in a replicating strategy.

\subsection{Limitations of the approach}

It remains open whether replicating or non-replicating spacetree
decomposition are superior. 
Non-replicating schemes tend to synchronize tighter, replicating schemes come at
the cost of more data to be held consistent.
They nevertheless seem to be more straightforward to program as synchronization
does not block other ranks to continue their traversal. 
It thus might be reasonable, in the future, to offer both alternatives.
Overlapping schemes fit into the presented mindset and have
successfully been applied to spacetrees and SFCs.
It however is unclear, to the best of the author's knowledge, how the overlaps
interplay with the data flows of a scheme that remains
non-replicating wherever possible.

\newA{
 It is in this context important to observe that our distinction of bottom-up
 and top-down decomposition starts to blur depending on the usage:
 If solely data on the finest grid is held, DFS in its unmodified form does not 
 serialize the distributed tree traversal.
 While we can filter out grid events on coarser levels on a per-rank
 basis, i.e.~for the non-distributed algorithm part, our code base also
 allows us to switch off reduction.
 This can be seen as parallel counterpart to the filtering.
 It makes the traversal pick up characteristics of leaf-only codes as detailed 
 in the experiments.
 We finally remark that bottlenecks in reductions often are
 mitigated by the introduction of a tree structure and replication of coarse
 grid nodes.
%  The tree in our code is automatically introduced.
%  Replication is currently out of scope.
} 

\section{Experiments}
\label{section:experiments}

 \begin{figure}[htb]
   \begin{center}
   \includegraphics[width=.25\textwidth]{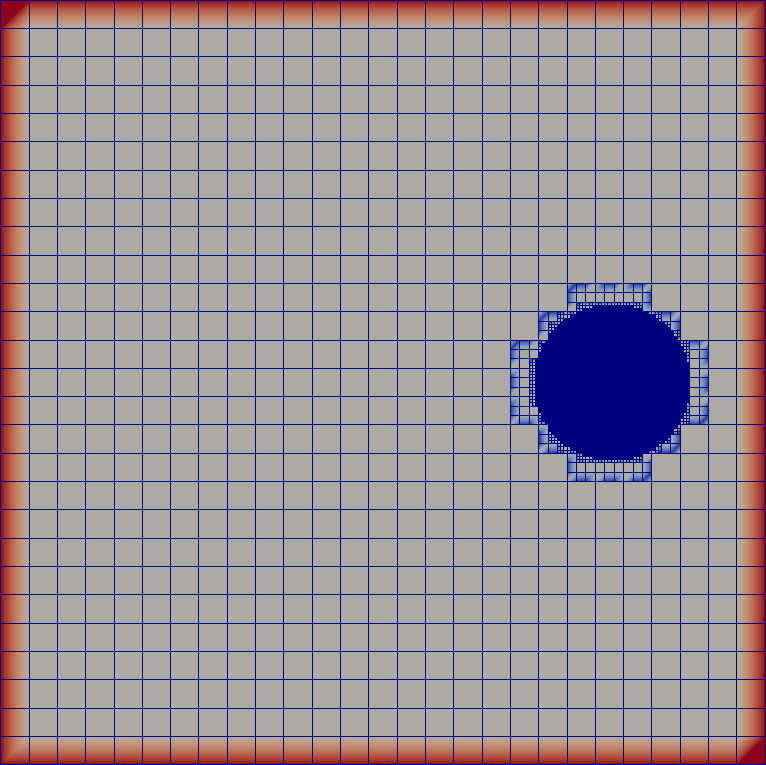}
   \hspace{0.4cm}
   \includegraphics[width=.25\textwidth]{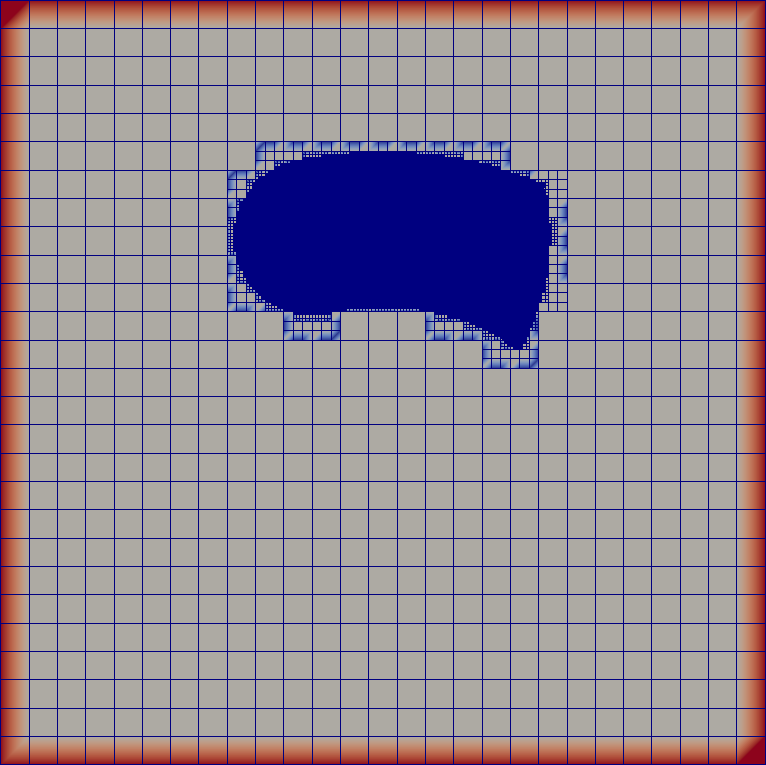}
   \hspace{0.4cm}
   \includegraphics[width=.22\textwidth]{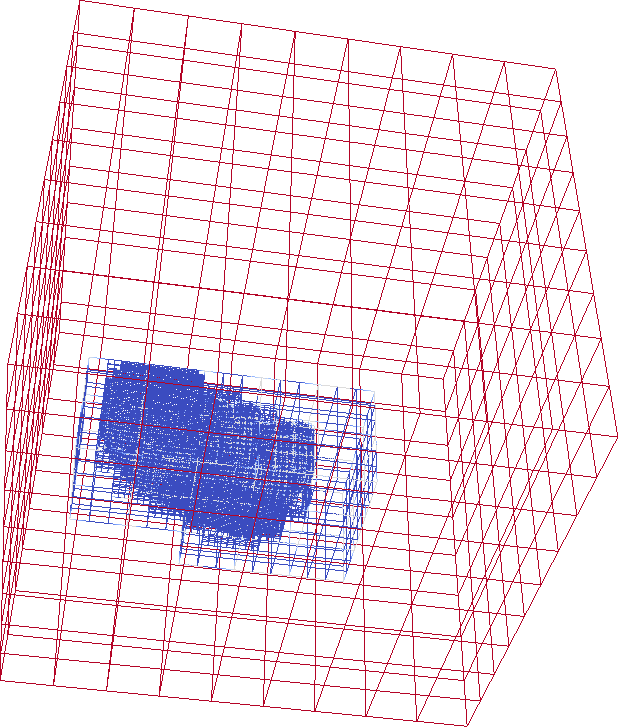}
   \caption{
     \newB{
     Grid snapshots of our $d=2$ test setup at startup (left) and after few
     time steps (middle) with $h_{max}=3^{-4}$ and $h_{min}=3^{-8}$,
     i.e.~$\Delta \ell =4$.
     The circular region of refinement runs anticlockwise.
     Right: Latter situation for $d=3$ mesh with $h_{max}=3^{-3}$ and
     $h_{min}=3^{-6}$.
     }
     \label{figure:results}
   }
   \end{center}
 \end{figure}

We close our discussion with benchmarks to uncover some of 
the performance characteristics tied to the grid management,
traversal and programming paradigm.
Our tests restrict to properties studied separately from each
other.
\newB{
Real-world examples bringing together spacetrees with applications and thus
also interweaving the individual properties are beyond scope.
}
Examples for such configurations are discussed in the appendix
\newB{or respective papers}.
Despite the lack of application context, we report runtime cost per grid
entity---a quantity which has to change if an application plugs into Peano.

Throughout our experiments working with grids illustrated in Figure
\ref{figure:results} we use six Peano instances holding 2, 11, 28, 768, 2,304
or 3,864 doubles per vertex.
\newB{
  All floating point operations are removed, as they depend strongly on the
  particular application.
  Without arithmetic load, our setup}
mirrors the Stream
\newB{COPY} benchmark \cite{McCalpin:95:Stream}.
% Other work-per-byte configurations resemble characteristic application
% profiles:
Two unknowns per vertex have to be stored at least for any matrix-free equation
system solve.
They hold the right-hand side and the solution.
%  which yields between 32
% and 128 flops per double for a cell-wise matrix-vector product.
If we store low-order discretization stencils per vertex, the memory footprint
grows to at least 11 doubles per vertex (9 for the stencil for $d=2$ plus the two
unknowns) or 28 for $d=3$ if we have an analytic right-hand side.
Patches within the spacetree can yield any memory footprint per grid entity, but
768 unknowns per cell or vertex, respectively, are a reasonable ``small'' count.
It arises for example from the shallow water code
\cite{Weinzierl:14:BlockFusion} with $16\times 16$ patches and three unknowns (velocity plus water height).
%embedded into the spacetree's fine grid cells.
Lattice Boltzmann with D2Q9 using $16 \times 16$
subgrids at the same time yields 2304 unknowns, while D3Q19 yields 3864
\cite{Neumann:13:Diss}.

Throughout the experiments, we manually prescribe the adaptivity: 
We first fix a minimum and a maximum grid resolution ($h_{min} \leq h_{max}$).
If they are equal, we study a regular grid.
Otherwise, we refine around a circle within the domain to the finest
spacetree level meeting $h_{min}$ and coarse outside of this circle up to the
coarsest mesh level meeting $h_{max}$.
\newB{
 All data quantifies the difference of $h_{max}$ and $h_{min}$ as difference
 $\Delta \ell$ in levels.
 $\Delta \ell = 0$ denotes the regular grid with mesh width $h_{min}$.
}
No balancing \cite{Isaac:12:Balancing,Sampath:08:Dendro,Sundar:08:BalancedOctrees} is imposed.
The ``refinement'' circle follows an ellipsoidal
trajectory,
\newB{
 i.e.~the mesh adaptivity changes in each mesh sweep (Figure
 \ref{figure:results}).
 The unit square or cube is our computational domain.
 If not stated otherwise, we choose $h_{max}=4.57 \cdot 10^{-4}$ for $d=2$ and 
 $h_{max}=4.12 \cdot 10^{-3}$ for $d=3$.
 The unit square or cube is the computational domain, and we start to refine
 from $h_{max}$ on.
}
% (Figure \ref{figure:results}).
% Third, we decide whether this circle moves within the domain or not,
% i.e.~we decide whether we study statically or dynamically adaptive grids.
% If the circle moves, we adapt the grid in each grid sweep.
%Tests are ran for both $d=2$ and $d=3$.

\newB{We use Durham's Intel
E5-2650V4 (Broadwell) cluster Hamilton with 24 cores per node.
They run at 2.4 GHz and are connected by Omnipath.
% Some experiments were redone on an Intel
% Knights Landing chip (Xeon Phi 7250) at 1.40GHz.
Upscaling data stems from 
}
SuperMUC hosting Sandy Bridge-EP
Xeon E5-2680 processors at 2.7 GHz.
All shared memory tests rely on Intel's Threading Building
Blocks (TBB).
% For the distributed memory parallelization, we use IntelMPI 5.0.3.
% Intel's C++ compiler 2015 version 2 translates all codes on the Ivy
% Bridges, while we use the 2017 version on the KNL.

\subsection{\newB{Persistency model and data management}}
\label{subsection:persistency-model-and-data-management}

\begin{table}[htb]
 \ifthenelse{\boolean{toms}}{
  \tbl{
    \newB{
    Performance counter data for \newC{a} $2$-dimensional benchmark setup. We
    compare stack- (top) to heap-based (bottom) data management.
    Bandwidth is given as MB/s while L2 denotes the L2 cache miss rate.
    }
    \newC{
    $\Delta \ell=3$ uses around 45,000 spacetree vertices per step, while the
    regular grid ($\Delta \ell =0$) reads and writes the data of 110,000
    vertices.
    }
    \label{table:results:performance_counters}
  }
 }
 {
  \caption{
    Performance counter data for a $2$-dimensional benchmark setup. We
    compare stack- (top) to heap-based (bottom) data management.
    Bandwidth is given as MB/s while L2 denotes the L2 cache miss rate.
    $\Delta \ell=3$ uses around 45,000 spacetree vertices per step, while the
    regular grid ($\Delta \ell =0$) reads and writes the data of 110,000
    vertices.
  }
  \label{table:results:performance_counters}
  \vspace{-0.2cm}
  \begin{center}
 }
  {\footnotesize
  \begin{tabular}{r|rr|rr|rr|rr|rr}
 & \multicolumn{2}{|c}{2} & \multicolumn{2}{|c}{11} & \multicolumn{2}{|c}{768} & \multicolumn{2}{|c}{2304} & \multicolumn{2}{|c}{3864} \\ 
$\Delta \ell $ & BW & L2  & BW & L2  & BW & L2  & BW & L2  & BW & L2  \\ \hline 
3 &  15.33 & 0.0004 &  28.67 & 0.0009 & 1,061.04 & 0.030 &   863.16 & 0.06 &   839.41 & 0.12 \\ 
2 &  13.54 & 0.0005 &  12.70 & 0.0012 & 2,694.28 & 0.037 & 2,020.64 & 0.13 & 1,806.46 & 0.45 \\ 
1 &  15.06 & 0.0004 &  27.07 & 0.0010 & 2,744.77 & 0.037 & 1,999.11 & 0.13 & 1,822.25 & 0.48 \\ 
0 &  55.04 & 0.0006 & 174.40 & 0.0010 & 2,853.33 & 0.036 & 1,365.75 & 0.14 & 1,798.71 & 0.48 \\ 
   \hline
3 &  15.62 & 0.0008 &  31.78 & 0.0009 & 1,819.75 & 0.010 & 3,327.44 & 0.021 &  4,068.98 &  0.027 \\ 
2 &   7.32 & 0.0009 &  10.74 & 0.0010 & 3,941.51 & 0.013 & 7,426.19 & 0.030 &  8,951.68 &  0.041 \\ 
1 &  48.29 & 0.0009 &  47.10 & 0.0010 & 4,156.80 & 0.013 & 7,556.84 & 0.030 &  9,179.52 &  0.041 \\ 
0 & 140.22 & 0.0009 & 215.27 & 0.0010 & 4,404.97 & 0.014 & 7,822.45 & 0.030 &  7,833.64 &  0.041 
  \end{tabular}
  }
 \ifthenelse{\boolean{toms}}{
 }{
  \end{center}
  }
\end{table}

We start our experiments with single core measurements.
Our deterministic, \newB{automaton-based} grid traversal of a linearised
spacetree pipes two data streams---one for the cells, one for the vertices---through the \newB{chip}
(Figure \ref{figure:use-cases:automata}).
Temporary, intermediate storage resides in the caches.
Refinement or coarsening triggers stream insertions or the removal of elements
from the output stream.
This yields, despite the dynamic adaptivity, data accesses with high spatial and
temporal locality \cite{Kowarschik:03:CacheTechniquesOverview} and, hence, high cache hit
rates \cite{Bungartz:10:PDEFramework,Mehl:06:MG,Weinzierl:09:Diss,Weinzierl:11:Peano}.
\newB{
Performance counter measurements confirm this advantageous cache behavior for
various application configurations (Table
\ref{table:results:performance_counters}).
}

%
% @todo 2d
%

\begin{figure}
  \begin{center}
    \includegraphics[width=0.4\textwidth]{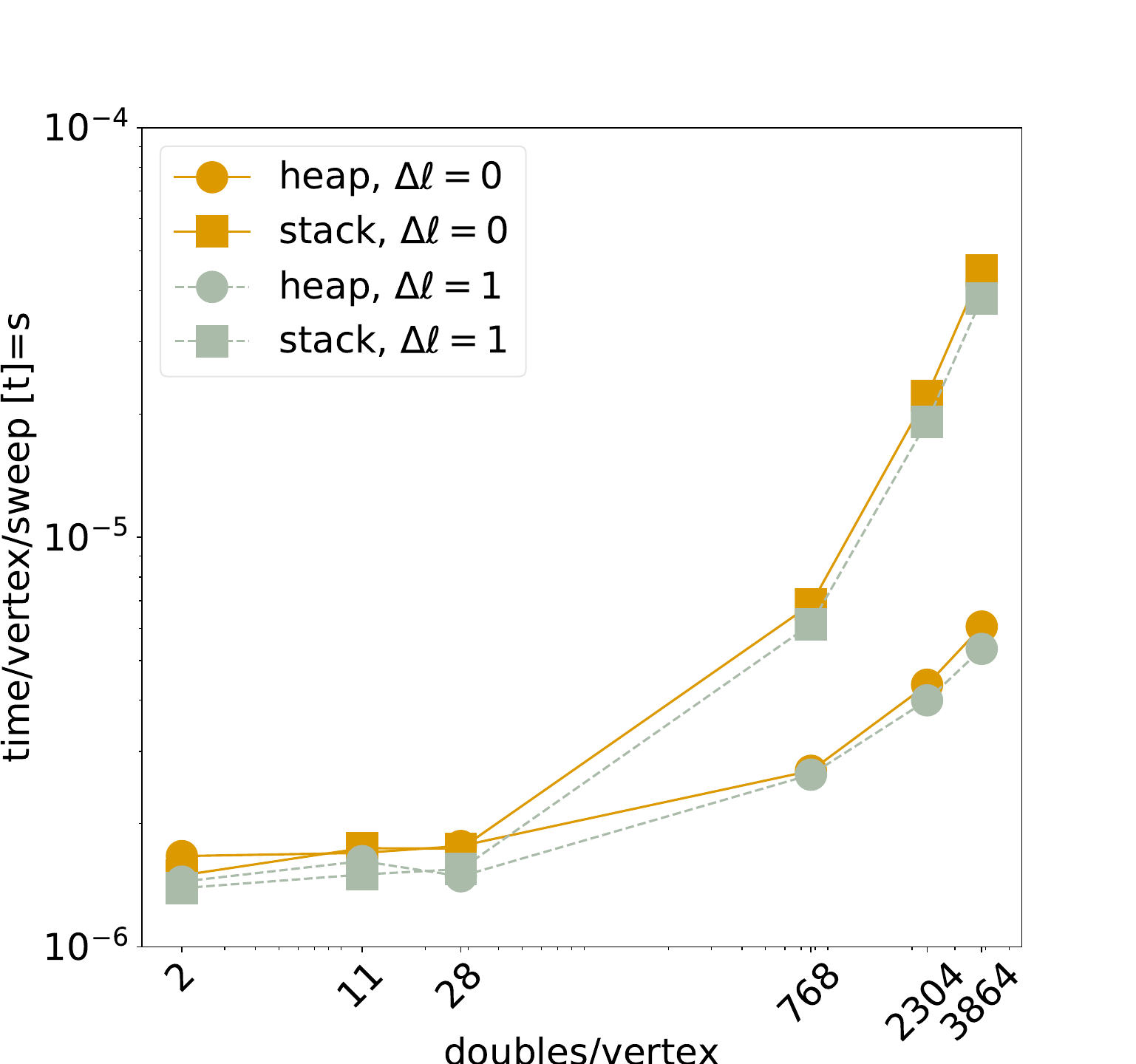}
    \includegraphics[width=0.4\textwidth]{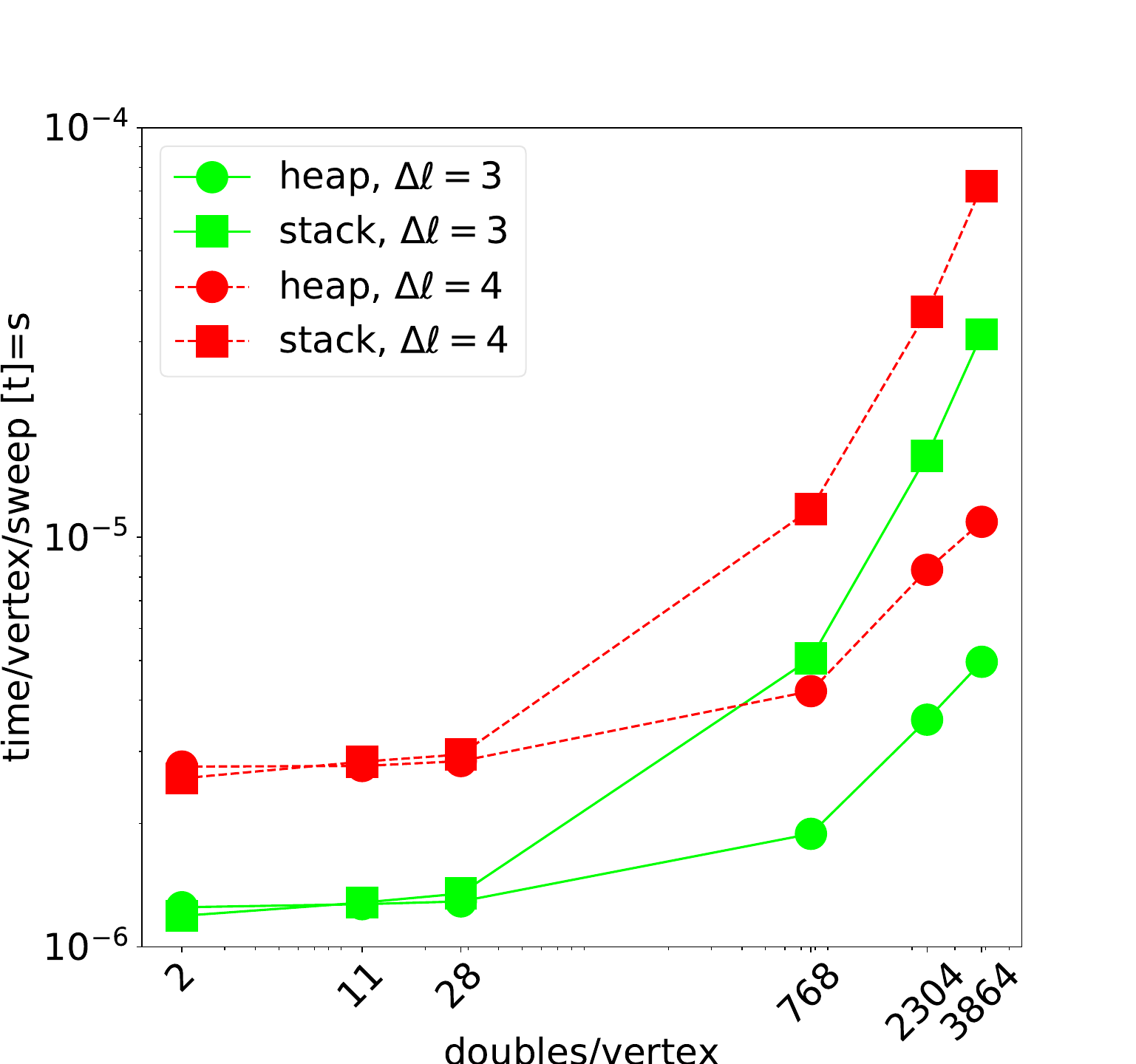}
  \end{center}
  \caption{
    \newB{
      Time per vertex per grid sweep for different level differences and
      dynamically adaptive $2d$ meshes.
      All floating point operations disabled. 
    }
    \label{results:storage-schemes:plain:hamilton}
  }
\end{figure}

%
% Characterization/explanation
%
% The $d=3$ grid administration is by at most a factor of two more expensive than
% its 2d counterpart, though the difference vanishes once the number of unknowns per
% grid entity grows large.
% The actual arithmetic cost per grid entity on one core does not make a
% difference here.
\newB{Our single core measurements tracking 
runtimes} (Figure \ref{results:storage-schemes:plain:hamilton}) compare stack-
to heap-based unknown storage.
For the setups with a small \newB{persistent} memory footprint per grid entity,
our approach is spacetree administration-bound as the
\newB{runtime profile is independent of the data stored per grid entity}.
From \newB{28} doubles per vertex on, the runtime increases with increasing data
cardinality.
The code's administration overhead team up with data
squeezing through the memory interconnect \newB{and caches}.
\newB{This holds for both storage paradigms.}

\begin{observation}
In the plain algorithm formulation, the \newB{administrative} cost per vertex is
almost agnostic of \newB{the} refinement pattern.
% \newB{For less than a}
% few hundreds of unknowns per grid entity, the multiscale
% spacetree administration cost dominate the runtime. 
\end{observation}

% \begin{figure}
%   \begin{center}
%     \includegraphics[width=0.4\textwidth]{experiments/storage-schemes/knl-2d-no-legend.pdf}
%     \includegraphics[width=0.4\textwidth]{experiments/storage-schemes/knl-3d-no-legend.pdf}
%   \end{center}
%   \caption{
%     Experiments from Figure \ref{results:storage-schemes:supermuc} reran on the
%     Intel KNL architecture.
%     \label{results:storage-schemes:knl}
%   }
% \end{figure}

% The computational work per grid entity plays a minor role here. 
% The spacetree administration overhead makes the three experiments with
% less 768 doubles per vertex give roughly the same performance.
% Only for bigger cardinalities, memory data transfer and compute workload
% co-influence the runtime.
% We validate this statement by a comparison to 
% Stream TRIAD which yields around 9,533 MB/s on the machine for a single core 
% and 68,961 MB/s if we saturate the bandwidth with eight cores.
% None of our single-core runs is bandwidth-bound.
% 
% 

% \marginpar{Triad baseline for Broadwell?}
% 
% All statements transfer qualitatively to the manycore 
% (Figure \ref{results:storage-schemes:knl}) though the results become more fuzzy.
% Notably, a heap-based unknown storage seems to pay off for $d=3$ 
% immediately.
% On the KNL, Stream TRIAD yields around 11,366 MB/s.
% For a single core the spacetree code exploits only half of the best-case single
% core throughput.
% This is the spacetree/adaptivity administration overhead.
% The difference in throughputs on the two architectures directly correlates with
% the difference in clock rates.
% 

%
% Observation
%

\noindent
\newB{Even in the ``worst case'' where (i) the mesh is extremely adaptive, (ii)
changes in each and every time step and (iii) each vertex has very small memory
footprint, the AMR overhead is not more than a factor of two.
However, memory allocations do slow down the code once the memory footprint per
vertex becomes reasonably high.
}

\newB{
Deploying user data to the heap yields a higher effective memory
throughput than sole stack-based storage.
It also yields a better cache hit rate.
As the heaps implicitly exploit the SFC/DFS locality through the tree access
keys, these observations imply that the stack-based approach suffers from
non-compulsory misses as data is written forth and back to temporary
stack containers.
Yet, the heap-based approach introduces one further memory access indirection
which renders a stack-based approach superior for very small memory footprints
(2 or 11 doubles per vertex only).
}
Embedding application-specific data into the
spacetree stream is advantageous if and only if the number of doubles per vertex
is small. 
As soon as we store more than a few doubles, it is advantageous to separate the
spacetree stream from the actual user data.
For large data cardinalities, we otherwise quickly loose up to an
order of magnitude of performance.
\newB{All observations also hold qualitatively for $d=3$.}

\begin{observation}
For very small data cardinalities per grid entity and rapidly
changing grids, it pays off to merge the compute data into the grid data.
Otherwise, it is better to separate the two containers.
\end{observation}

\noindent
\newB{
 These data confirm that effective grid implementations have to
 exploit any grid regularity or time invariance to bring
 administrative overhead \newC{down}.
 \newC{The other way round,}
 they suggest that effective codes based upon trees should
 exploit and introduce data regularity.
 Examples for this are block-structured or Structured AMR (SAMR) approaches
 \cite{Deiterding:05:AMR,Dubey:16:SAMR,Schornbaum:18:HPCBlockStructured,Weinzierl:14:BlockFusion}
 which strive for hybrids of free adaptivity and data regularity.
}
Overall, \newB{a code which is solely tree based}
remains significantly below the \newC{throughput} threshold 
\newC{of} the Stream \newB{COPY} benchmark \cite{McCalpin:95:Stream} on
a single core.
\newC{
 For the latter, we obtain 18,000--19,000~MB/s for setups which pick up
 memory footprints from the $\geq 768$ doubles-per-vertex configurations.
 With smaller per-vertex footprint, Stream COPY throughput becomes
 worse---down to 13,000 MB/s---until we reach a situation where the data
 footprint fits into a cache. 
 This holds for all memory footprints corresponding to 2 double-per-vertex and
 the adaptive 11 double-per-vertex choices.
 While Stream starts to exhibit throughputs significantly above 20,000~MB/s
 then, the tree traversal's administrative overhead stops us from
 benefitting from the fits into cache.
 For such tiny setups, a multiscale set of regular Cartesian meshes (arrays) is
 superior to the AMR tree idea.
}

\subsection{\newB{Transformation of DFS into BFS}}

\begin{figure}
  \begin{center}
%     \includegraphics[width=0.4\textwidth]{experiments/08b_transformation-of-dfs-into-bfs/heap-2d-2-log.pdf}
%     \includegraphics[width=0.4\textwidth]{experiments/08b_transformation-of-dfs-into-bfs/stack-2d-2-log.pdf}
%      \\
    \includegraphics[width=0.4\textwidth]{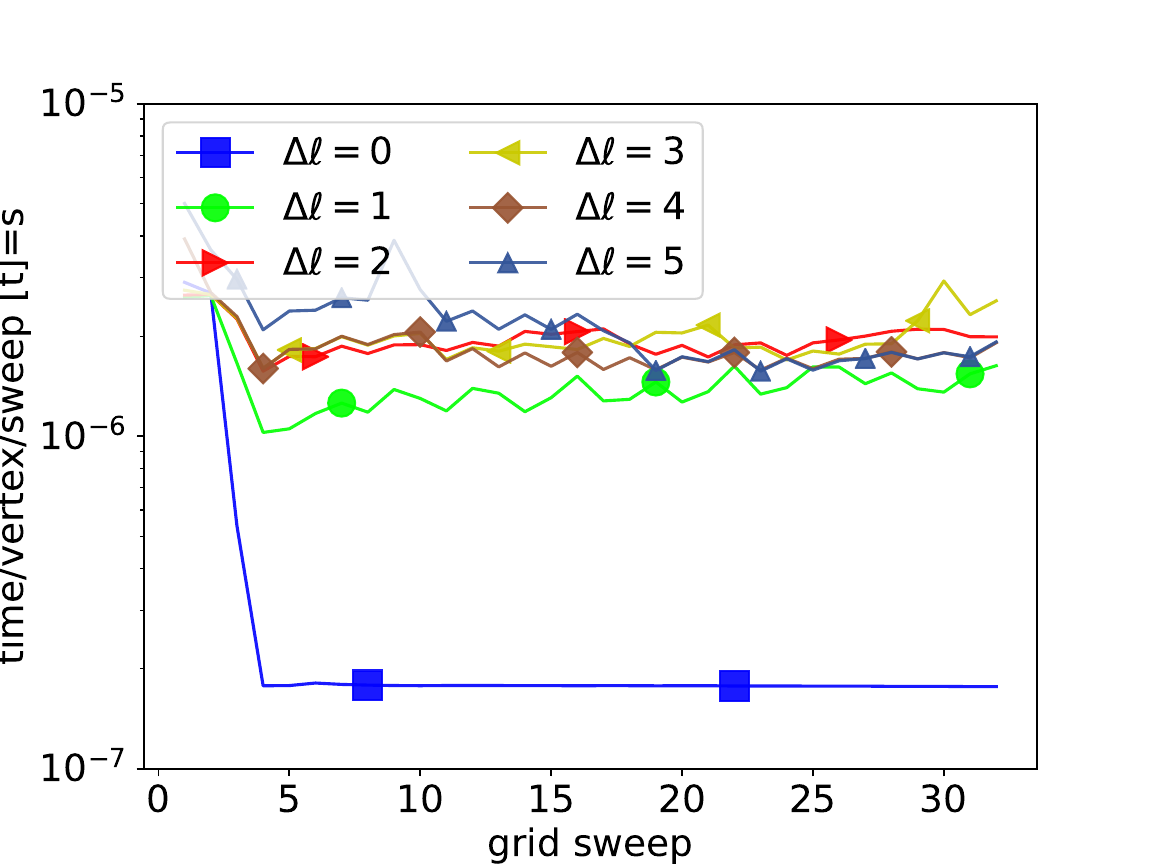}
    \includegraphics[width=0.4\textwidth]{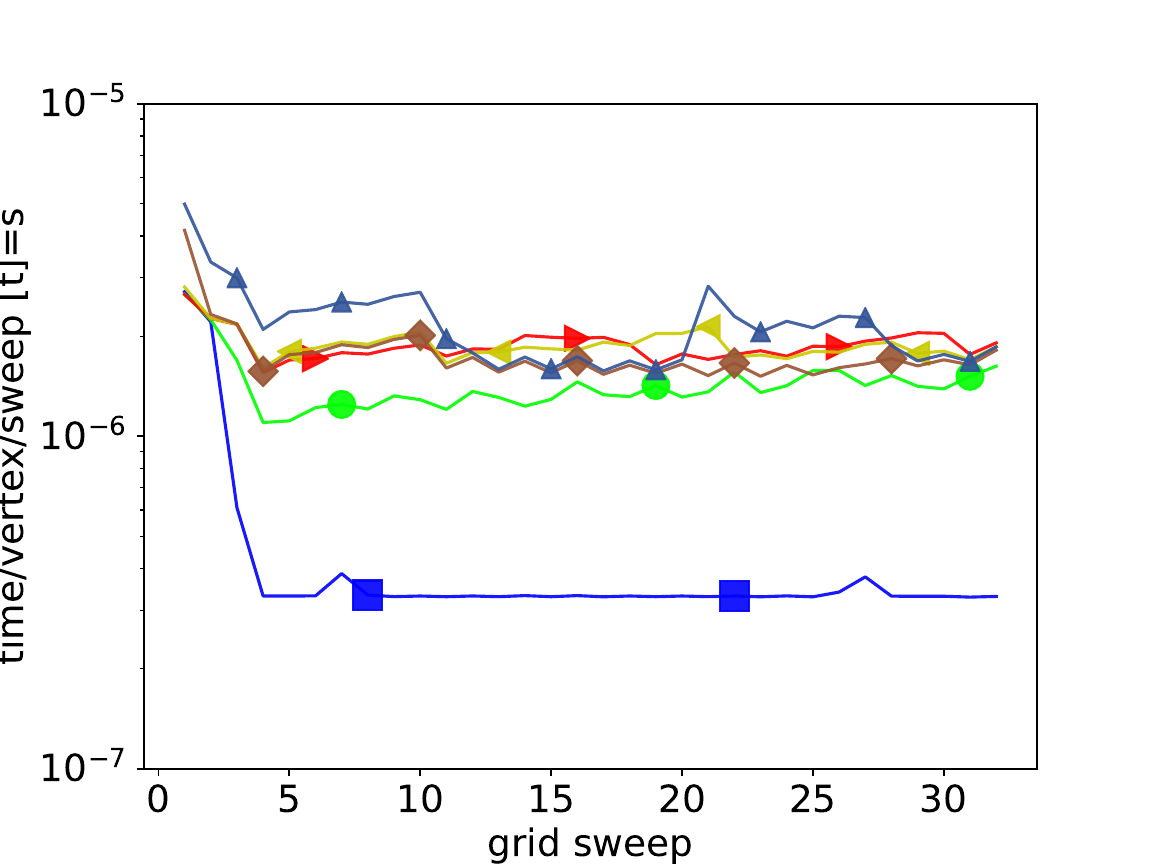}
     \\
%     \includegraphics[width=0.4\textwidth]{experiments/08b_transformation-of-dfs-into-bfs/heap-2d-2304-log.pdf}
%     \includegraphics[width=0.4\textwidth]{experiments/08b_transformation-of-dfs-into-bfs/stack-2d-2304-log.pdf}
%      \\
    \includegraphics[width=0.4\textwidth]{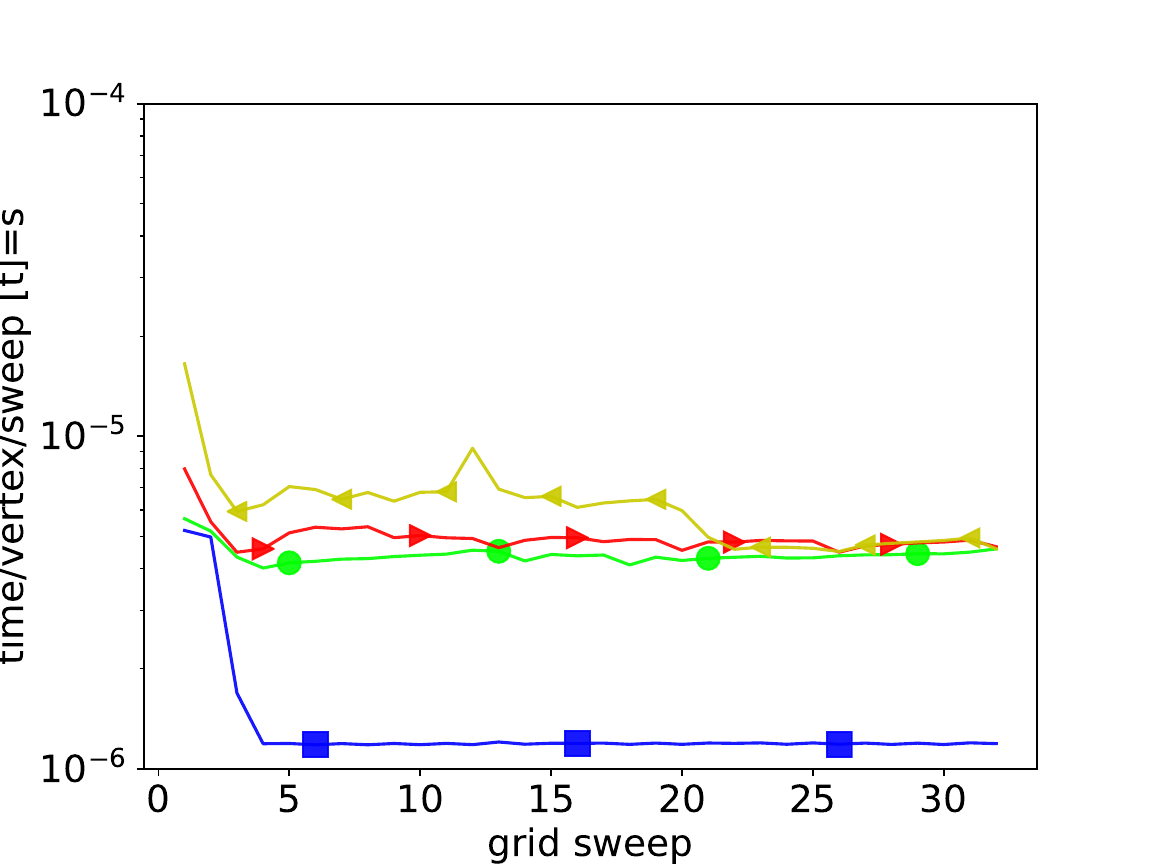}
    \includegraphics[width=0.4\textwidth]{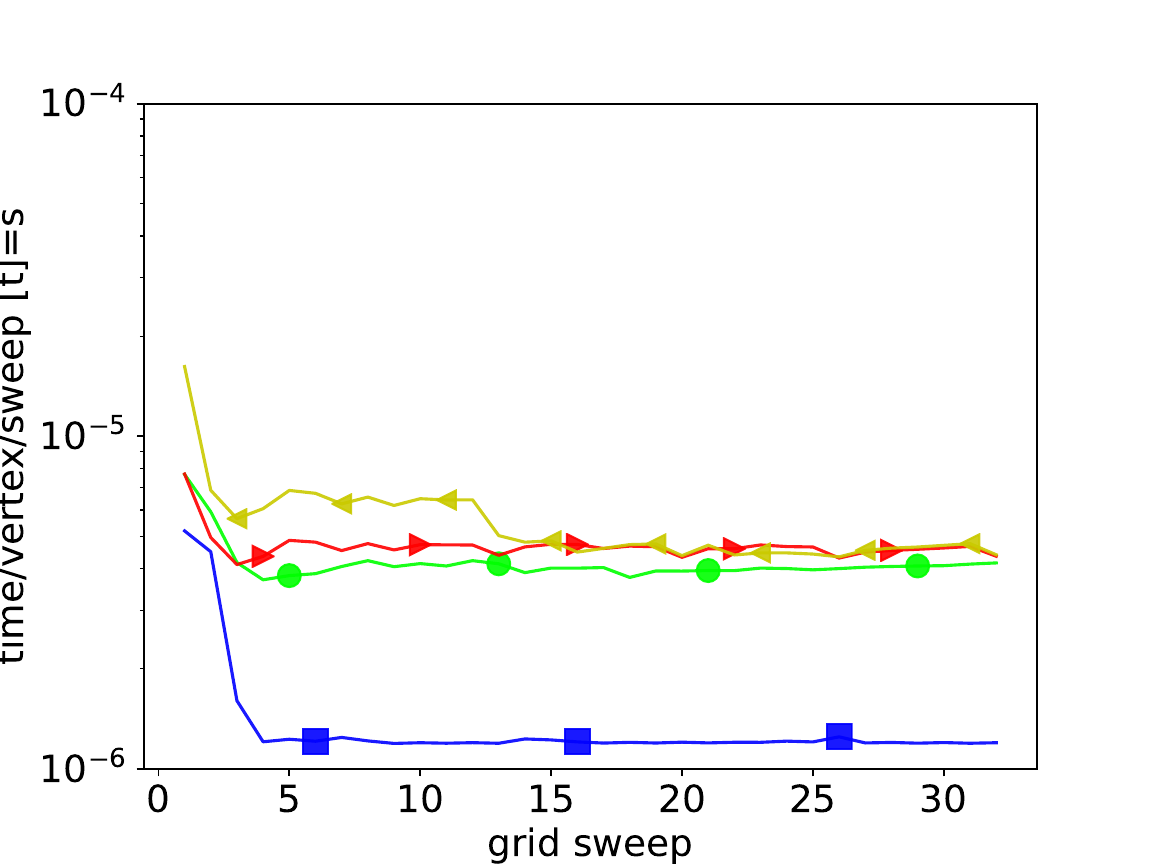}
  \end{center}
  \caption{
    Cost  \newA{(real time)} per grid sweep per vertex on a
    Hamilton node.
    $d=2$ \newB{(top) vs.~$d=3$ (bottom) with 28 doubles per vertex.
    We compare stack-based user data management (left) to heap-based storage
    (right).
    All floating point operations eliminated. 
    }
    \newC{
     $\Delta \ell $ denotes the AMR levels, so 
     $\Delta \ell =0$ identifies a regular grid.
    }
    \label{results:storage-schemes:persistent-subtrees}
  }
\end{figure}

%
% Problem
%
\newB{
 Despite good memory access characteristics, stream-based data management
 suffers from data movements.
 The pressure on the caches is high as data is permanently written and read from
 and to the caches.
 Notably vertices are written forth and back up to $2^d$ times.
 The more user data we hold on heaps, i.e.~the fewer data is embedded into the
 tree stream, the smaller the impact of this overhead.
%  If the actual user data is ``sourced out'' to a heap, this leverages the
%  movement penalty further.
 Still, the whole tree administration is not negligible.
 We therefore propose in Section
 \ref{section:traversal-optimization:regular-subtrees} 
 to cut out regular subgrids from the global grid.
 If all user data is stored on heaps, i.e.~no data is embedded into the
 (linearized) stream, no user data is unnecessarily moved by the grid traversal
 anymore within the regular subtrees.
}

%
% Description
%
Runs on the \newB{Broadwell} (Figure
\ref{results:storage-schemes:persistent-subtrees}) compare a plain realization
to a realization exploiting the grid regularity through an on-the-fly
switch into BFS.
\newB{This realization also stores} regular subtrees separately.
\newB{
If a part of the spacetree is stationary and regular, 
the traversal requires one} kick-off \newB{grid sweep where}
(\ref{equation:recursion-unrolling-grammar}) is determined.
\newB{If $f\geq 0$, the next iteration switches locally to BFS.
If a grid segment remains stationary and regular after the first BFS sweep, it
is cut from the global grid and stored separately.
The three-step pattern is clearly observed for the regular grid.
We note that the very first grid sweep is the grid construction where $f = \bot$
everywhere by definition.
Besides the three-step pattern,
} 
% The third sweep benefits \newB{most} from $f\geq 0$.
runtime peaks are caused by memory allocations for the regular subtrees
\newB{and whenever regular subtrees cut out of the mesh are reintegrated into
this very one due to dynamic mesh refinement or coarsening.
}

%
% Interpretation/evaluations
%
An evaluation of (\ref{equation:recursion-unrolling-grammar}) and an on-the-fly
switch to BFS for regular subtrees speeds \newB{up} traversals by \newB{up to a
factor of ten.
For the regular subgrids, it is advantageous to embed the user data into the
tree.
This is a surprise as heap-based storage for regular subgrids means that the
grid data is solely read but no data is written.
Heap-based storage however introduces scattering.
The regular grid levels stored as continuous block link to scattered data on the
heap.
% With stack-based storage where the whole subtree data is basically held as one
% continuous memory block, the single core performs at Stream TRIAD speed. 
} 
% remains independent of the adaptivity pattern.
% Refinement and erase are localized per grid sweep.
Unaltered grid regions \newB{are key to improve performance.} 
For them, there is an overhead through the traversal reordering that is compensated by the simplified event invocation and the
elimination of case distinctions.

\begin{observation}
  On-the-fly tracking of regular subtrees pays off. 
  Replacing regular grid regions within the tree 
  \newB{is advantageous}
  if the grid does not
  change often, if the regular subtrees held separately are
  sufficiently large, and few doubles are held per grid vertex.
\end{observation}

% we let the automaton identify regular grids, and if we process those BFS
% skipping many case checks for hanging nodes, e.g.
% We gain a factor of almost 20 if we store these subtrees separate from
% the tree stream.
% Both statements hold for a regular grid and quantify the DFS/SFC overhead.

\noindent
Cutting out regular subgrids from the overall grid is delicate
as we have to keep the redundant vertices between the
grid regions consistent. 
This administrative overhead, \newB{which arises mainly for the stack-based
storage,} 
is only amortized if the persistent
subregions are large \newB{$\Delta \ell \in \{0,1\}$}.
% 
%  ($h=3^{-9}$ here) or the grid is stationary (not
% shown).
% Large is to be read as ratio of number of grid entities per
% subtree to number of unknowns per vertex.
% 
% 
% We reiterate that the selling point of the regular subtree
% identification is not primarily serial speed but an increase of the concurrency
% level.
% As such, our measurements establish a performance baseline and show how
% this baseline correlates to unaltered grid traversals.
% Recursion unrolling with properly configured thresholds from which on it pays
% off to replace subtrees is used from hereon.
% 
% 
Our observations suggest that it is, in practice, unavoidable to
use some kind of patches or high order methods with lots of unknowns per
grid cell if high FLOP rates are to be obtained.
With the full multiscale tree formalism, we need a significant workload
to mitigate the spacetree administration overhead.
% Embedding all data into the spacetree stream is in practice inferior to 
% scattered data storage using a heap realized through a hash map, e.g.
% We however emphasise that large data cardinality often goes hand in hand with
% grid regularity constraints (patch-based programming), and that a cost
% assessment exercise (invested operations and bytes per scientific accuracy) then
% is non-trivial.
\newB{
 This notably holds for $d\geq 3$ where huge regular subgrids are less likely to
 arise.
}
\subsection{Concurrency impact of the DFS-BFS transformation}

We continue with multicore experiments handling solely the finest grid level---we
label those with \texttt{finegrid}---or \newB{traversing} all grid levels (label 
\texttt{multiscale}).
The \newB{access patterns simulate}
 stencil (matrix-free operator) evaluations.
The latter\newB{---mirroring data accesses of an additive multigrid
scheme---avoids} race conditions between multiscale vertex accesses through
proper mesh colouring with $7^d$ colours, while the finegrid variant succeeds with four colours in total.
Again, we run the setups for various combinations of maximum and minimum mesh
sizes.
All experiments not employing regular grids work with dynamic
adaptivity changing each grid sweep.

%
% User kann jederzeit selber Tasks absetzen und sich um synchronization
% kuemmern. In vielen Faellen muss er genau das wohl tun. 
%

% > Design decision 10
% > ``In Peano, traversal follaws the task dependencies ... task graph is never
% > set up explicitly.''
% > Useful point and feature. I would like to read the details (*?). Again, as
% > with the previous point, how is this done and what's the algorithms behind it?
% > This would also improve the role of this paper in helping others with their
% > research. As is, I would not know how to reimplement the feature.

Various shared memory parallelization features of Peano's traversal
automaton can be switched on or off by the user code, 
\newA{and its traversal automaton using static problem partitioning for all
parallel loops can also be fed with situation-specific grain sizes}.
We thus face a large range of parameter choices: which features are to be switched on and off
and which grain sizes are to be chosen?
For the present manuscript, we \newA{use Peano's hard-coded default values}.

% \begin{figure}
%   \begin{center}
%   \includegraphics[width=0.48\textwidth]{experiments/08h_multicore/dummy-oracle/flop-comparison-plot-2d-2-bytes-per-vertex-coarsest-level-8-heap-dynamic-multiscale-log-no-legend.pdf}
%   \includegraphics[width=0.48\textwidth]{experiments/08i_multicore-with-clusters/flop-comparison-plot-2d-2-bytes-per-vertex-coarsest-level-8-heap-dynamic-multiscale-log.pdf}
%   \\
%   \includegraphics[width=0.48\textwidth]{experiments/08h_multicore/dummy-oracle/flop-comparison-plot-2d-2-bytes-per-vertex-coarsest-level-7-heap-dynamic-multiscale-log-no-legend.pdf}
%   \includegraphics[width=0.48\textwidth]{experiments/08i_multicore-with-clusters/flop-comparison-plot-2d-2-bytes-per-vertex-coarsest-level-10-heap-dynamic-multiscale-log-no-legend.pdf}
%   \end{center}
%   \caption{
%    \newB{Scaling curves on Hamilton for $d=2$ setups with different arithmetic
%     load per vertex.
%     Two doubles are held per vertex, and we process all levels
%     (\texttt{multiscale}).
%     Left: We use solely colouring plus DFS-to-BFS transformations on regular
%     grids.
%     Right: We process the regular subtrees independent of each other.
%     Top: Regular grid with 5,413,808 vertices.
%     Bottom: Same start grid plus $\Delta \ell=2$ additional adaptive levels that
%     move every grid sweep.
%    }
%     \label{results:multicore:hamilton}
%   }
% \end{figure}

\begin{figure}
  \begin{center}
  \includegraphics[width=0.48\textwidth]{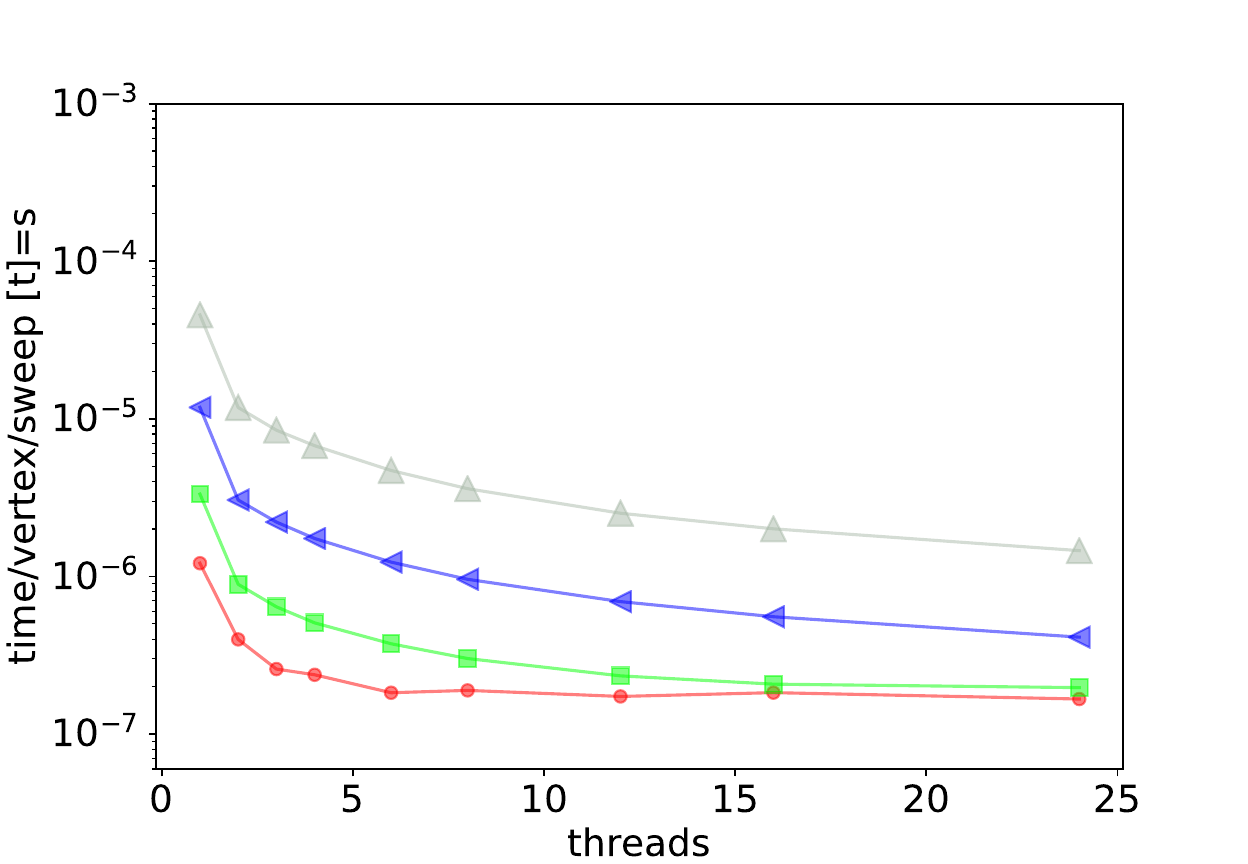}
  \includegraphics[width=0.48\textwidth]{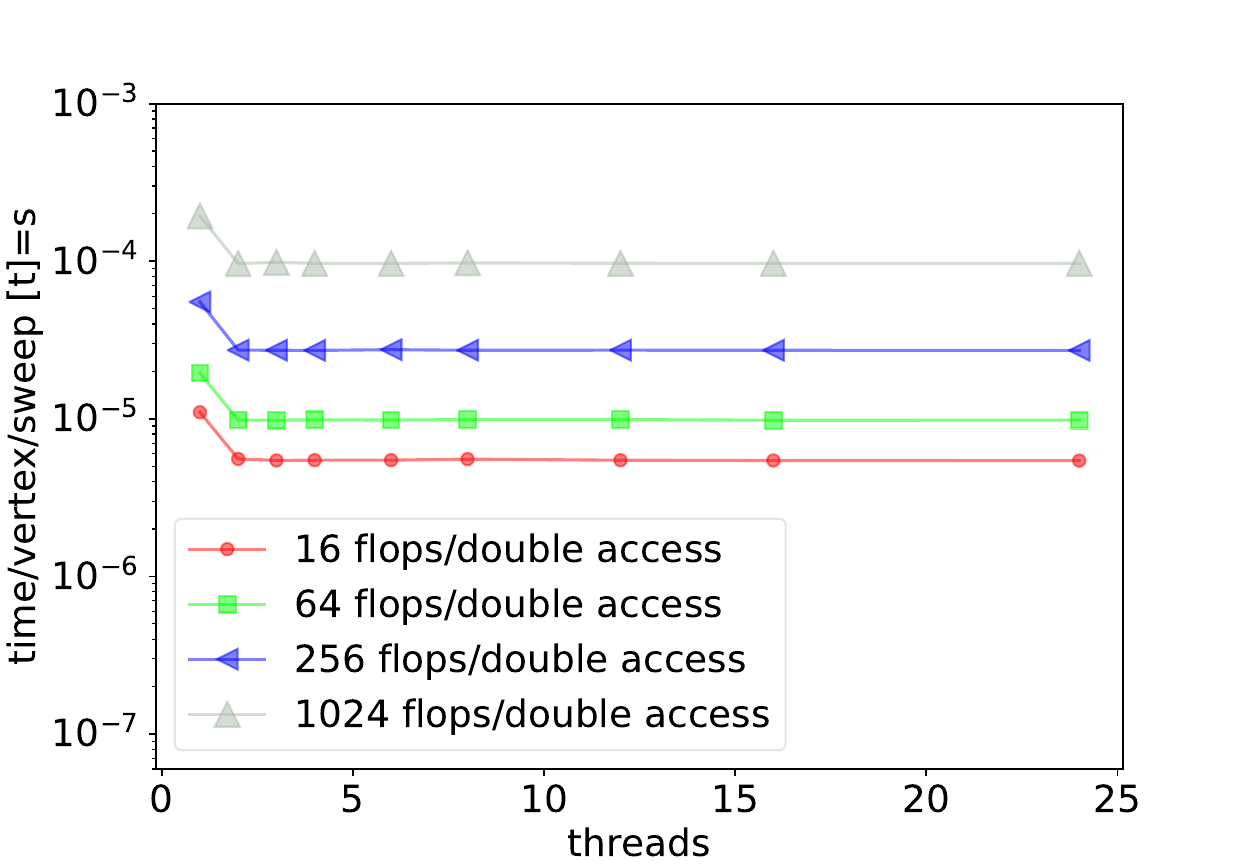}
  \end{center}
  \caption{
   \newB{Scaling curves on Hamilton for $d=2$ setups with different arithmetic
    load per vertex.
    Two doubles are held per vertex, and we process all levels
    (\texttt{multiscale}).
    Regular grid (left) vs.~dynamically adaptive grid (right) with $\Delta \ell=2$.
   }
    \label{results:multicore:hamilton}
  } 
\end{figure}

Its outcome can qualitatively be summarized
as follows:
Significant scalability is \newB{obtained} for the colouring of
the BFS traversal phases
\newB{where we loop over the regular subtrees' levels with a parallel for}.
\newB{The more computational load per grid entity the better the scaling}.
The parallel efficiency \newB{is lost} 
as soon as we run into dynamically adaptive grids (Figure
\ref{results:multicore:hamilton}).
Only if large regular sub-grids are encountered, we observe scaling.
Variants with only fine grid manipulations scale
better if the work per vertex is sufficiently high---an obvious property given
the weaker concurrency constraints.
Furthermore, we observe scaling for adaptive subpatterns if and only if the
adaptivity pattern remains stationary
\newB{The last two properties are not shown here.}
Decomposing the data load and store process along the surface of regular
subtrees into tasks is robustly superior to a sequential load and store.
Once we load data for $f \geq 1$, we may furthermore hide the handling of coarse
levels behind the load and stores of finer levels.
\newB{
 Both techniques \cite{Schreiber:13:Cluster} yield a parallel speedup of
 two.
 The effect is visible best for the dynamically adaptive grid which otherwise
 does not scale at all, i.e.~it benefits solely from the parallelised
 loads/stores and the overlapping of processing and memory accesses.
}

Finally, we \newB{remark} that a proper grain size should scale with the problem
size:
If the grain size is too small relative to the problem size, our academic
decomposition into tiny tasks yields a too high task administration overhead.
Peano offers a plug-in point to inject proper problem size-dependent into any
application.
If a manual identification is too cumbersome, 
its toolbox collection provides a generic machine learning algorithm to
derive proper grain size choices on-the-fly \cite{Charrier:17:Autotuning}.

Strong dynamic adaptivity makes our DFS/BFS-based parallelization deteriorate
from a parallelization strategy into a minor speed improvement.
This teaches \newA{five} lessons:
First, we have to employ classic domain decomposition working with
separated trees to obtain good scalability in
our AMR context.
This \newA{allows threads to work on separate memory regions and }
can be done either via MPI as studied next or with low-overhead 
shared memory (\newB{compare} \cite{Schreiber:13:Cluster,Schreiber:13:sfc-based}, e.g.).
Data decomposition has to be applied on-node, \newA{too.}
Second, scaling \newA{a} Peano code \newA{is simplified if} the basic
operations per grid entity \newA{themselves} exploit multiple cores
(\newB{compare} to the concepts of inter-patch and intra-patch concurrency in
\citet{Weinzierl:14:BlockFusion}).
A pure grid-based parallelization \newA{easily} falls short of exploiting all
cores.
Third, it makes sense to (artificially) increase the grid 
regularity.
The loss in efficiency measured by work per accuracy can be compensated by an
increased scalability. 
\newB{Fourth,} it makes sense in a multicore environment to fix the grid over
multiple grid sweeps.
\newB{Through temporal blocking}, we can eliminate thread synchronization.
\newA{Finally,} it is reasonable to evaluate where tasks can decouple from the
grid traversal.

\newB{
Splitting of tree processing, i.e.~working in a logically distributed
environment, is next discussed in a (logically) distributed memory context.
Decoupling of grid traversal and tasking naturally leads to a producer-consumer
pattern:
The grid traversal then creates tasks which are processed by other threads.
In \cite{Eckhardt:15:SPHCompression}, we use such a mechanism to deploy data
conversions after a cell processing to background threads.
The workstream pattern from \cite{Turcksin:2016:WorkStream} establishes a
similar data flow.
In such a producer pattern, the (limited) scalability of the mesh traversal
becomes advantageous.
The little bit of scaling ensures that tasks are created reasonably quickly,
while the limited concurrency ensures that the node does not hinder all
cores to process actual work.
Again, it becomes obvious that for very large regular (sub)trees, it is
convenient to disable such data flow patterns and instead fall back to direct
DFS-BFS transformation.
}

\subsection{Distributed memory scalability}

\begin{figure}
  \begin{center}
    \includegraphics[width=0.4\textwidth]{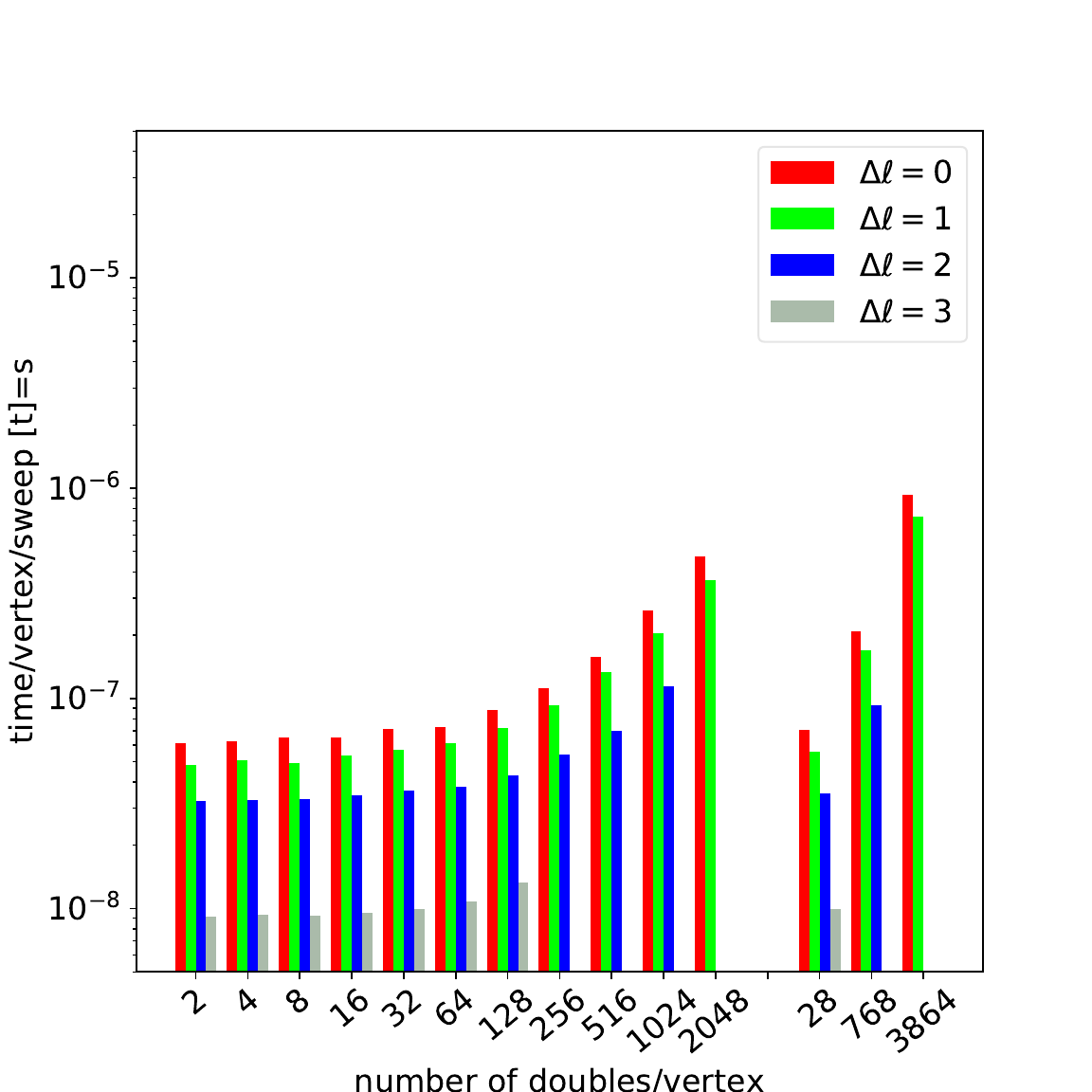}
    \includegraphics[width=0.4\textwidth]{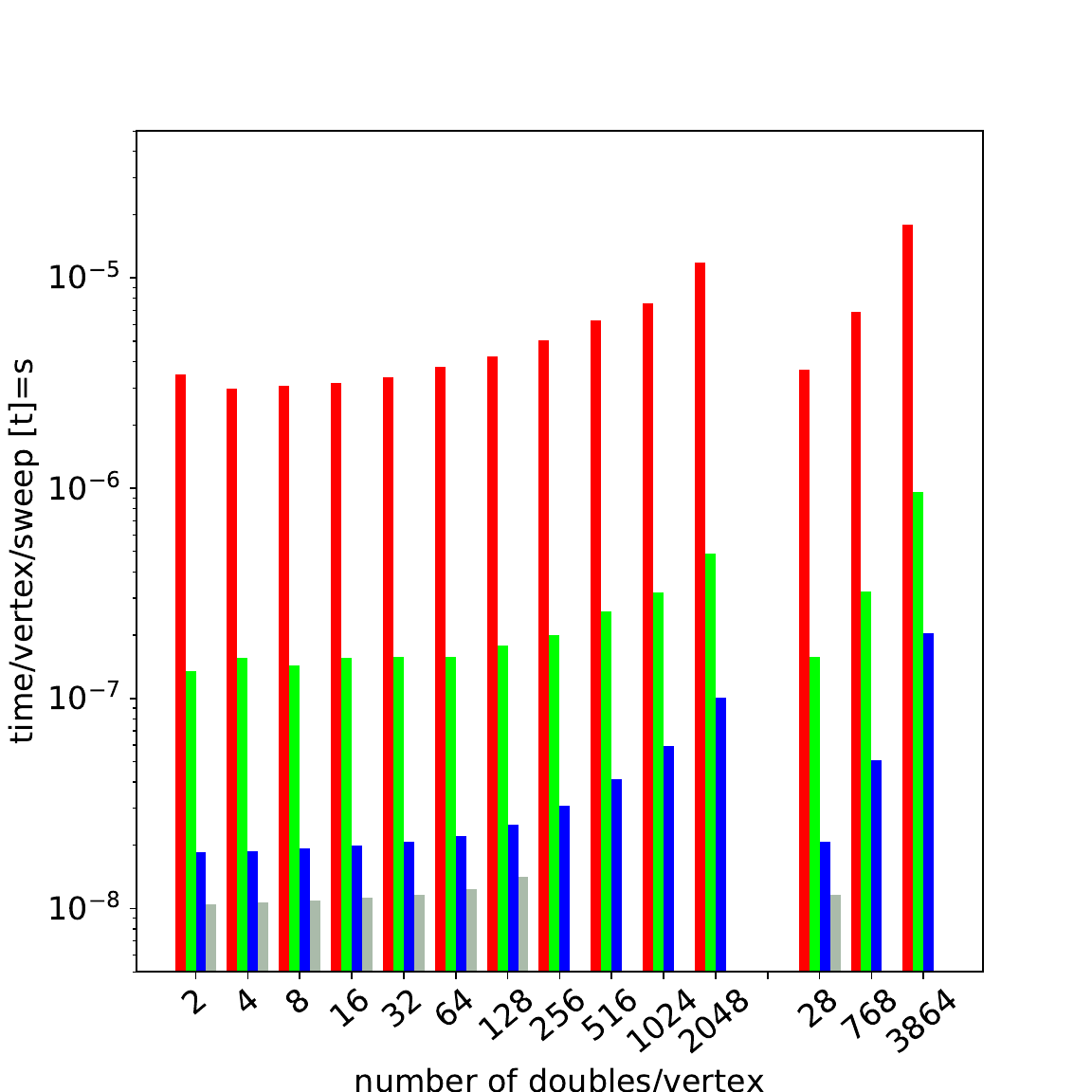}
  \end{center}
  \caption{
    Average runtime per grid sweep on SuperMUC for $d=2$ (left; ten
    \newB{ranks}) and $d=3$ (right; 28 ranks) \newB{against memory footprint
    per vertex}. The three measurements to the right in each graph
    (28, 768, 3864) pick up typical parameter choices from other experiments
    \newB{from the paper}.
    \label{results:mpi:hide} 
  }
\end{figure}

Our tree decomposition experiments enable the BFS/DFS
transformations but switch off any shared memory feature.
The tree decomposition is realized with MPI.
% On the standard Intel processors, we furthermore deploy eight MPI ranks on every
% node of 16 cores; assuming that even without proper application our shared
% memory parallelization exploits the remaining cores.
We start with a \newB{setup with} ten ranks ($d=2$) or 28 ranks ($d=3$)
where we configure the code to hold different number of unknowns per vertex.
\newC{Though we remove all computations from the code,}
% Though the code is stripped off all computations, 
it does exchange all data
along the domain boundary. 
This yields a communication worst-case setup.

Qualitatively similar to the serial data, we see an almost constant
runtime per sweep (Figure~\ref{results:mpi:hide}) as long as we hold up to 64
doubles per grid entity.
Once we go beyond 64 doubles per vertex, the
runtime starts to depend on the data held per vertex.
It slowly grows into a linear relation.
For $d=3$, we obtain comparable behavior.

For small relative memory footprints, our code is spacetree
administration-bound;
it is not bandwidth-influenced, neither \newB{with respect to}~memory bandwidth nor
\newB{with respect to}~network bandwidth.
All data exchange can hide behind the traversal because of the discussed SFC
stream paradigm that allows the automaton to send out data while it still runs
through the grid.
% The performance is solely determined by rank synchronization.
% We refer to this as algorithmic latency.
For large memory footprints, our code can not hide data transfer behind the
traversals anymore.
The fact that the time-to-memory footprint relation is transitioning into a 
linear relation slowly shows that the code continues to succeed in hiding
some data exchange behind the actual grid traversal.
However, the more data is to be transferred the less significant this hiding.

% It instead exhibits algorithmic latency (overhead) behaviour where the runtime
% does not depend on the actual data transferred or handled, respectively.

\begin{observation}
The deterministic, automaton-based traversal allows the automaton to hide data
exchange automatically behind the user-defined events.
This is behavior known for codes that work solely on 
$\Omega _h$.
Our data reveal that we can \newB{recover} the characteristics for multiscale
data. 
\end{observation}

\begin{figure}
  \begin{center}  
    \includegraphics[width=0.48\textwidth]{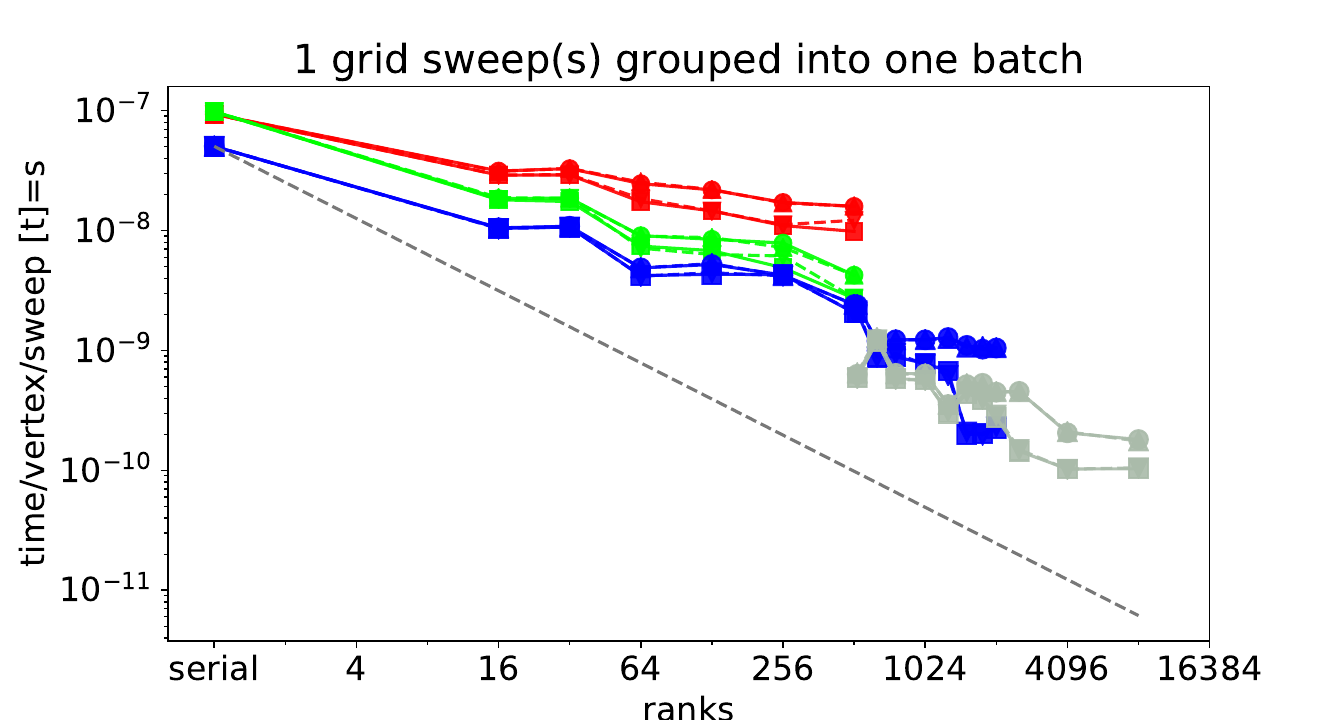}
    \includegraphics[width=0.48\textwidth]{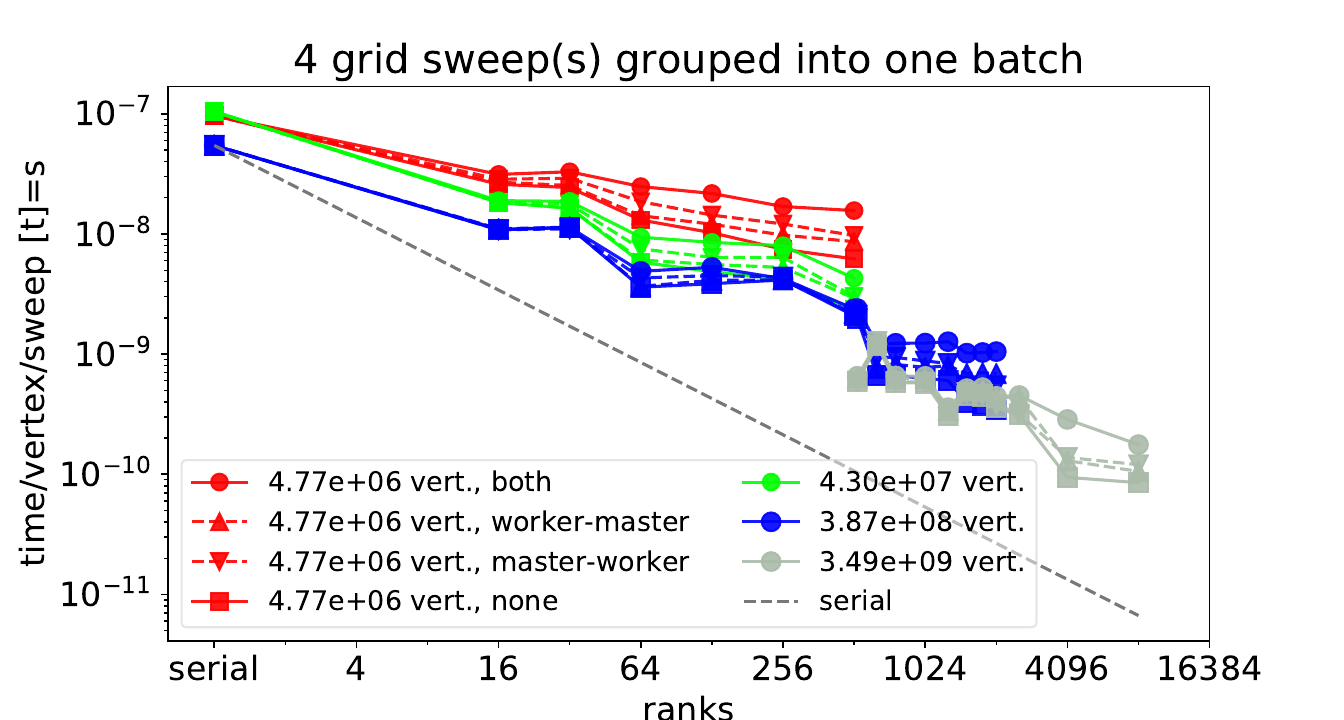}
    \\
    \includegraphics[width=0.48\textwidth]{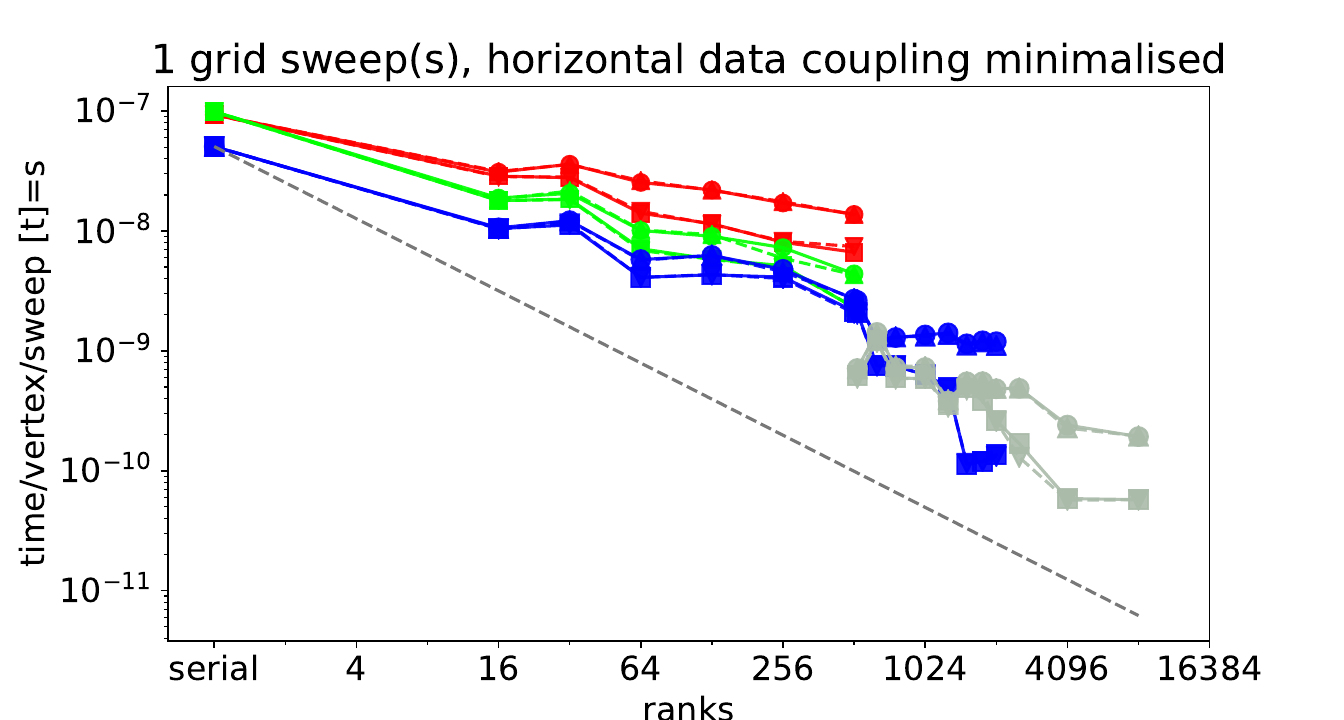}
    \includegraphics[width=0.48\textwidth]{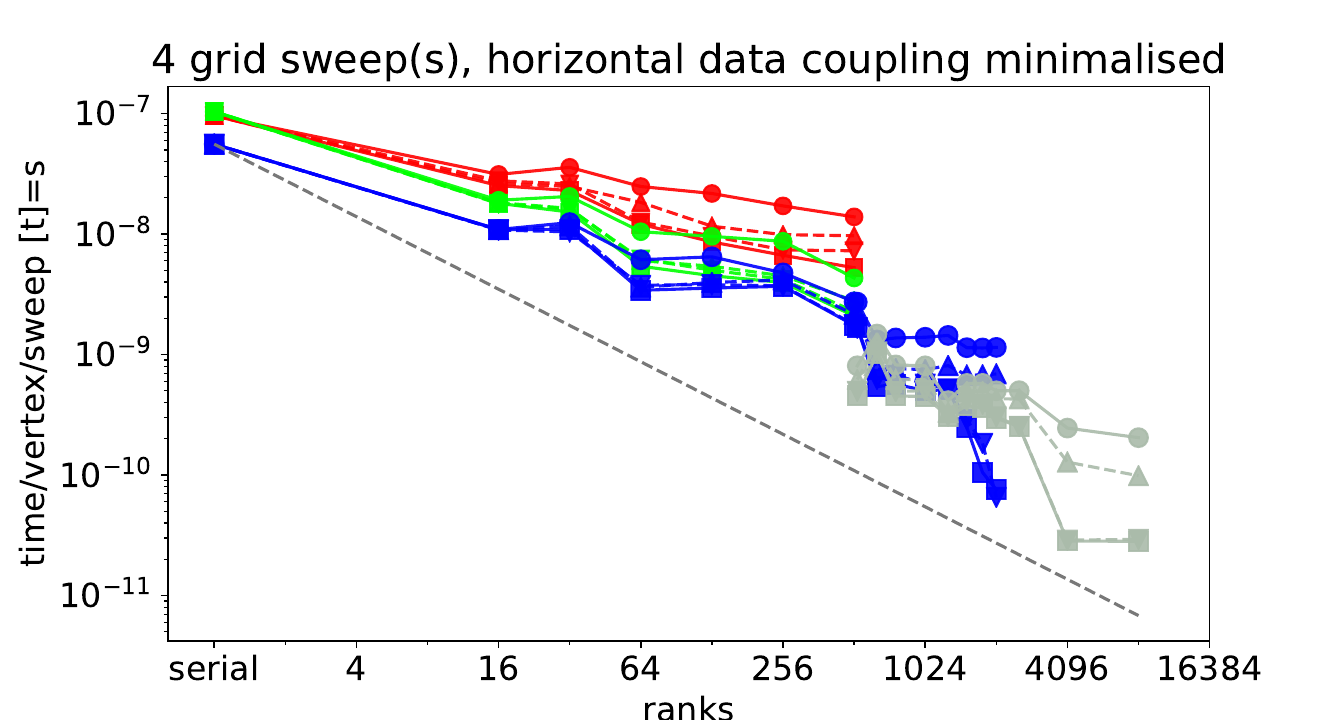}
  \end{center}
  \caption{
    $d=2$ scalability \newA{as runtime per sweep per vertex} on SuperMUC. Eight 
    doubles are assigned to each vertex.
    In the top row, each and every vertex along the domain boundary is exchanged
    with all neighboring partitions.
    In the middle and lower row, vertices are only exchanged when they change
    their state.
    We either switch on vertical communication both ways, send only
    data from the masters to the workers, the other way round, or eliminate all vertical
    data exchange per batch completely, i.e.~exchange vertical data only at the
    end of the batch. No arithmetic work is done.
  }
  \label{figure:mpi:vertical-2d-8}
\end{figure}

\noindent
We continue with classic upscaling and launch a sequence of \newA{$n$} grid
sweeps.
\newA{Then we run the same simulation again} but trigger always two grid
traversals as one batch.
Then, we do four sweeps in one batch, and so forth.
% The experiments again strip the code off the arithmetics and thus study the sole
% communication behaviour.
\newA{Each experiment, i.e.~$n$-configuration,} is done four times: We either
preserve all vertical data exchange, 
\newA{or we} switch off the wake-up from master to worker \newA{within one
batch}---the worker-master synchronization then results from the exchange
of the multiscale domain boundary data---skip worker-master reductions or
eliminate both; all elimination is only done in-between batches.
As soon as a batch
of grid sweeps terminates, we restrict all data vertically.
When a batch is
kicked off, we run all master-worker data sends.

Our measurements reveal strong scaling effects in combination with an amortization
of administrative overhead (Figure \ref{figure:mpi:vertical-2d-8}):
All measurements stagnate the later the
higher the vertex count.
Furthermore, the higher
the vertex count the lower the cost per vertex for a fixed rank number.
Grouping grid sweeps into batches alone does not pay off. 
\newB{Minimizing} horizontal data exchange, i.e.~to exchange only
vertices that are updated, improves the performance slightly.
Both techniques combined yield fast code once we also skip vertical data
exchange within a batch.
% Natural use cases for such a technique are multigrid solvers with multiple
% smoothing steps, e.g., or particle-in-cell algorithms with limited
% particle movement in-between grid resolution levels
% \cite{Weinzierl:16:PIC}.
% To eliminate data transfer in both directions yields the fastest code.
If we compare an elimination of worker-master to worker-master information
exchange, a skip of the reductions contributes more to good scalability.
This diversification \newB{with respect to}~vertical data exchange becomes observable when the
graph enters the strong scaling stagnation regime.
For several setups it can invert the classic strong scaling behavior, i.e.~the
expectation that a simulation scales the better the more detailed the grid
\newA{that is used is}.
This behavior is reasonable once we emphasize that finer grids
couple individual ranks stronger through horizontal data exchange than shallow
grids.

\begin{observation}
  The elimination of vertical data exchange in combination with batching allows
  us to decouple ranks that do not exchange a significant amount of data through
  the domain boundaries.
\end{observation}

%
% Fuer letzte Rank-Zahl sieht man das Problem
\noindent
The advantageous behavior with the data exchange skips 
\newB{does not eliminate the fact that there are massive}
steps in the scalability graphs:
\newB{Our na\"ive, static dummy load balancing cuts}
through the spacetree in a top-down fashion:
Optimal scalability \newB{thus} is
obtained if and only if the number of ranks matches the grid structure. 
\newB{
Notably, whenever the
mesh is very close to regular while the rank count is from $\{k^{id}
| i \in \mathbf{N}^+ \}$, this approach yields perfect partitions. 
These are the best-cases in the graph.
}
For a $d=2$ regular grid, a rank count of nine for example allows us to split up
\newB{a regular} grid properly. 
If only two ranks are available \newB{only}, a proper splitting
\newB{would} exhibit an ill-balancing of 4:5.
The master-worker topology however permits only ratios of 1:8.
The effect \newA{becomes} stronger for $d=3$ (not shown) and explains the
plateau at 4.096 ranks.

\begin{observation}
  The master-worker MPI topology restricts the admissible fine grid partitions.
  \newB{Many}
  reasonable fine grid splittings cannot be mapped onto a
  master-worker topology
  \newB{and can only be approximated}. 
  While the non-replicating scheme minimizes the data and work done per rank,
  it struggles to compete with replicating schemes.
\end{observation}

\begin{table}
 \ifthenelse{\boolean{toms}}{
  \tbl{
   \newA{
    Some speedups for various (averaged) vertex counts in $\Omega _h$ where we
    use the Peano SFC to derive proper domain decompositions and deploy six ranks
    per 24 core compute node.
    The speedups are normalized to the smallest problem setup.
    We present data for 2 (top) and 768 (bottom) doubles per vertex.
%     \newB{The code is stripped off all computations.}
    \newC{All computation is stripped off from the code.}
    \label{figure:mpi:2d-3d}
   }
  }
 }
 {
  \caption{
    Some speedups for various (averaged) vertex counts in $\Omega _h$ where we
    use the Peano SFC to derive proper domain decompositions and deploy six ranks
    per 24 core compute node.
    The speedups are normalized to the smallest problem setup.
    We present data for 2 (top) and 768 (bottom) doubles per vertex.
    All computation is stripped off from the code.
  }
  \label{figure:mpi:2d-3d} 
  \vspace{-0.2cm}
  \begin{center}
 } 
  {\footnotesize
  \begin{tabular}{l|rrrr|rrrr}
   nodes/
    &
    \multicolumn{4}{|c|}{$d=2$} &
    \multicolumn{4}{c}{$d=3$} 
   \\
   ranks & 
   $5.90 \cdot 10^4$ & $5.31 \cdot 10^5$ & $4.78\cdot 10^6$ &
   $4.30 \cdot 10^7$ &
   $5.31 \cdot 10^5$ & $1.43\cdot 10^7$ & $3.87\cdot 10^8$ & 
   $1.05 \cdot 10^{10}$
   \\
   \hline 
   1/6    & 2.89 & 4.39 &      && 		2.13 & 3.99 \\ 
2/12   & 10.1 & 13.4 &      && 		3.04 & 4.79 \\ 
4/24   & 10.4 & 12.4 & 20.9 && 		3.07 & 13.7 & 17.0 \\ 
8/48   & 10.7 & 21.2 & 32.2 && 		3.05 & 9.35 & 23.6 \\ 
12/72  & 11.3 & 24.2 & 34.4 && 		3.05 & 9.28 & 24.5 \\ 
16/96  &      & 29.6 & 85.2 && 		     & 9.26 & 24.6 & 79.7 \\ 

   \hline
%    \input{experiments/mpi-with-sfcs/data-28.table}
%    \hline
    1/6    & 1.59 & 2.48 & 3.52 &      & 		0.84 & 2.13 & 4.05 \\ 
2/12   & 1.35 & 8.74 & 12.7 &      & 		0.78 & 3.03 & 4.84 \\ 
4/24   & 2.72 & 8.39 & 11.0 & 19.4 &        0.77 & 3.05 & 14.0 & 17.2  \\ 
8/48   & 3.00 & 8.85 & 17.9 & 28.1 & 	         & 3.04 & 9.38 & 23.9  \\ 
12/72  &      & 8.97 & 19.5 & 30.7 & 	 \\ 
16/96  &      &      & 40.0 & 86.9 & 	 \\ 

%    \hline
%    \input{experiments/mpi-with-sfcs/data-3864.table}
  \end{tabular}
  }
  \ifthenelse{\boolean{toms}}{
 }{
  \end{center}
 }
\end{table}

\noindent
There are two straightforward solutions to this challenge.
We either can combine a non-replicating scheme on coarser resolutions with a
replicating domain decomposition for finer scales.
This is subject of future work\newB{---besides the development of better
non-greedy load balancing schemes which approximate the best-case partitionings
better. 
The latter is a subject independent of Peano as it is connected through a
well-defined, small API.
Complementary,}
 we can deploy multiple MPI ranks per compute node, derive a proper domain
splitting---in Peano's case using the Peano SFC is a natural candidate---and
ensure that workers responsible for subdomains neighboring along the SFC are
deployed to the same rank.
This technique smoothes out the scalability to some degree (Table
\ref{figure:mpi:2d-3d}) once the ratio data per vertex relative to the mesh
size becomes reasonably small.
It seems to be particularly promising for applications that cannot exploit all
cores through shared memory parallelization.
It seems not to be promising for $d=3$ setups with massive data
per vertex and, thus, massive communication demands.
We however emphasize that our results in Table
\ref{figure:mpi:2d-3d} are biased as we strip the
code off any computation.
% We may assume that the arithmetic intensity grows
% significantly once we sitch from $d=2$ to $d=3$.

 \begin{figure}
   \begin{center}
     \includegraphics[width=0.48\textwidth]{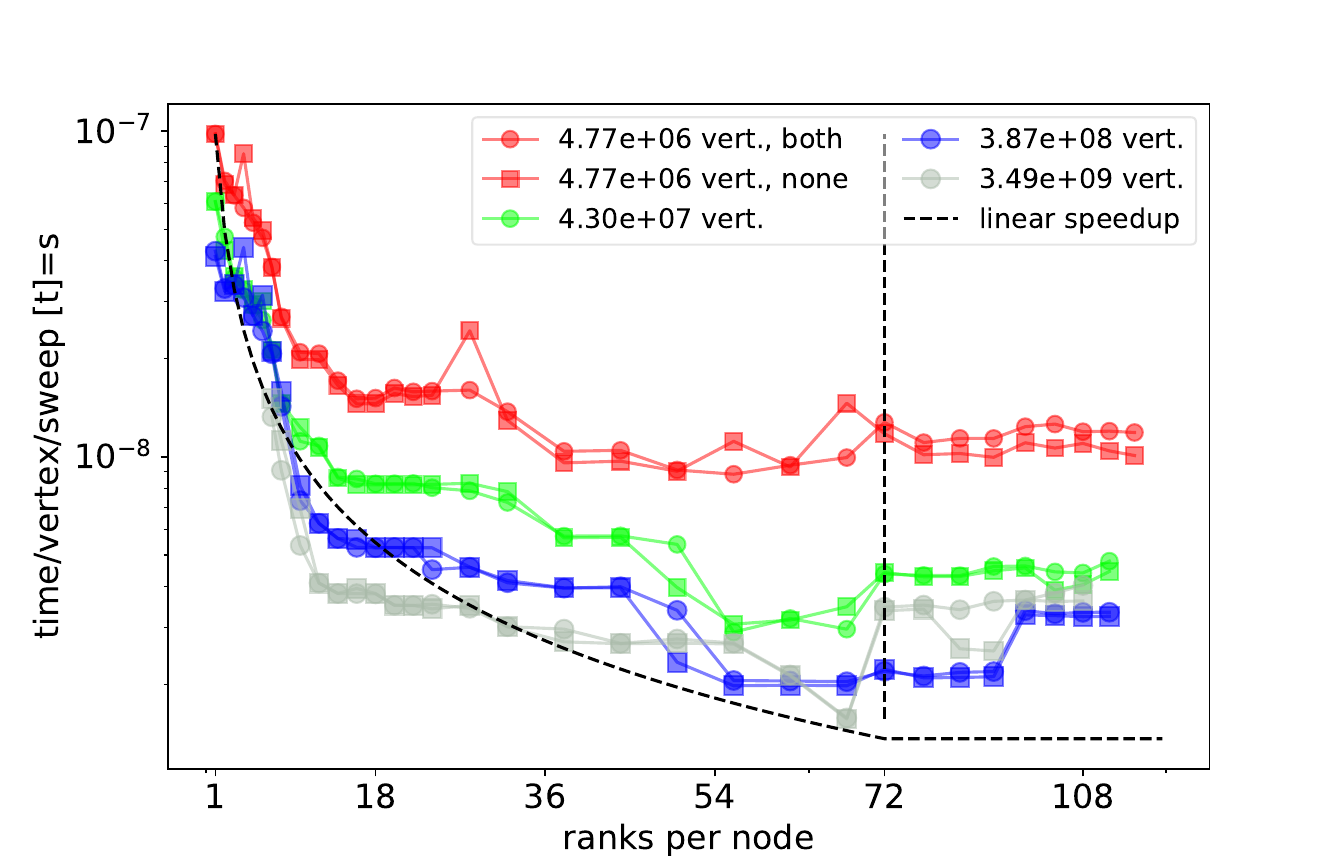}
     \includegraphics[width=0.48\textwidth]{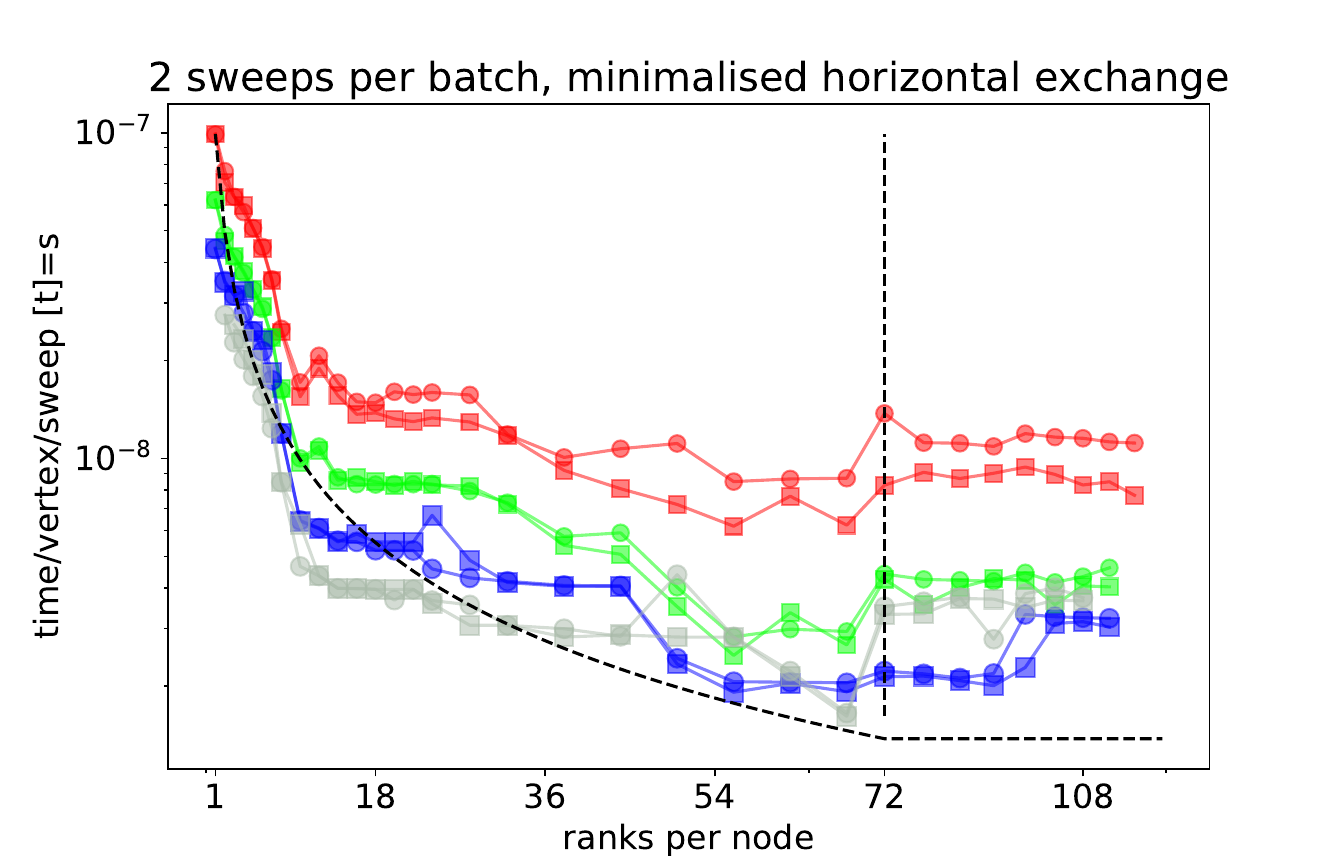}
   \end{center}
   \caption{
     Cost \newA{as time per sweep} per vertex for $d=2$ on one
     KNL node with 72 cores.
     The spacetree is decomposed into subtrees and distributed via MPI. No
     BFS parallelization is enabled.
%      The exactly same
%      composition is ran below on two KNL
%      nodes.
   }
   \label{figure:mpi:knl}
 \end{figure}

% 
%
% Red halt von Drops statt Steps
% Offensichtlich ist eine Kombination der sinnvolle Weg
% Letzen Core sollte man evtl. lieber net nehmen

Cutting a tree into distributed memory pieces has the advantage that it
suits both distributed and shared memory machines.
We close our distributed tree discussion by running the tree decomposition 
on a manycore architecture as alternative to our DFS/BFS transformation.
The latter is orthogonal, i.e.,~it can be combined with the decomposition.
Our results (Figure \ref{figure:mpi:knl}) suggest that such an on-node strategy
can yield another speedup of
close to a factor of almost 20 with high efficiency.
%  by cutting a tree into pieces with different
% memory regions.
Going beyond a factor of 20 or  
overloading do not pay off. We notably 
should refrain from booking all cores for the computation.
% One core at least should be spared.
Vertical data exchange or its skipping, respectively, do not play a
major role for larger trees.
They unleash their power in-between the nodes.

% 9 ist halt einfach perfekt offensichtlich
% Fit vs. DFS/BFS transformation
% Due to the top-down greedy splitting used here, the speedup is significant for
% the first $3^d$ ranks used per node: the children of the local spacetree's root
% then are distributed among the cores.
% The speedup afterwards deteriorates.
% Notably for a dynamically adaptive scheme, an on-the-fly BFS/DFS
% transformation here is a more powerful mechanism.

\begin{observation}
 A mixture of tree decomposition with on-the-fly BFS/DFS transformations is a
 promising strategy to exploit manycore architectures.
\end{observation}

%
% Computational load required
% Not a silver bullet
% data synchronisation
% hiding
% efficiency
%

\noindent
\newC{
We conclude that Peano's ``vanilla'' grid
traversal has upscaling potential, though it
is limited. Yet, 
}
\newB{
all \newC{presented} setups are worst-case: 
They change the grid in each and every grid sweep, they process all resolution
levels, and they couple these levels with each other.
Notably, they do only move data through the system but don't compute.
}
\newC{
 Under these constraints, even the combination of space-filling curves with
 traversal optimisation techniques does not guarantee for scalability---despite
 the fact that data exchange can be hidden behind the
 traversal to are large extent.
}

\newC{
 The limited scalability does neither allow to draw conclusions about the
 well-suitedness of trees and SFCs for parallel computing in general nor on the
 potential of Peano; notably as many publications rely successfully on the
 SFC ingredients
 \cite{Bader:13:SFCs,Bangerth:11:dealiiwithp4est,Bungartz:06:Parallel,Burstedde:11:p4est,Deiterding:05:AMR,Dreher:05:Racoon,Griebel:99:SFCAndMultigrid,Meister:12:Software,Rahimian:10:BloodFlow,Sampath:08:Dendro,Schornbaum:18:HPCBlockStructured,Schreiber:13:sfc-based,Schreiber:13:Cluster}.
 It allows to draw a conclusion on their application:
 Scaling a code requires us to weaken
}
\newB{
one or several of the constraints \newC{and} we \newC{have introduce significant
computation.}
}

\newB{
It seems to be necessary that either very high computational load per grid
entity is applied or the grid acts as meta data structure hosting reasonable
large subproblems that scale internally.
Examples for such applications are very high order approaches alike
\cite{Charrier:18:EfficientADERDG} or hybrid approaches as we find them in
\cite{Bergen:04:HHG,Weinzierl:14:BlockFusion}.
Here, regular subgrids are embedded into the actual (spacetree) meta mesh.
}
\newC{
 Orthogonally, it is recommendable for scaling codes that
}
\newB{
 we do not couple all resolution levels in each and every sweep or
 that \newC{we} keep the grid stationary over multiple grid sweeps.
}
\newC{
 We then can exploit temporary and spatial grid regularity.
}

% \noindent
% \newB{
% We conclude that the code has upscaling potential.
% All studied setups are worst-case: 
% They change the grid in each and every grid sweep, they process all resolution
% levels, and they couple these levels with each other.
% Notably, they do only move data through the system but don't compute.
% As soon as one or several of these worst-case configurations/constraints are
% weakened---we do not couple all resolution levels in each and every sweep or
% keep the grid stationary over multiple grid sweeps---or we eliminate
% synchronization between the ranks, the scalability improves.
% }
% 
% 
% \newB{
% Nevertheless, the inherent scalability of \newC{Peano's} grid
% traversal\newC{---without any computational load---}is very limited.
% It seems to be necessary that either very high computational load per grid
% entity is applied or the grid acts as meta data structure hosting reasonable
% large subproblems that scale internally.
% Examples for such applications are very high order approaches alike
% \cite{Charrier:18:EfficientADERDG} or hybrid approaches as we find them in
% \cite{Bergen:04:HHG,Weinzierl:14:BlockFusion}.
% Here, regular subgrids are embedded into the actual (spacetree) meta mesh.
% }

\section{Conclusion and outlook}
\label{section:conclusion}

%
% What have we done, what was it good for; focus on software
%
The present paper introduces a software framework for dynamically
adaptive multiscale grids that fuses data management and data traversal as well
as, if not explicitly outsourced to a heap, application data storage.
The underlying programming model is formalized with two automata yielding an
event-based algorithm development approach.
Besides \newA{the} fact that the software delivers a merger of multiple
state-of-the-art features such as support for arbitrary
dimensions, multiscale data representation and low memory footprint capabilities, 
it is resonably simple to handle due to its restrictive,
minimalistic programming model.  
This makes the software well-suited for fast prototyping as well as bigger codes
with a clear separation of concerns as long as they accept the
restrictive academic  programming paradigm which might imply that not each
legacy code can directly use the AMR features.

%
% Take away and character of follow-up work
%
The manuscript characterizes, classifies and motivates several design decisions
made while the framework was \newA{written}.
It also explicitly highlights shortcomings and open questions notably in
comparison with other approaches found in literature.
The references to alternative pathways towards realization are backed up with
some experiments that highlight that the software is capable to code
sophisticated applications running in parallel.
\newA{Yet, the discussion and the results also highlight that other spacetree
code developers might favour other realization variants that are superior for
particular challenges}.

%
% Next work
%
There are two \newA{natural} directions of future \newA{work}.
\newA{
Both start from the observation that the present discussion is very academic
and computer science-centric. It lacks challenges from real-world
computational simulation codes that either rely on the
present software stack, some of its components or discussed paradigms.
}
On the one hand, it is important to track their code maturity, maintainabiliy
and clarity
\newA{once we ``scale'' up the code \newB{with respect to}~simulation detail complexity}.
This will allow us to assess and understand the limitations and drawbacks
imposed by the present programming model from a usability point of view. 
It also might help to identify ways to make the software easier accessible
without giving up its clarity and clear separation of concerns.

On the other hand, it is important to track the computational and algorithmical
efficiency of application codes.
Our case studies reveal that there is scalability potential arising from the
present code base.
Yet, this scalability, though we restrict to worst-case studies that will not be
found this way in applications, is far from optimal and will drive future developments.
We see notably potential in mergers of replicating and non-replicating
tree decomposition schemes and in the combination of our grid-based task
model with tasking approaches where tasks are decoupled from the grid entities
and can run in parallel to the grid traversal.

\section*{Acknowledgements}

The author appreciates support received from the European Union’s Horizon 2020
research and innovation programme under grant agreement No 671698 (ExaHyPE).
This work made use of the facilities of the Hamilton HPC Service of Durham
University.
The author furthermore gratefully acknowledges the Gauss Centre for
Supercomputing e.V. (www.gauss-centre.eu) for funding this project by providing
computing time on the GCS Supercomputer SuperMUC at Leibniz Supercomputing Centre (www.lrz.de).
Finally, this manuscript particularly has been benefitting from the support of
the RSC Group who granted us early access to their KNL machines.
All underlying software is open source \cite{Software:Peano}.

\ifthenelse{\boolean{appendSupplementMaterialAsAppendix}}{
  \ifthenelse{\boolean{toms}}{
    \pagebreak
    \section*{APPENDIX}
    \setcounter{section}{0}
    \section{Software base}

Peano is freely available from \cite{Software:Peano}.
We offer both tarballs and repository access through Subversion.
Support is provided through a \newB{gitlab discussion board}.
\newA{An} extensive guidebook discusses how to implement solvers in Peano.
\newA{It} also comprises case studies.

The code baseline offers support for 
adaptive Cartesian grids
\cite{Akcelik:03:EarthquakesAtSC,BangerthEtAl:07:dealII,Burstedde:11:p4est,Gadeschi:15:HierachicalCartesianGrids,Griebel:99:SFCAndMultigrid,Jeong:01:PhaseField,Khokhlov:98:FullyThreadedTree,Kolobov:16:PICAMR,Lashuk:12:MultipoleOnHeterogeneous,Robey:13:Hashing,Sampath:08:Dendro,Sundar:12:ParallelMultigrid,Sundar:08:BalancedOctrees,Teunissen:17:Afivo,Tumblin:15:CompactHashing}
as discussed in the manuscript.
It thus follows an academic approach to AMR programming and lacks the
flexibility and generality of software alike
\citet{Software:Chombo,Davison:00:Uintah,Software:dealii,BangerthEtAl:07:dealII,Bastian:08:Dune1,Bastian:08:Dune2}.

Extensions of the sole grid and its traversal are available via a template
mechanism and small routine collections (toolkits) that inject plotting
facilities for Paraview/VisIt, \newA{realize} shared memory autotuning
\cite{Charrier:17:Autotuning} or add dynamic load balancing based upon graph partitioning \newA{or the underlying Peano space-filling curve}. 
Further examples for toolkits are routine collections for matrix-free multigrid
\cite{Reps:16:Helmholtz,Weinzierl:17:BoxMG}, Particle-in-Cell features
\cite{Weinzierl:16:PIC} similar to \cite{Kolobov:16:PICAMR} or patch-based PDE
solvers \cite{Weinzierl:14:BlockFusion}.
The latter allows users to compose the AMR grid as assembly of regular
patches and makes the software resemble block-structured AMR
\newA{alike \citet{Software:Chombo,Davison:00:Uintah}, e.g.}

The software currently offers bindings to TBB for
task-based parallelization. 
These bindings are hidden through a software layer.
With task approaches having successfully been used in various spacetree
and non-spacetree codes
such as \cite{Davison:00:Uintah,Meister:12:Software,Weinbub:14:ViennaX}, the
bindings could be replaced by a more sophisticated task subsystems.

\newB{
 Peano supports native C user data natively.
 If users want to model more complex data structures per cell or vertex,
 respectively, Peano relies on DaStGen
 \cite{Bungartz:08:Dastgen,Bungartz:10:Precompiler}.
 DaStGen is a C++ augmentation which automatically equips C++ classes with MPI
 data types, data compression, different class variants distinguishing temporary
 from persistent object attributes, and so forth.
 Peano's internal data structures all are modelled with DaStGen.
 If users rely on stack data management, the (maximum) data cardinality per
 vertex or cell, respectively, has to be known at compile time.
 If users rely on heaps, the number of objects per vertex or cell---either
 native C types or DaStGen objects---is flexible, i.e.~may grow and shrink.
}

    \section{Remarks on some Peano applications}

In Section \ref{section:api:supported-application-types}, we sketch which
types of applications do fit to our restrictive API.
The present section substantiates these statements with examples from recent
projects
\newB{and further remarks}.

\newB{
\paragraph{Tripartitioning}
From an application's point of view bi- and tripartitioning both
come along with pros and cons.
Aggressive $k=3$ coarsening can pose a challenge to geometric multigrid
algorithms, e.g., if a too large spectrum of solution modes is thrown away per
coarsening step.
In return, fine grid cell centres
coincide with coarse ones for tripartitioning.
This simplifies the coding of cell-centred discretizations.
Another pro/con pair is the transition from one resolution $h$ into another
resolution $\hat h>h$ if 2:1-balancing is enforced: 
Our tri-partitioning creates a denser communication/data-exchange pattern than
bipartitioning.
In return, very fine resolutions $h$ do not spread out that significantly into
coarser regions with $\hat h$.
Different	considerations might arise for different applications. 
}

%
% Welches Problem gehen wir an
% Was ist context/major findings
% Bild
% Wie schauen die Datenstrukturen aus
% Was sind lessons learned bzgl. API and memory organization
% Remarks/addenda
%
\paragraph{Low-order, $d$-linear finite element examples}

In \cite{Reps:16:Helmholtz,Weinzierl:17:BoxMG}, we study variants of
geometric-algebraic multigrid solvers for Helmholtz and convection-diffusion
equations.
% (Figure \ref{figure:use-cases:multigrid}).
The Helmholtz setup studies high-dimensional grids with complex-valued grid
spacing, while the convection-diffusion studies derive mergers of algebraic and
geometric multigrid algorithms that combine the geometric efficiency and
memory modesty with the algebraic robustness.
Both papers introduce single-touch algorithms and both study low order
discretizations, i.e.~compact 9-point stencils ($d=2$), 27-point stencils
($d=3$), \ldots 
They solve the PDEs
\begin{eqnarray}
  -\Delta u + k^2 u & = & b \qquad \mbox{and} \nonumber \\
  -\nabla ( \epsilon \nabla) u + v \cdot u & = & b
  \label{equation:use-case:multigrid}
\end{eqnarray} 
with $u \approx \sum _i u_i \varphi _i$ through a Ritz-Galerkin formulation
where $ \varphi _i$ is a $d$-linear shape \newB{function} centred around a
non-hanging vertex on $\Omega _{h,\ell} $ and spanning $2^d$ cells.

 Both papers use full approximation storage (FAS), i.e.~they store a real
 representation of \newB{the} solution on each grid level.
 They work on generating systems.
 This renders the treatment of hanging vertices straightforward.
 It exploits the fact that Peano holds the individual grids $\Omega _{h,\ell}$
 separately, i.e.~each vertex is unique through its level plus position and
 multiple vertices coinciding spatially can hold different data: fine grid
 vertices membering $\Omega _h$ encode the solution while coarser vertices at
 the same position hold FAS data plus multigrid correction terms.
 As all stencils are compact, they decompose additively over the $2^d$ adjacent
 cells and thus can be evaluated element\newB{-wise}.

 We augment each vertex by a residual holding a vertex's $r=b-Au$ contribution 
 if $A$ is the system matrix arising from (\ref{equation:use-case:multigrid}).
 Our application automaton plugs into \texttt{touchVertexFirstTime} to set $r
 \gets 0$.
 Within \texttt{enterCell}, it reads in the $2^d$ $r,u$ and $b$ values and
 contributes the residual contributions from the respective cells.
 The residual is accumulated.
 When the traversal automaton triggers \texttt{touchVertexLastTime}, we know
 that the residual for one vertex is accumulated completely, and we can trigger
 a smoothing step or residual restrictions.
 As Peano offers the opportunity to plug into the traversal
 automaton's steps up and down in the tree, inter-grid transfer operators can be realized
 ``element\newB{-wise}'', too. 
 \newC{
 Traversing the whole spacetree in one rush mirrors
 the behaviour of additive multigrid schemes.
 If we run through the levels sweep by sweep, i.e.~mask out some levels, the
 traversal mirrors the data access of multiplicative multigrid.
}

 The Helmholtz setup stores all quantities directly within the vertices, i.e.~it
 fuses spacetree grid data and solution values.
 Each vertex holds one double being the
 weight $u_i$ within (\ref{equation:use-case:multigrid}) plus the discretised
 right-hand side.
 Additional temporary variables holds the residual, hierarchical residuals,
 error estimators, and so forth.
 The convection-diffusion code starts
 from this data representation, too.
 Yet, it also stores inter-grid transfer stencils and discretization stencils,
 and those quantities are held on Peano's heap as they consist of at
 least $2\cdot 5^d+3^d$ doubles and we furthermore study techniques how to
 convert them on-the-fly into non-IEEE representations where the resulting total
 memory footprint is not known a priori.
 \newB{
  With the work meandering around additive multigrid and BPX as preconditioner,
  Peano's DFS approach vertically integrating the operations on various levels
  \cite{Adams:15:Segmental} yields excellent memory access characteristics per
  level plus in-between the level updates.
 }

 Our Helmholtz studies not only study higher-dimensional grids ($d \geq 4$),
 they also discuss how the solution of multiple PDEs on one grid can be merged
 to yield higher arithmetic intensity.
 In this case, multiple PDEs are stored within one grid, and the arising
 stencils not only tackle the individual PDEs, they also couple them which
 yields additional flops.
 Our convection-diffusion studies compare the derived solvers to a setup where
 each vertex holds only one integer.
 This integer serves as local index to PETSc \cite{Software:PETSc}, i.e.~we
 compare a matrix-free, monolithic solver with Peano to a solver where Peano yields solely the
 discretization, the user automaton realizes the assembly, and the actual
 equation system is administered and solved by a black-box solver.
 \newB{
 Our results suggest that Peano's matrix-free/integrated approach is superior to
 explicit assembly with PETSc if the grid changes frequently throughout the
 solve, i.e.~if we realize for example full multigrid cycles which
 build up the grid setp by step or if we integrate a dynamic refinement criterion into the solve.
 If the grid is stationary, an explicit assembly seems to be significantly
 faster.
 }

\paragraph{Higher-order DG discretizations}

In the ExaHyPE project \cite{Software:ExaHyPE}, we study Discontinuous
Galerkin (DG) discretizations of first-order hyperbolic equations
\begin{equation}
 \frac{\partial}{\partial t} Q
 +
 \nabla \cdot F(Q)
 +
 \sum_i \mathcal{B}_i\,\frac{\partial Q}{\partial x_i}
 = 
 S + \sum \delta
  \label{equation:use-case:ader-dg}
\end{equation}
on dynamically adaptive grids. 
They are subject to the explicit ADER-DG timestepping scheme.
$Q$ here is defined cell\newB{-wise} as high-order ($p \in \{3,4,\ldots,9\}$)
polynomial over the cells of $\Omega _h$.

We make each Peano cell hold a pointer to an entry in Peano's heap.
Each heap entry stores, per quantity in $Q$, $(p+1)^d$ weights of the Gauss
Legendre \newB{shape functions}.
The spacetree acts as meta data structure.
The actual compute data, i.e.~$Q$, are held in the heap's hash map.
ADER-DG is a predictor-corrector scheme:
First, the automaton runs over $\Omega _{h,\ell}$ and computes a predicted
solution to (\ref{equation:use-case:ader-dg}).
This is done within \texttt{enterCell}.
Second, we equip each vertex with $2^d$ pointers to the respective adjacent
cells.
This allows us to traverse the grid, i.e.~all faces, and solve Riemann problems
to the predicted solution there.
The idea here is to plug into \texttt{touchVertexFirstTime} and to equip each
face with one marker bit. 
The bit is set to zero by \texttt{touchVertexLastTime}.
Once we enter \texttt{touchVertexFirstTime}, we analyse the $2\cdot 2^{d-1}$
adjacent faces.
For each face where our flagging bit is not set, we set it and evaluate the
Riemann problem.
Finally, we traverse the grid's cells and bring together the
predicted solution with the Riemann solves.
We exploit (\ref{equation:order-on-events}): it ensures that each vertex has
been ``touched'' prior to \texttt{enterCell}.
Consequently, all adjacent faces to a cell have been processed, too.
It is one insight of the underlying project that this works with single-touch
semantics, i.e.~one grid sweep per time step.

\newA{
  Hanging vertices inherit the pointers from their parent vertices when the
  automaton triggers \texttt{createHangingVertex}.
  \newC{
   The parents of a vertex are all vertices on the next coarser level which are
   adjacent to a cell which in turn is a parent cell to a cell adjacent to the
   vertex of interest.
   With this definition,
  }
  we can determine neighbouring cells of any hanging vertex
  and handle resolution transitions.
  ADER-DG quickly becomes unstable in the presence of shocks.
  For shocks, we apply a Finite Volume solver as limiter.
  If limiters are present, our grid carries two solvers per cell.
}

\paragraph{Patch-based finite volume solvers}

\newA{
  In \cite{Weinzierl:14:BlockFusion}, we study a standard Finite Volume solver 
  for shallow water equations
  \begin{equation}
\frac{\partial}{\partial t} 
\left(\begin{array}{c}
h\\hu\\hv\\ b
\end{array}\right)
+
\frac{\partial}{\partial x}
\left(\begin{array}{c}
hu\\
hu^2 + 0.5gh^2\\
huv\\
0\\
\end{array}\right)
+
\frac{\partial}{\partial y}
\left(\begin{array}{c}
hv\\
huv\\
hv^2 + 0.5gh^2\\
0\\
\end{array}\right)
+
\left(\begin{array}{c}
0\\
hg \cdot b_x\\
hg \cdot b_y\\
0\\
\end{array}\right)
= 0.
\label{equation:application:swe}
\end{equation}
}

  In the simplest variant (Figure \ref{figure:overhead-reduction:blocks}), we apply a
  simple Rusanov flux. 
  We store three unknown quantities per cell.
  In an original variant they are held within the spacetree stream.
  Per face, we have to hold the fluxes.
  As Peano does not natively provide face \newB{unknowns}, we store the
  quantities from the positions $(x+h/2,y)$ and $(x,y+h/2)$ within the vertex at $(x,y)$.
  $h$ is the cell size.
  We map the staggered degree of freedom layout onto vertex and cell storage
  locations.
  Within each cell, we have, through the $2^d$ adjacent vertices, all data
  available to run the Finite Volume updates.

 Few codes solving equations \newB{like} (\ref{equation:application:swe}) use
 strongly adaptive meshes where single cells resolve particular properties.
 Adaptivity criteria here typically yield areas of refinement.
 We therefore next embed whole $n\times n$ patches into the individual cells,
 equip them with a halo layer of width one, and hold both the current and the 
 previous solution within the patch.
 When the automaton triggers \texttt{enterCell}, we first fill the ghost layer
 and then trigger the Finite Volume scheme on the patch.
 This block-structured AMR also yields reasonable complex compute kernels per
 cell that can be vectorised.
\newC{
 If patch-based processing first interpolates and then starts from the finest
 level, i.e.~plugs into the DFS traversal's backtracking, it mirrors the
 data flow of classic local time stepping for hyperbolic equation systems with
 subcycling (cf.~references in \cite{Dubey:16:SAMR,Deiterding:05:AMR}).
}

  There is an obvious trade-off between patch size $n$---the bigger $n$, the
  lower the spacetree's administration overhead and the higher the impact of
  classic stencil optimization techniques and vectorization---and degree of
  freedom per accuracy.
  In \cite{Weinzierl:14:BlockFusion} we apply a technique along the lines of 
  (\ref{equation:recursion-unrolling-grammar}) to our patches:
  Whenever we identify an assembly of $3 \times 3$ patches on a level
  $\ell$, we can replace this patch by one $3n \times 3n$ patch embedded into
  level $\ell -1$.
  The argument applies recursively.
  Our spacetree's multilevel nature thus allows us to use tiny patches with
  Finite Volume kernels that cannot optimise too aggressively.
  These patches are (temporarily, as long as refinement criteria allow us to do
  so) replaced by larger patches on coarser tree levels.
  For these, optimised kernels can be written.
  Once an adaptivity criterion requires us to refine, we first break up the
  fused patches, and then continue with classic block-structured AMR with tiny
  patches.
  This technique can be read as a predecessor to the present paper's
  optimizations in Section \ref{section:overhead-reduction}.

\paragraph{Particle-grid data structures for Particle-in-Cell, the discrete
element method and SPH}

 Lagrangian methods and particle models often employ meshes as helper data
 structure to speed up operations such as collision detection or the evaluation
 of formula which are subject to a cut-off radius.
 In \cite{Weinzierl:16:PIC}, we derive a multiscale particle administration
 algorithm that scales.  
 This algorithm is used for an SPH application in
 \cite{Eckhardt:15:SPHCompression} and as base of a discrete element method code
 in \cite{Krestenitis:17:DEM} (Figure \ref{figure:use-cases:automata}).

  Both codes exploit the fact that each particle has a cut-off radius or a
  neighbourhood where we have to search for possible collisions, respectively.
  Both quantities determine a mesh size $h$ if we embed the objects into a 
  grid. 
  Given a certain nonstationary particle distribution, we create a spacetree that
  can accommodate the particles:
  it \newB{is} refined down to the finest mesh size $h$, but particles are
  embedded into the level that suits their cut-off/search radius.
  To store them, each vertex is given a pointer to \newB{a dynamic
  range vector of particles}.
  Each particle is stored to the vertex on the respective level next to its
  centre.
  The efficient on-the-fly administration of the respective particle lists is
  subject of discussion in \cite{Weinzierl:16:PIC}.
  On one level, we compare particles to each other within \texttt{enterCell}.
  Inter-grid operators in return compare particles living on various scales.

\newB{
\paragraph{Limitations}
While pointers from vertices to neighbouring cells allow the automaton per cell
to access the neighbouring cells,
Peano's strict element-wise automaton paradigm forbids codes to access neighours
of neighbours or vertices of neighbouring cells.
It might be possible to permit such freedom of access through additional helper
structures along the lines of the vertex-to-cell pointers, but it then might be
reasonable to assess completely alternative software solutions anyway.
Therefore, we haven't investigated into this option yet.
}

\newB{
Peano focuses on grid traversals through the whole tree.
More anarchic temporal access patterns (process only cells of a certain
resolution or carrying markers) as we find them in multiplicative multigrid for
elliptic setups or local time stepping for hyperbolic equation system solvers
can be realised by a marker approach where the grid traversal automaton process
the whole grid but updates only particular grid entities.
If the actual grid operations are expensive and can be deployed to dedicated
threads, we end up with a producer-consumer pattern:
Peano efficiently traverses the tree and spawns tasks which are then evaluated
on separate threads.
If the spawned tasks are reasonably expensive, such a pattern can scale, and
Peano provides the required infrastructure for this.
}

\newB{
The realisation of complex multilevel algorithms that require information
propagation from coarse to fine levels and the other way round can
straightforwardly be realised through multiple grid sweeps.
Again, it is not clear however whether other grid organisation paradigms then
would be more appropriate.
As an alternative to multiple sweeps, one can investigate pipelining along
the lines of
\cite{Charrier:17:Autotuning,Reps:16:Helmholtz,Weinzierl:13:Hybrid}:
Pipelining introduces helper variables per entity.
These helper variables store additional information (about unknown updates,
e.g.) such that grid traversals can be eliminated or reduced.
}

    \section{Balancing}
\label{section:appendix:balancing}

\newC{
 While grid balancing is not built into Peano's algorithms by construction, we
 offer a balancing extension.
 It can be compiled into any application without
 further need for user customization.
 It helps to ensure that the level of no two cells which are adjacent to each
 other, i.e.~vertex-connected, differs by more than one.
} 

%
% How does it work
%
\newC{
 The extension exploits Peano's or-based refinement
 convention \cite{Weinzierl:11:Peano}:
 Let a spacetree cell be refined if and only if at least one of its non-hanging
 adjacent vertices holds a refinement marker.
 The other way round, a marker on a vertex makes all of the vertex's adjacent
 cells refine.
 When the grid traversal automaton backtracks, it plugs into the storage of the
 persistent, i.e.~non-hanging vertices (\texttt{touchVertexLastTime}).
 In our spacetree, a refined cell spans a $3 \times 3 \times \ldots$ patch on
 the next level.
 Any vertex on this next level consequently has a position $p \in
 \{0,1,2,3\}^d$ relative to the coarser refined cell.
 This enumeration is not unique---for vertices along the coarse grid faces, it
 depends on whether we look ``from the right or the left''---but the
 nondeterminism does not affect the algorithm's outcome.
 If a fine grid vertex at $p$ holds a refinement marker and if any parent vertex
 of the vertex at $\hat p$ with $\hat p_i = \min (3(p_i-1),3),\ i \in
 \{1,,2,\ldots,d\}$ does not hold a refinement marker, then the grid is not balanced.
 We refine this parent vertex if it is not a hanging.
}
  
%
% Effizienz
%

\newB{
Peano's out-of-the-box 2:1 balancing eliminates too aggressive refinement
incrementally.
\newC{
 The above algorithm eliminates one level of ill-balancing between two adjacent
 cells.
}
It thus can lead to rippling which propagates a refinement through the grid over
several grid sweeps.
}
\newC{
 For the aforementioned demonstrator applications, we rarely observed
 ill-balanced grids with more than two levels of resolution difference between
 adjacent cells, and we always have been able to accept that this ill-balancing
 is immediately resolved in the subsequent grid traversal.
 Any rippling integrates into the grid sweeps.
 If an algorithm requires immediate well-balancing after each grid modification,
 users manually have to hold their computation once the premanufactured
 balancing identifies ill-balancing and have to insert iterative
 re-balancing grid sweeps.
 The cost per sweep corresponds to an AMR grid traversal without any
 computation.
 Alternatively, it might be reasonable to investigate into 
}
\newB{
sophisticated grid balancing along the lines of 
\cite{Tu:2005:Octor,Sundar:08:BalancedOctrees}.
}
\newC{
 This is however not offered out-of-the-box.
}

%
% Boundary
%
\newC{
 While the iterative, lazy elimination of ill-balancing so far has been
 sufficient for most of our applications, we found that elliptic solvers with
 non-homogeneous boundary conditions benefit from an additional constraint:
 Eliminate hanging nodes on the domain boundary.
 Whenever hanging vertices along the boundary are found, it is convenient for
 user codes to trigger refinement next to the boundary on the next coarser
 level.
 This way, they avoid that we underresolve the boundary and thus suffer from
 pollution effects from inaccurate boundary conditions.
}

    \section*{Thanks}

Thanks are due to all the scientists and students who contributed to the
software in terms of software fragments, applications, extensions and critical
remarks.
Notably, thanks are due to Hans-Joachim Bungartz and his group at Technische
Universit\"at M\"unchen who provided the longest-term environment for the
development of this code.
\newB{The paper is dedicated to Christoph Zenger} who \newB{kicked off and
supervised the implementation of the first generation of Peano.}
% 
% came up with the
% fundamental ideas concerning the space-filling curve and the usage of a stack-based
% automaton.
% 
% The trigger to write the present paper has been the H2020 project ExaHyPE
% \cite{Software:ExaHyPE} for which Peano has become a code base.
% The author thus appreciates on the one hand support received from the European
% Union’s Horizon 2020 research and innovation programme under grant agreement No 671698 (ExaHyPE).
% On the other hand, the feedback from the involved consortium members is highly
% appreciated.

  }{
    \appendix
    
    \section{Examples on statements from paper}

This section comprises \newA{a} few examples illustrating statements from the
manuscript.
To compact the presentation, examples stick to bipartitioning ($k=2$) and two
dimensions ($d=2$) if not stated otherwise.
Most examples illustrate concepts with the help of Figure
\ref{figure:spacetrees:tree-vs-grid}.
\newA{
 The examples clarify statements from the text but also clarify that most
 manuscript concepts are not tied to tripartitioning which is used in Peano's
 code base.
}

 \begin{figure}[htb]
   \begin{center}
   \includegraphics[width=.7\textwidth]{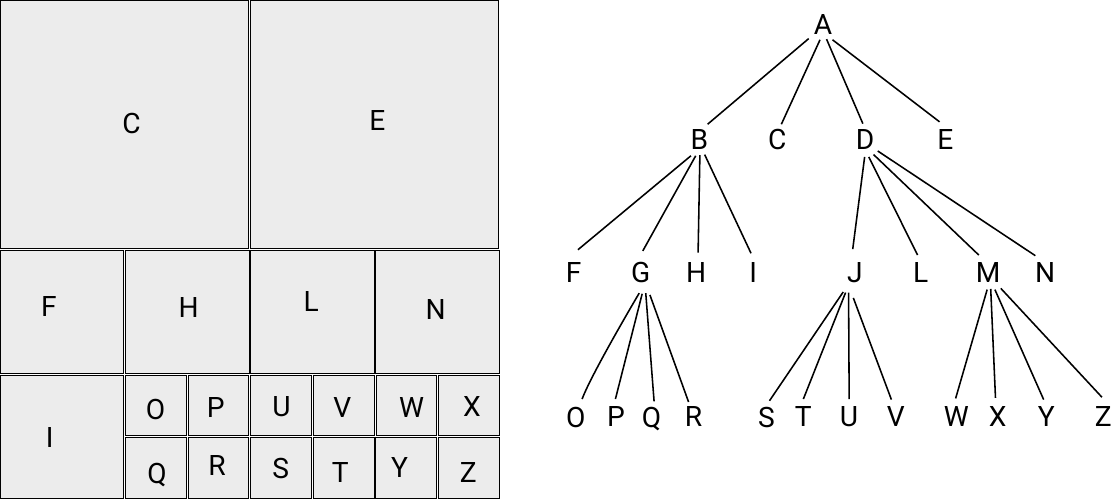}
   \caption{
     Illustration of an adaptive Cartesian spacetree grid for $d=2, k=2$ (left)
     with its tree (right).
     The labels do not follow any particular SFC but are arbitrarily chosen
     along a DFS.
     \label{figure:spacetrees:tree-vs-grid}
   }
   \end{center}
 \end{figure}

Section 2
%\ref{section:spacetree} 
picks up classic tree enumerations applied here
to adaptive Cartesian grids. 
We illustrate some classic textbook orderings with the help of the example tree.

\begin{example}
   For the spacetree in Figure \ref{figure:spacetrees:tree-vs-grid}, we 
   write down following orderings:
   \begin{itemize}
     \item Depth-first, Morton/z-curve ordering: \\ 
     \texttt{A,B,I,G,Q,R,O,P,F,H,D,J,S,T,U,V,M,Y,Z,W,X,L,N,C,E} 
     \item Breadth-first, Morton/z-curve ordering: \\
     \texttt{A,B,D,C,E,I,G,F,H,J,M,L,N,Q,R,O,P,S,T,U,V,Y,Z,W,X}
     \item Depth-first, Hilbert ordering: \\
     \texttt{A,B,I,G,Q,R,P,O,H,F,C,E,D,N,L,J,V,U,S,T,M,Y,W,X,Z}
     \item Breadth-first, Hilbert ordering: \\
     \texttt{A,B,C,E,D,I,G,H,F,N,L,J,M,Q,R,P,O,V,U,S,T,Y,W,X,Z}
   \end{itemize}
\end{example}

\noindent
The core spacetree enumeration in Peano is never computed explicitly.
Instead, all numbering is given implicitly by the ordering within the streams.
We nevertheless quickly reiterate explicit tree orderings, as they help to
digest the \newA{linearization} discussions and can be of help to study heap
storage techniques where the spatial plus level position of a grid entity acts as
preimage to the hashing function.
\newA{A (historical) overview of SFC numberings, related publications and their
relation to text replacement systems can be found in \cite{Bader:13:SFCs}, e.g.}

\begin{example}
  \label{example:spacetrees:DFS-codes}
  For the spacetree in Figure \ref{figure:spacetrees:tree-vs-grid} and
  depth-first Morton order, we study the cell $P$. In a binary basis, it
  is encoded by \newA{\texttt{00|01|11} (use the first cell (00) on level 1,
  then descend into the second (01) cell along the SFC on level 2; finally
  pick the third cell on level 3 within the cell chosen before)} and each digit
  $d$-tuple describes a subcell.
  We can derive a child's position within its parent from the tuple's $d$
  entries.
  Furthermore, the code allows us both to determine the
  cell's spatial position and size within the grid as well as neighbour
  relations, i.e.~to look up a neighbour on any level, we can manipulate the
  code and directly check whether a cell with such a code exists.
  For $k=3$, we have to choose a ternary base.
\end{example}

 \begin{figure}[htb]
   \begin{center}
   \includegraphics[width=.8\textwidth]{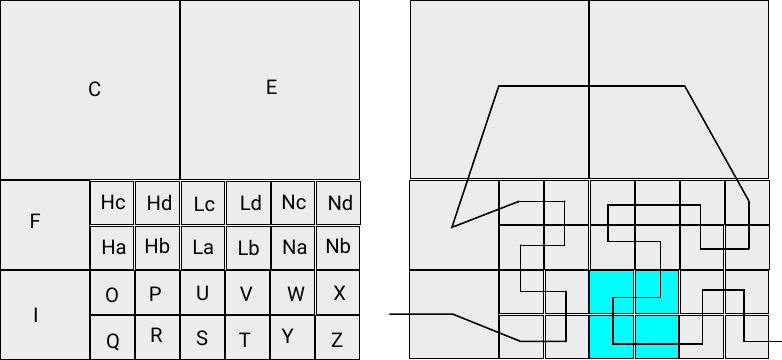}
   \caption{
     Left:
     Illustration of a non-balanced tree where cells (\texttt{Lc} and
     \texttt{E}, e.g.) are adjacent that differ by more than one level.
     Right: 
     Traversal of the grid along the Hilbert SFC.
   }
   \end{center}
   \label{figure:spacetrees:non-balanced-and-hanging}
 \end{figure}

\noindent
Section 3 
%\ref{section:api} 
and other statements in the manuscript clarify that
Peano supports unbalanced grids (compare the 2:1-balanced grid in
Figure \ref{figure:spacetrees:tree-vs-grid} to the unbalanced grid in Figure
\ref{figure:spacetrees:non-balanced-and-hanging}).
Proper grid balancing might simplify and speed up any explicit assembly for
example as it constrains the matrix bandwidth.
Another application domain of interest are explicit hyperbolic DG solvers where
grid balancing reduces discretization errors.
Both benefit from 2:1 balancing.
Yet, there is no need to enforce it by the grid code always.
Users may switch on balancing which identifies on-the-fly whether any resolution
transition between two neighbouring cells exceeds one level of refinement and, if this is
the case, refines the coarser cell in the subsequent grid sweep.
Such a primitive approach causes a delayed rippling, i.e.~grid balancing is not
immediately \newA{enforced} but propagates successively through the grid.
If perfect balancing is mandatory, any grid modification has to be
followed by a while loop converging the grid into a balanced variant.
More sophisticated algorithms are known
\cite{Isaac:12:Balancing,Sundar:08:BalancedOctrees} yet have
not been implemented yet.
Peano offers a toolbox to inject the non-sophisticated balancing into the grid.

Adaptive Cartesian grids yield hanging nodes such as the vertex adjacent to
I,O,Q. 
As Peano makes a vertex unique through its level plus its position, also the
vertex adjacent to I,O,F,H is hanging, while its parent vertex adjacent to
F,G,H,I is not hanging.
Peano's event signature from Table \newA{\ref{table:events}} 
ensures that a pair of
\texttt{createHangingVertex} and \texttt{destroyHangingVertex} events is
triggered as least once per traversal per hanging vertex.
However, the framework is free to trigger create-destroy pairs more often,
i.e.~to remove hanging vertices temporarily throughout the grid sweep.

\begin{example}
In the following example, the tuple \newA{subscript} of indices identifies all
adjacent cells of a vertex. 
We show \newA{exemplarily} how the lifecycle of hanging
vertices integrates into \newA{(\ref{equation:order-on-events})}.
For the grid from Figure \ref{figure:spacetrees:non-balanced-and-hanging} we
obtain
\begin{eqnarray*}
  \mbox{\texttt{touchVertexFirstTime}}(v_{I,G,F,H})
    & \sqsubseteq & 
    \mbox{\texttt{enterCell}}(I) \\
  \mbox{\texttt{enterCell}}(G)
    & \sqsubseteq & 
    \mbox{\texttt{enterCell}}(O) \\
  \mbox{\texttt{touchVertexFirstTime}}(v_{I,G,F,H})
    & \sqsubseteq & 
    \mbox{\texttt{createHangingVertex}}(v_{I,O,F,H}) \\
  \mbox{\texttt{touchVertexFirstTime}}(v_{I,G,F,H})
    & \sqsubseteq & 
    \mbox{\texttt{createHangingVertex}}(v_{I,Q,O}) \\
  \mbox{\texttt{createHangingVertex}}(v_{I,Q,O}) 
    & \sqsubseteq & 
    \mbox{\texttt{enterCell}}(O) \\
  \mbox{\texttt{leaveCell}}(O)    
    & \sqsubseteq & 
    \mbox{\texttt{destroyHangingVertex}}(v_{I,Q,O}) \\
  \mbox{\texttt{destroyHangingVertex}}(v_{I,Q,O}) 
    & \sqsubseteq & 
    \mbox{\texttt{touchVertexLastTime}}(v_{I,G,F,H})  
\end{eqnarray*}
\end{example}

\noindent
The last constraint in the example is not completely describing the grid
behaviour. 
As long as no other constraint is harmed, the traversal automaton may destroy
hanging nodes (and invoke the corresponding destruction) and then on-the-fly
recreate them invoking the creational event.
Hanging nodes are never persistent.
Yet, a user always can use a hanging \newA{node's} position plus level to hold
hanging vertex data on the heap.

% \begin{example}
%   We discuss matrix-free additive and multiplicative multigrid in
%   \cite{Mehl:06:MG,Reps:16:Helmholtz,Weinzierl:17:BoxMG} and introduce a PIC
%   code in \cite{Weinzierl:16:PIC} while SPH are subject of \cite{Eckhardt:15:SPHCompression}.
%   Patch-based formalisms with ghost layers are subject of discussion in
%   \cite{Weinzierl:14:BlockFusion}, while Peano's guidebook provides simple examples also for
%   PETSc-based explicit assembly.
%   A larger project relying on Peano is the ExaHyPE code \cite{Software:ExaHyPE}
%   for which an extensive guidebook is available.
% \end{example}

The on-the-fly computation of (\ref{equation:recursion-unrolling-grammar}) 
in Section \ref{section:overhead-reduction} 
is one of the pillars of Peano's
traversal optimization, 
as it facilitates recursion unrolling.
Recursion unrolling in turn yields cascades of regular Cartesian grids that on
the one hand enable us to traverse these sub-data structures along BFS or to
outsource the regular subtrees in the memory.

\begin{example}
  Let $S,T,U,V$ from Figure \ref{figure:spacetrees:tree-vs-grid}
  each be refined
  once more.
  After one traversal, $f(J)=2$ \newA{following 
  (\ref{equation:recursion-unrolling-grammar})}.
  When the traversal automaton runs into $J$, i.e.~loads it from the input
  stream, it can load the following $\left( 2^1 \right) ^d+\left( 2^2 \right)
  ^d$ cells en block from this stream.
  A similar reasoning holds for the accompanying vertices.
%    and their data
%   embedded into the stream.
  The automaton then invokes \texttt{touchVertexFirstTime} for all vertices
  adjacent to $S,T,U,V$ that have not been touched before, it then invokes
  \texttt{enterCell} for $S,T,U,V$, it then invokes
  \texttt{touchVertexFirstTime} for all vertices adjacent to children of $S,T,U,V$, and so forth.
  Finally, the whole subtree \newA{is} piped into the output streams.
\end{example}

\noindent
Studying Figure \ref{figure:spacetrees:non-balanced-and-hanging}, we can
illustrate all statements on the boundary exchange between regular subtrees and
the spacetree.
We assume that the coloured region of cells $S,T,U,V$ plus their parent $J$ are
identified as regular subgrid and held separate from the remaining subtree.
The remaining subtree is an adaptive Cartesian spacetree grid with holes.

\begin{example}
If we have an SFC traversal of the grid as illustrated in
\ref{figure:spacetrees:non-balanced-and-hanging} and outsource the marked grid
region as regular subtree, the following properties hold:
\begin{itemize}
  \item Peano holds the vertex $v_{R,S,P,U}$ redundant, i.e.~within the holed
  spacetree\newA{, i.e.~the skeleton grid remaining after we have cut out
  regular subtrees,} and as part of the outsourced regular subtree.
  \item Within the subtree, Peano reorders the grid traversal BFS
  lexicographically ($J,S,T,U,V$) or subject to given constraints/colouring
  schemes.
  \item The consistency updates between regular subgrid and holed spacetree
  exploit the fact that the SFC orders all vertices along the interface. Let the
  holed spacetree enumerate the vertices with $v_{R,S,P,U} \sqsubseteq
  v_{P,U,Hb,La}$ and temporarily dump the vertices along this order. Once the
  traversal automaton enters the regular subtree, it is aware of the SFC and can
  read in/copy the subregion's surface vertices along this order.
\end{itemize}
\end{example}

\noindent
The exchange between regular subtrees and the holed spacetree grid here is
illustrated for the Hilbert SFC.
For $d=2$, such an illustration is straightforward.
The extension to $d=3$ is cumbersome for Hilbert.
Peano however uses the Peano SFC which has the advantageous property that the
SFC's projection onto the surface of the regular subgrid is a Peano SFC of
dimension $d-1$ \cite{Weinzierl:11:Peano}:
\newA{
 The vertex enumeration along the subdomain's boundary is continuous and
 the enumeration itself ``meanders'' over the surface again, while two adjacent
 ranks exchange all data in the same order, i.e.~no reordering is required.
}

\begin{figure}
  \begin{center}
   \includegraphics[width=0.8\textwidth]{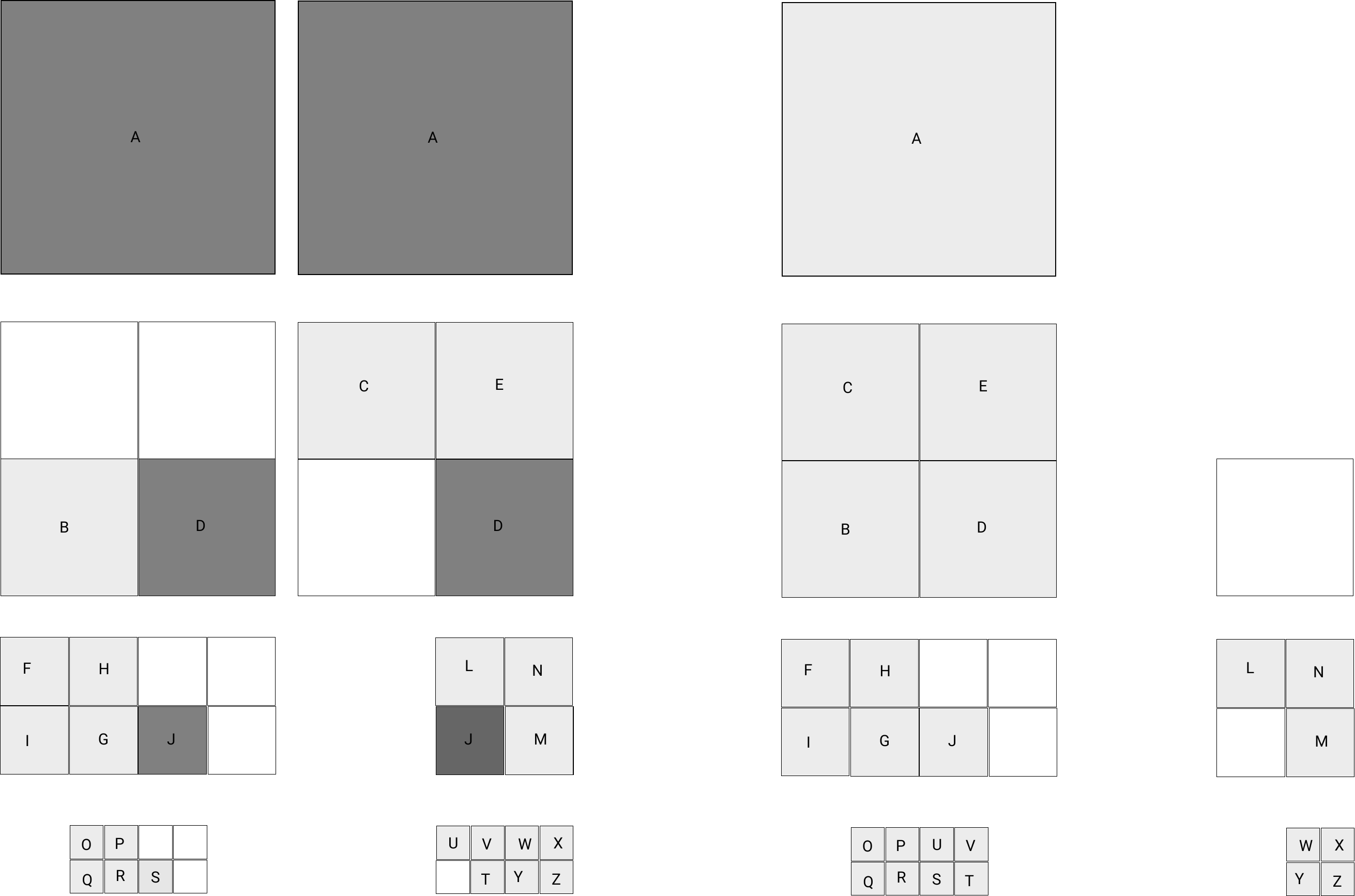}
  \end{center}
  \caption{
    Left: Bottom-up splitting of tree from Figure
    \ref{figure:spacetrees:tree-vs-grid}\newA{---a variant Peano does not
    support at the moment}. Dark cells \newA{here} are cells held redundantly
    be both ranks. Their data has to be kept consistent.
    White cells are cells not handled by the local rank.
    Their sole purpose is to make the local data structure a proper spacetree.
    Right: Top-down splitting \newA{as used by Peano} with the left rank
    being a master of the right rank. No data is held redundantly.
  }
  \label{figure:mpi:lw-dfs}
\end{figure}

We continue with statements on Section \ref{section:mpi}, 
notably with
references to other software solutions relying on a bottom-up SFC partitioning 
\cite{Akcelik:03:EarthquakesAtSC,Burstedde:11:p4est,Granding:14:ParallelHighDimAMR,Grandin:15:HighDimAMR,Griebel:98:HashStorage,Griebel:99:SFCAndMultigrid,Meister:12:Software,Sampath:08:Dendro,Sundar:08:BalancedOctrees,Sundar:12:ParallelMultigrid}.
For our examples, 
let $J,L,D \in \mathcal{T}$ with $J,L \sqsubseteq _{child\ of} D$ and $L$
handled by a worker of the rank responsible for $D$ and $J$. 
Let $D$ descend into $L$ and trigger the traversal of the subtree rooted by $L$.
If the worker-master consistency check from (\ref{equation:order-on-events})
integrates into the recursion unrolling, the rank may not continue with $J$
before the traversal of the remote automaton handling $L$ terminates
(\newA{Figure
\ref{figure:mpi:lw-dfs}}).

\begin{example}
  \label{example:mpi:one-level-unrolling}
  The Morton order from Figure \ref{figure:spacetrees:tree-vs-grid}
  
  \texttt{A,B,I,G,Q,R,O,P,F,H,D,J,S,T,U,V,M,Y,Z,W,X,L,N,C,E} 
  
  is rewritten level\newB{-wise} depth-first into

  \texttt{A,B,D,C,E,I,G,F,H,Q,R,O,P,J,M,L,N,S,T,U,V,Y,Z,W,X}.
\end{example}

\begin{example}
  \label{example:mpi:parallel-traversals}
  We study Morton orders from Figure \ref{figure:spacetrees:tree-vs-grid} on two
  processes that split up after cell
  \texttt{S}.
  The tree is processed by two ranks with a replicating DFS scheme as
  \\  
  {\footnotesize
  Rank R1: 
  \texttt{A,B,I,G,Q,R,O,P,F,H,D,J,S,T(e),U(e),V(e),M(e),L(e),N(e),C(e),E(e),sync(J,D,A)}
  \\
  Rank R2: 
  \texttt{A,B(e),D,J,S(e),T,U,V,M,Y,Z,W,X,L,N,C,E,sync(J,D,A)} 
  }
  \\
  cells with addendum \texttt{(e)} are empty, i.e.~replicated to create a valid
  octree though they hold no data.

  For a non-replicating scheme, we deploy the cells \texttt{M},\texttt{L} and
  \texttt{N} along the SFC---they are all on one level and thus preserve
  trivially the logic tree topology between the ranks---and run level\newB{-wise} DFS
  through the tree:
  \\
  {\footnotesize
  Rank R1 (master): 
  \texttt{A,B,D,C,E,I,G,F,H,Q,R,O,P,J,M(e),L(e),N(e),start-R2,S,T,U,wait-for-R2} 
  \\
  Rank R2 (worker):
  \texttt{wait-for-R1,J(e),M,L,N,Y,Z,W,X,sync-with-R1}
  }
\end{example}

\begin{example}
 \newA{
  The fine grid from Figure \ref{figure:spacetrees:tree-vs-grid} has 19 fine
  grid cells in $\Omega _h$. 
  An optimal load balancing for four ranks $r_0,r_1,r_2,r_3$ assigns four
  or five cells to each rank. 
  If we use the Hilbert ordering, a splitting could read
  \begin{eqnarray*}
    r_0 \mapsto \{I,Q,R,P,O\} && r_1 \mapsto \{H,F,C,E,N\} \\
    r_2 \mapsto \{L,V,U,S,T\} && r_3 \mapsto \{Y,W,X,Z\}. 
  \end{eqnarray*} 
  This decomposition of the tree does not yield a master-worker topology. 
 }
\end{example}

\noindent
\newA{
An admissible decomposition would be
\begin{eqnarray*}
  r_0 \mapsto \{I,H,F\}\ \mbox{plus}\ B 
  && 
  r_1 \mapsto \{Q,R,P,O\}\ \mbox{plus}\ G
  \\
  r_2 \mapsto \{N,L,V,U,S,T,Y,W,X,Z\}\ \mbox{plus}\ D 
  && 
  r_3 \mapsto \{C,E\}\ \mbox{plus}\ A
\end{eqnarray*} 
\noindent
which is not well-balanced. One solution to this is to use up to four
ranks per node, i.e.~to subdivide the problem logically further and to construct
\begin{eqnarray*}
    r_{0a} \mapsto \{A,B\} 
    \quad
    r_{1a} \mapsto \{C\} 
    &&
    r_{2a} \mapsto \{E\} 
    \quad
    r_{3a} \mapsto \{D\} 
    \\ 
    r_{1b} \mapsto \{H\} 
    &&
    r_{1c} \mapsto \{F\} 
    \\
    r_{2b} \mapsto \{N\} 
    \quad
    r_{2c} \mapsto \{L\}.
\end{eqnarray*} 
\noindent
which yields a more advantageous partition that both fits to the master-worker
paradigm on the tree, obeyes a SFC-based partitioning, yields---per
node---face-connected partitions, and is reasonably balanced.
}

%     r_{0a} \mapsto \{I\} 
%     &&
%     r_{0b} \mapsto \{Q,R,P,O\} 

%     && r_1 \mapsto \{H,F,C,E,N\} \\
%     r_0 \mapsto \{I,Q,R,P,O\} && r_1 \mapsto \{H,F,C,E,N\} \\
%     r_2 \mapsto \{L,V,U,S,T\} && r_3 \mapsto \{Y,W,X,Z\}. 

%    \input{a3_experimental-setups}
    \section*{Thanks}

Thanks are due to all the scientists and students who contributed to the
software in terms of software fragments, applications, extensions and critical
remarks.
Notably, thanks are due to Hans-Joachim Bungartz and his group at Technische
Universit\"at M\"unchen who provided the longest-term environment for the
development of this code.
\newB{The paper is dedicated to Christoph Zenger} who \newB{kicked off and
supervised the implementation of the first generation of Peano.}
% 
% came up with the
% fundamental ideas concerning the space-filling curve and the usage of a stack-based
% automaton.
% 
% The trigger to write the present paper has been the H2020 project ExaHyPE
% \cite{Software:ExaHyPE} for which Peano has become a code base.
% The author thus appreciates on the one hand support received from the European
% Union’s Horizon 2020 research and innovation programme under grant agreement No 671698 (ExaHyPE).
% On the other hand, the feedback from the involved consortium members is highly
% appreciated.

  }
}{
}

\ifthenelse{\boolean{arxiv}}{
  \bibliographystyle{siam}
}{}
\ifthenelse{\boolean{sisc}}{
  \bibliographystyle{siamplain}
  \footnotesize
}{}
\ifthenelse{\boolean{toms}}{
  \bibliographystyle{ACM-Reference-Format-Journals}
}{}

\bibliography{paper}

% \bibliographystyle{siam}
% \bibliography{paper}

\end{document}